\documentclass[12pt]{article}

\usepackage{jheppub}

\toccontinuoustrue
\input{packages-extra.sty}
\input{styling-extra.sty}

\title{\boldmath  A natural mechanism for approximate Higgs alignment in the 2HDM}

\author[a]{Patrick Draper}
\author[b]{Andreas Ekstedt}
\author[c]{Howard~E.~Haber}

\affiliation[a]{Department of Physics, University of Illinois, Urbana, IL 61801, USA}
\affiliation[b]{Institute of Particle and Nuclear Physics, Charles University, Prague, Czech Republic}
\affiliation[c]{Santa Cruz Institute for Particle Physics, University of California, Santa Cruz, CA 95064, USA}

\emailAdd{pdraper@illinois.edu}
\emailAdd{andreas.ekstedt@ipnp.mff.cuni.cz}
\emailAdd{haber@scipp.ucsc.edu}

\abstract{The $2$HDM possesses a neutral scalar interaction eigenstate whose tree-level properties coincide with the Standard Model (SM) Higgs boson.
In light of the LHC Higgs data which suggests that the observed Higgs boson is SM-like, it follows that the mixing of 
the SM Higgs interaction eigenstate with the other neutral scalar interaction eigenstates of the 2HDM should be suppressed, corresponding to the so-called Higgs alignment
limit. The exact Higgs alignment limit can arise naturally due to a global symmetry of the scalar potential. If this symmetry is softly broken,
then the Higgs alignment limit becomes approximate (although still potentially consistent with the current LHC Higgs data).
In this paper, we obtain the approximate Higgs alignment suggested by the LHC Higgs data as a consequence of a softly broken global symmetry
of the Higgs Lagrangian. However, this can only be accomplished if the Yukawa sector of the theory is extended. We propose an extended 
2HDM with vector-like top quark partners, where explicit mass terms in the top sector provide the source of the soft symmetry breaking of
a generalized CP symmetry. In this way, we can realize approximate Higgs alignment without a significant fine-tuning of the model parameters.
We then explore the implications of the current LHC bounds on vector-like top quark partners for the success of our proposed scenario.}

\begin{document}
\maketitle
\flushbottom

%%%%%%%%%%%%%%%%%%%%%%%%%%%%%%%%%%%%%%%%
\section{Introduction}\label{sec:intro}
%%%%%%%%%%%%%%%%%%%%%%%%%%%%%%%%%%%%%%%%I

Since the discovery of the Higgs boson at the LHC in $2012$~\cite{Aad:2012tfa,Chatrchyan:2012ufa} the ATLAS and CMS Collaborations have 
embarked on a detailed study of the properties of the Higgs bosons 
(e.g., total cross sections, differential cross sections, decay branching fractions, decay angular distributions, etc.)
in order to verify the predictions of the Standard Model (SM)
and perhaps uncover deviations from SM predictions that would require the presence of new physics beyond the SM (BSM).
After analyzing data from the Run $1$ and Run $2$ data sets, the LHC experimental collaborations have determined that 
the properties of the Higgs boson coincide with those of the SM Higgs boson to within the current accuracy of the
accumulated data, typically in the range of 10\%--20\% depending on the observable~\cite{Sirunyan:2018koj,Aad:2019mbh,CMS:2020gsy}.

One possible conclusion of the LHC experimental precision Higgs studies is that the Standard Model is confirmed and
there is no evidence for BSM physics. However, it is perhaps surprising that the fundamental theory of particles and
their interactions at the energy scale of electroweak symmetry breaking (EWSB) consists of a scalar sector that is of minimal form.
Namely, the SM Higgs boson comes from a single electroweak complex-scalar doublet that yields precisely 
one physical degree of freedom after electroweak symmetry breaking. This should be contrasted with 
the non-minimal structures inherent in a fermion sector
that consist of three generations of quarks and leptons and a gauge sector based on a direct product of three separate
gauge groups.   Having now discovered the first state of an (apparently) elementary spin $0$ scalar sector, the naive expectation
would be to anticipate a non-minimal structure here as well.  

However, one cannot simply add additional scalar bosons to the model at will, since experimental constraints limit the
structure of any extended Higgs sector.   For example, the observation of the electroweak $\rho$-parameter close to $1$
strongly suggests that the scalar sector must be comprised of electroweak doublets and perhaps singlets~\cite{Gunion:1989we}.  One of 
the simplest extensions of the SM Higgs sector posits the existence of additional electroweak scalar doublets (of the
same hypercharge as that of the SM Higgs doublet).   The two-Higgs doublet model (2HDM) provides a nontrivial
extension of the SM that introduces new physical phenomena (e.g. charged scalars and CP-odd scalars) that can be
searched for at the LHC.\footnote{Of course, extended Higgs sectors that add additional doublets or singlet scalars are also possible.
Adding additional doublets makes the analysis less tractable analytically without adding significantly new observable phenomena.
The 2HDM has also been motivated by the fact that it is a necessary part of the minimal supersymmetric extension of the
Standard Model~\cite{Fayet:1974pd,Dimopoulos:1981zb}, which has been advocated as a possible solution to the gauge hierarchy problem~\cite{Susskind:1982mw}.}
Comprehensive reviews of the 2HDM can be found in
Refs.~\cite{Gunion:1989we,Branco:2011iw}.  

Nevertheless, even the 2HDM must be constrained in light of the LHC Higgs data, since one must be able to explain why
the properties of the observed Higgs boson at the LHC is SM-like.   In any extended Higgs sector that contains at least one
complex scalar doublet (with the $\ug_{\rm Y}$ hypercharge of the SM Higgs boson), after EWSB there exists a
neutral scalar eigenstate whose properties coincide with those of the SM Higgs boson.  But, such a scalar eigenstate
will in general mix with other neutral scalar eigenstates that are present in the extended Higgs sector.  Thus, generically one
would not expect there to be a physical (mass eigenstate) neutral scalar that is SM-like, in conflict with the LHC Higgs data.

In the so-called Higgs alignment limit~\cite{Craig:2013hca,Haber:2013mia,Carena:2013ooa,Asner:2013psa}, there exists one neutral scalar mass eigenstate that is aligned with the direction of
the Higgs vacuum expectation value in field space.  This direction corresponds precisely to the interaction eigenstate with the
tree-level properties of the SM Higgs boson.  In light of the LHC Higgs data, if an extended Higgs sector
exists then the Higgs alignment limit must be approximately realized, which then implies that the mixing of
the SM Higgs interaction eigenstate with other neutral scalar mass eigenstates is suppressed.

How is this suppressed mixing realized in a realistic model?  There are two possible mechanisms.  One possibility, called
the decoupling limit~\cite{Haber:1989xc,Gunion:2002zf}, posits that all neutral scalar states (excluding the observed Higgs boson) are significantly heavier than
the scale of electroweak symmetry breaking (which can be taken to be the vacuum expectation value
of the Higgs doublet in the SM, denoted by $v\simeq 246$~\si{\giga\electronvolt}).  If the scale of the heavy scalars is $\Lambda$, then
one can formally integrate out these states below the scale $\Lambda$, which results in an effective theory corresponding
to the SM with one Higgs doublet.  Deviations from SM-like behavior of the observed Higgs boson would be of $\mathcal{O}(v^2/\Lambda^2)$, 
which are consistent with the observed Higgs data if $\Lambda$ is sufficiently large.  Of course, in this scenario it might be very challenging
to discover experimental evidence for the presence of the heavier scalars at the LHC.  In particular, if $\Lambda$ is sufficiently large then it may not
be possible to discover such heavy scalars above SM backgrounds.

A second possibility is to simply fine-tune the parameters of the 2HDM in such a way that the mixing of
the SM Higgs interaction eigenstate with other neutral scalar mass eigenstates is suppressed at the level required by the LHC Higgs data.
This can always be done, and allows for the possibility of new scalar states whose masses are not significantly larger than that of the
observed Higgs boson, thereby presenting opportunities in future LHC runs for their discovery.   However, the arbitrary fine-tuning
required to achieve this scenario seems completely ad hoc and is not particularly appealing from a theoretical point of view.

In this paper, we will consider a third possibility in which the Higgs alignment limit is realized as the result of a symmetry.   The simplest
example of such a scenario is known as the inert doublet model (IDM)~\cite{Ma:2006km,Barbieri:2006dq}, in which a second complex scalar doublet is added to
the SM that is odd under a discrete $\mathbb{Z}_2$ symmetry, whereas all SM fields are $\mathbb{Z}_2$-even. 
It follows that the Higgs alignment limit is
exactly realized, since the $\mathbb{Z}_2$ symmetry forbids the mixing of the first Higgs doublet (which contains the SM Higgs field)
and the second Higgs doublet. Consequently, the tree-level properties of the neutral CP-even scalar field that resides inside the
first Higgs doublet coincides precisely with those of the SM Higgs boson.  In practice, the observed Higgs boson of this model
deviates from the SM Higgs boson in its loop induced properties.  For example, the amplitude for $h(125)\to\gamma\gamma$
would include contributions from a loop of charged Higgs bosons.
However, such corrections are typically too small to be seen in the present Higgs data, and could very well lie beyond the reach of
the precision Higgs program at the LHC.  

If deviations from SM-like Higgs properties are revealed in future experimental Higgs studies, then one would conclude that the 
Higgs alignment limit is only approximately realized.  In this case, a natural explanation for the observed SM-like Higgs boson could be 
attributed to an approximate symmetry.   In such a case, if the symmetry breaking is soft (generated by dimension two or three
terms in the Lagrangian), then the deviations from SM-like Higgs behavior would be naturally small.  In contrast, if the 
symmetry breaking is hard then one can only ensure small deviations from SM-like Higgs behavior by fine-tuning the size of the
hard symmetry breaking terms of the Lagrangian.
In the case of the IDM, it is not possible to break the $\mathbb{Z}_2$ symmetry softly, since a $\mathbb{Z}_2$-breaking squared mass term of the Higgs potential must be accompanied by a hard $\mathbb{Z}_2$-breaking
dimension-four parameter of the scalar potential due to the scalar potential minimum conditions.

Thus, our primary goal in this paper is to introduce a global symmetry beyond that of the IDM that can be softly broken in order
to provide a natural explanation for
approximate Higgs alignment.  There exist a number of possible global symmetries
that can be imposed on the scalar potential of the 2HDM that enforce the exact Higgs alignment limit~\cite{Dev:2014yca,Pilaftsis:2016erj}.  However
(with the exception of the IDM), it is not possible to extend these symmetries to the Yukawa Lagrangian that describes the
interactions of the scalars with the quarks and leptons.  That is, the Yukawa Lagrangian, which consists of dimension-four terms
(and dimensionless couplings) constitutes a hard breaking of the global symmetry that is imposed to yield exact Higgs
alignment.   This means that it is not possible to naturally preserve the global symmetry in the scalar potential.   
The authors of Ref.~\cite{Dev:2014yca,Pilaftsis:2016erj} proposed that the global symmetry of the scalar potential is exactly realized at a very high energy scale (e.g., the Planck scale), and
assumed that some unknown dynamics is responsible for generating the symmetry breaking Yukawa interactions at the same scale.
Then, they employed renormalization group (RG) evolution of the model parameters from the high energy scale down to the low energy scale to determine the 
effective 2HDM parameters at the electroweak scale.   Thus, RG evolution generates a departure from the Higgs alignment limit,
which can then be compared with the properties of the Higgs boson that are measured at the LHC.

Our strategy is different and is inspired by the work of Ref.~\cite{Draper:2016cag}, which proposed to extend
the Yukawa sector by adding vector-like top partners.\footnote{A more complete model would
introduce vector-like partners for all quarks and leptons.  But, we shall demonstrate that the effect of the top partners dominates, so one
can simplify the analysis by focusing on top partners alone.}
The motivation of Ref.~\cite{Draper:2016cag} was to construct a 2HDM in which no additional fine-tuning was required beyond the
one fine-tuning of the SM that sets the scale of EWSB.   In this work, we have repurposed this idea to provide a natural explanation
for approximate Higgs alignment.  As in Ref.~\cite{Draper:2016cag}, the addition of the vector-like top partners
allows us to extend the global symmetry transformation laws imposed on the scalar potential to the Yukawa sector.  At this stage,
the Higgs alignment would be exact as it is protected by the global symmetry.   However, the masses of the vector-like top partners that are generated by EWSB
would yield top partners with masses that are easily excluded by LHC searches.   To avoid this problem, we add gauge invariant dimension-three terms to the 
Yukawa Lagrangian that generate additional contributions to the masses of the vector-like top partners that are sufficiently large to avoid 
the limits on vector-like quark masses deduced from LHC searches.  Such terms necessarily provide a soft breaking to the global symmetry
and thus will generate deviations from the exact Higgs alignment limit.  Nevertheless, the soft nature of the symmetry breaking allows for the possibility
that the deviations from exact alignment are in a range consistent with the present LHC Higgs data.

The model that we describe is not ultraviolet complete.   Thus, we imagine that there is an ultraviolet (UV) cutoff scale $\Lambda_c$ that is well above the TeV scale.
The physics that lies above this scale is ultimately responsible for generating the symmetry-breaking dimension-three terms that appear in the Yukawa
Lagrangian.  In order to avoid excessive fine-tuning, $\Lambda_c$ cannot be arbitrarily large.   In addition, the mass terms are assumed to be large enough to avoid the LHC limits on top quark partner masses while small compared to $\Lambda_c$
to ensure the validity of the effective theory that includes the top quark partners.   By imposing limits on the amount of fine-tuning that we are
prepared to tolerate, we can obtain an upper limit on the top quark partner masses.   Hence, the goal of our analysis is to map out the
region of parameter space in which the deviations from the Higgs alignment limit and the absence of observed top quark partners is
consistent with LHC data with a requirement of at most a moderate of fine-tuning of model parameters.   If this program is successful,
it would provide a correlation between the predicted deviation from SM-like Higgs behavior and the masses of top quark partners that
could be revealed in future runs at the LHC.

In \sect{modelreview}, we begin with a brief review of the theoretical structure of the 2HDM.  The enhanced global symmetries of 
2HDM are enumerated, and we identify those symmetries that ensure the exact Higgs alignment limit.
Two possible generalized CP symmetries of the $2$HDM~\cite{Ferreira:2009wh} (denoted as GCP$2$ and GCP$3$) provide 
compelling models for exact Higgs alignment.   Since we anticipate that these symmetries will be softly broken, we 
also include  soft-symmetry-breaking squared-mass terms in the scalar potential.  In general
the softly-broken GCP$2$ scalar potential includes CP-violating effects in the scalar sector, whereas a softly-broken GCP$3$ scalar
potential is CP-invariant.   Thus, in order to simplify our analysis, we focus on the softly-broken GCP$3$ scalar potential for the remainder of
the paper.

Details of the softly-broken GCP$3$-symmetric 2HDM scalar potential are provided in \sect{gcpthree}.   By an appropriate
change of the scalar field basis (details are relegated to Appendix~\ref{appequiv}), the 
dimension-four terms of the scalar potential when expressed in terms of the new basis fields is invariant under a direct product of
a Peccei-Quinn $\ug$ global symmetry~\cite{Peccei:1977ur} and a $\mathbb{Z}_2$ symmetry~\cite{Fayet:1974fj}.  
Our analysis simplifies considerably in this new basis, so all
results are henceforth presented under the assumption of a softly-broken $\ug\otimes\mathbb{Z}_2$-symmetric $2$HDM scalar potential.

In \sect{cp3yukawa}, the softly-broken $\ug\otimes\mathbb{Z}_2$ symmetry is extended to the Yukawa sector by introducing
a vector-like top quark partner.   Due to mixing between the interaction eigenstate top quark and partners, one must 
determine the appropriate mass eigenstates of the top sector.  This is accomplished by performing a singular value decomposition of
a real $2\times 2$ matrix (details of which are provided in Appendix~\ref{app:svd}).  The computation is performed in two steps, where
EWSB effects are only taken into account in the second step.  (Of course, one can derive the same result in one single step as outlined
in Appendix~\ref{twostep}.)   Using the soft masses introduced in the Yukawa sector, we estimate the magnitudes of the squared-mass
parameters of the scalar potential that softly break the $\ug\otimes\mathbb{Z}_2$ symmetry, and we discuss the implications for
the degree of fine-tuning that is associated with the soft symmetry breaking effects.

Finally in \sect{analysis}, we survey the parameter space of 
our model and identify those parameter regimes that are 
consistent with the LHC Higgs data, the searches for non-SM-like neutral Higgs scalars and charged Higgs scalars, and the
searches for vector-like top quarks.   Conclusions of this work are presented in \sect{conclusions}.

%%%%%%%%%%%%%%%%%%%%%%%%%%%%%%%%%%%%%%%%
\section{The scalar sector of the \texorpdfstring{$2$}{2}HDM}
\label{modelreview}
%%%%%%%%%%%%%%%%%%%%%%%%%%%%%%%%%%%%%%%%

%--------------------------------------------------------------------------
\subsection{The \texorpdfstring{$2$}{2}HDM scalar potential}
\label{sec:higgsscalarpotential}
%--------------------------------------------------------------------------

Let $\Phi_1$ and $\Phi_2$ denote two complex hypercharge $Y=1$, SU($2$)$\ls{L}$ doublet scalar fields.
The most general gauge invariant renormalizable scalar potential is given by
\beqa
\mathcal{V}&=& m_{11}^2\Phi_1^\dagger\Phi_1+m_{22}^2\Phi_2^\dagger\Phi_2
-[m_{12}^2\Phi_1^\dagger\Phi_2+{\rm h.c.}]
 +\half\lambda_1(\Phi_1^\dagger\Phi_1)^2
+\half\lambda_2(\Phi_2^\dagger\Phi_2)^2
+\lambda_3(\Phi_1^\dagger\Phi_1)(\Phi_2^\dagger\Phi_2)\nonumber\\[8pt]
&&\qquad\qquad\,\,
+\lambda_4(\Phi_1^\dagger\Phi_2)(\Phi_2^\dagger\Phi_1)
+\left\{\half\lambda_5(\Phi_1^\dagger\Phi_2)^2
+\big[\lambda_6(\Phi_1^\dagger\Phi_1)
+\lambda_7(\Phi_2^\dagger\Phi_2)\big]
\Phi_1^\dagger\Phi_2+{\rm h.c.}\right\}\!. \label{lambdapotential}
\eeqa
In general, $m_{12}^2$, $\lambda_5$, $\lambda_6$ and $\lambda_7$ can be complex.
In order to avoid tree-level Higgs-mediated flavor changing neutral currents (FCNCs), we shall
impose a Type I, II, X and Y structure on the Higgs-quark and the Higgs-lepton interactions~\cite{Hall:1981bc,Barger:1989fj,Aoki:2009ha}.
These four types of Yukawa couplings can
be naturally implemented~\cite{Glashow:1976nt,Paschos:1976ay} by imposing a softly-broken
$\mathbb{Z}_2$ symmetry, $\Phi_1\to +\Phi_1$ and $\Phi_2\to -\Phi_2$, which implies that $\lambda_6=\lambda_7=0$, whereas $m_{12}^2\neq 0$ is 
allowed.\footnote{The absence of
tree-level Higgs-mediated FCNCs is maintained in the presence of a soft breaking of the $\mathbb{Z}_2$ symmetry (due to $m_{12}^2\neq 0$), and the FCNC effects generated at one loop
are small enough to be consistent with phenomenological constraints 
over a significant fraction of the $2$HDM parameter space~\cite{Haisch:2008ar,Mahmoudi:2009zx,Gupta:2009wn,Arbey:2017gmh}.}
In this basis of scalar doublet fields (denoted as the $\mathbb{Z}_2$-basis),
the discrete $\mathbb{Z}_2$ symmetry of the quartic terms of \eq{lambdapotential} is manifest.
The scalar fields can then be rephased such that
$\lambda_5$ is real, which leaves $m_{12}^2$ as the only potential complex parameter of the scalar potential.

The scalar fields will
develop non-zero vacuum expectation values (vevs) if the Higgs mass matrix
$m_{ij}^2$ has at least one negative eigenvalue.
Moreover, we assume that only the neutral Higgs fields acquire
non-zero vevs, i.e.~the scalar potential does not admit the possibility
of stable charge-breaking minima~\cite{Barroso:2005sm,Ivanov:2006yq}.
Then, the doublet scalar field vevs
are of the form
\beq
\langle \Phi_1 \rangle=\frac{v}{\sqrt{2}} \left(
\begin{array}{c} 0\\ c_\beta \end{array}\right), \qquad \langle
\Phi_2\rangle=
\frac{v}{\sqrt{2}}\left(\begin{array}{c}0\\  e^{i\xi} s_\beta
\end{array}\right)\,,\label{potmin}
\eeq
where $c_\beta\equiv\cos\beta=v_1/v$, $s_\beta\equiv\sin\beta=v_2/v$ and $v^2\equiv v_1^2+v_2^2\simeq (246~{\rm GeV})^2$.   By convention we take
$0\leq\beta\leq\half\pi$ and $0\leq\xi<2\pi$.   

The parameters $v$, $\beta$ and $\xi$ (or equivalently, $v_1$, $v_2$ and $\xi$) are determined by minimizing the scalar potential.   The minimization conditions in the case of $\lambda_6=\lambda_7=0$ and real $\lambda_5$ are 
given by,
\beqa
m_{11}^2 v_1&=& \Re(m_{12}^2 e^{i\xi})v_2-\half\lambda_1 v_1^3-\half\lambda_{345} v_1 v_2^2\,, \\
m_{22}^2 v_2&=& \Re(m_{12}^2 e^{i\xi})v_1-\half\lambda_2 v_2^3-\half\lambda_{345} v_2 v_1^2\,, \\
\Im(m_{12}^2 e^{i\xi})v_1&=& \half\lambda_5 v_1^2 v_2\sin 2\xi\,, \\
\Im(m_{12}^2 e^{i\xi})v_2&=& \half\lambda_5 v_2^2 v_1\sin 2\xi\,, 
\eeqa
where
\beq \label{threefourfive}
\lambda_{345}\equiv\lambda_3+\lambda_4+\lambda_5\cos 2\xi\,.
\eeq
Assuming that $v_1\neq 0$ and $v_2\neq 0$, the minimization conditions simplify to,
\beqa
m_{11}^2 &=& \Re(m_{12}^2 e^{i\xi})\tan\beta-\half\lambda_1 v^2 c^2_\beta-\half\lambda_{345} v^2 s^2_\beta\,,\label{min1} \\
m_{22}^2 &=& \Re(m_{12}^2 e^{i\xi})\cot\beta-\half\lambda_2 v^2  s_\beta^2-\half\lambda_{345} v^2 c_\beta^2\,, \label{min2} \\
\Im(m_{12}^2 e^{i\xi})&=& \half\lambda_5 v^2 s_\beta c_\beta\sin 2\xi\,.\label{min3}
\eeqa

In contrast, if one of the two vevs vanishes, then the minimization conditions are 
\beqa
m_{12}^2 &=&0\,, \qquad m_{22}^2=-\half \lambda_2 v^2\,,\qquad \text{if $v_1=0$ and $v_2=v$}, \label{inert1} \\
m_{12}^2 &=&0\,, \qquad m_{11}^2=-\half \lambda_1 v^2\,,\qquad \text{if $v_2=0$ and $v_1=v$}.\label{inert2}
\eeqa
 
Of the original eight scalar degrees of freedom, three Goldstone
bosons ($G^\pm$ and~$\go$) are absorbed (``eaten'') by the $W^\pm$ and
$Z$.  The remaining five physical Higgs particles are: three neutral scalars ($h_1$, $h_2$ and $h_3$)
and a charged Higgs pair ($\hpm$).  If CP is conserved in the scalar sector, then the neutral scalars consist of  two CP-even
scalars ($\hl$ and $\hh$) and one CP-odd scalar
($\ha$).   It is straightforward to identify the scalar mass eigenstates and their interactions.   In general, none of the neutral scalars will possess
the properties of the Standard Model (SM) Higgs boson, due to mixing of the would-be SM Higgs state with the additional neutral scalar degrees
of freedom.

As discussed in \sect{sec:intro}, we seek a symmetry beyond the $\mathbb{Z}_2$ symmetry already imposed above in order to 
provide a natural explanation for the approximate Higgs alignment observed in the LHC Higgs data.
In particular, we shall employ an approximate symmetry by allowing the symmetry to be softly broken by mass terms in the scalar potential.

We begin by considering the possible enhanced symmetries of the scalar potential.   
It will be convenient to analyze the scalar potential in the Higgs basis~\cite{Donoghue:1978cj,Georgi:1977gs,Botella:1994cs,Branco:1999fs,Davidson:2005cw,Haber:2006ue}, 
which is introduced in the 
next subsection.

%--------------------------------------------------------------------------
\subsection{Enhanced symmetries of the \texorpdfstring{$2$}{2}HDM scalar potential} 
%--------------------------------------------------------------------------
\label{sec:enhanced}

The scalar potential given in \eq{lambdapotential} is expressed in the $\mathbb{Z}_2$-basis of scalar doublet fields in which the $\mathbb{Z}_2$ discrete symmetry of the quartic terms is manifest.  
It will prove convenient to re-express the scalar doublet fields in terms of Higgs basis fields $H_1$ and $H_2$, which are defined 
by the linear combinations of $\Phi_1$ and $\Phi_2$ such that $\langle H_1^0\rangle=v/\sqrt{2}$ and $\langle H_2^0\rangle=0$.  That is,
\beq \label{invhiggs}
H_1\equiv c_\beta\Phi_1+s_\beta e^{-i\xi}\Phi_2\,,\qquad\quad H_2=e^{i\eta}\bigl[-s_\beta e^{i\xi}\Phi_1+c_\beta \Phi_2\bigr]\,,
\eeq
where $e^{i\eta}$ accounts for the fact that Higgs basis is not unique since one is always free to rephase the Higgs basis field $H_2$~\cite{Boto:2020wyf}.
In terms of the Higgs basis fields defined in \eq{invhiggs}, the scalar potential is given by, 
 \beqa
 \mathcal{V}&=& Y_1 H_1^\dagger H_1+ Y_2 H_2^\dagger H_2 +[Y_3 e^{-i\eta}
H_1^\dagger H_2+{\rm H.c.}] \nonumber 
\\
&&\quad 
+\half Z_1(H_1^\dagger H_1)^2+\half Z_2(H_2^\dagger H_2)^2
+Z_3(H_1^\dagger H_1)(H_2^\dagger H_2)
+Z_4(H_1^\dagger H_2)(H_2^\dagger H_1) \nonumber  \\
&&\quad
+\left\{\half Z_5 e^{-2i\eta}(H_1^\dagger H_2)^2 +\big[Z_6 e^{-i\eta} (H_1^\dagger
H_1) +Z_7 e^{-i\eta} H_2^\dagger H_2)\big] H_1^\dagger H_2+{\rm
h.c.}\right\}\,.\label{higgspot}
\eeqa
The scalar potential minimum conditions are,
\beq \label{YZ}
Y_1=-\half Z_1 v^2\,,\qquad\quad Y_3=-\half Z_6 v^2\,.
\eeq
The charged Higgs mass is given by,
\beq
m^2_{H^\pm}=Y_2+\half Z_3 v^2=
\frac{2\Re(m_{12}^{2}e^{i\xi})}{s_{2\beta}}
-\half v^2(\lambda_4+\lambda_5\cos 2\xi)\,.
\eeq

The squared-masses of the neutral Higgs bosons are given by the eigenvalues of
the neutral Higgs squared mass matrix, which is presented with respect to the neutral scalar field basis, $\{\sqrt{2}\Re H_1^0-v\,,\,\Re H_2^0\,,\,\Im H_2^0\}$,
\beq  \label{matrix33}
\mathcal{M}^2=v^2\left( \begin{array}{ccc}
Z_1&\quad \Re(Z_6 e^{-i\eta}) &\quad -\Im(Z_6 e^{-i\eta})\\
\Re(Z_6 e^{-i\eta})  &\quad \half\bigl[Z_{34}+\Re(Z_5 e^{-2i\eta})\bigr]+Y_2/v^2 & \quad
- \half \Im(Z_5  e^{-2i\eta})\\ -\Im(Z_6  e^{-i\eta}) &\quad - \half \Im(Z_5  e^{-2i\eta}) &\quad
\half\bigl[Z_{34}-\Re(Z_5 e^{-2i\eta})\bigr]+Y_2/v^2\end{array}\right),
\eeq
where $Z_{34}\equiv Z_3+Z_4$.
The would-be SM Higgs state is $h_{\rm SM}\equiv \sqrt{2}\Re H_1^0-v$. The Higgs alignment limit then corresponds to $Z_6=0$, in which case the mixing
of $h_{\rm SM}$ with $\Re H_2^0$ and $\Im H_2^0$ is completely absent.  The tree-level properties of $h_{\rm SM}$ then coincide with those of the SM Higgs boson.

It is straightforward to compute the corresponding Higgs basis parameters in terms of the parameters of \eq{lambdapotential}.
The $Y_i$ are given by,
\beqa
Y_1&=& m_{11}^2 c_\beta^2+m_{22}^2 s_\beta^2-\Re(m_{12}^2 e^{i\xi})s_{2\beta}\,,\\
Y_2&=& m_{11}^2 s_\beta^2+m_{22}^2 c_\beta^2+\Re(m_{12}^2 e^{i\xi})s_{2\beta}\,,\\
Y_3&=&\bigl[\half(m_{22}^2-m_{11}^2)s_{2\beta}-\Re(m_{12}^2 e^{i\xi})c_{2\beta}-i\Im(m_{12}^2 e^{i\xi})\bigr]e^{-i\xi}\,.\label{why3}
\eeqa
In light of \eq{YZ}, the Higgs alignment limit is realized if $Y_3=0$.   One way of satisfying $Y_3=0$ is to set $m_{12}^2=0$, in which case one must also require that either
$s_{2\beta}=0$ or $m_{11}^2=m_{22}^2$.   
Note that the condition $m_{12}^2=0$ is enforced if the $\mathbb{Z}_2$ symmetry imposed above is unbroken.  
If $s_{2\beta}=0$, then the $\mathbb{Z}_2$ symmetry is unbroken by the vacuum. This case yields the inert doublet model (IDM), which
is known to possess a neutral scalar state with the tree-level properties of the SM Higgs boson.
Although the IDM is consistent with the LHC Higgs data over a significant part of its parameter space, one cannot break the  $\mathbb{Z}_2$ softly since
$Y_3\neq 0$ would yield $Z_6\neq 0$ due to \eq{YZ} and would thus constitute a hard breaking of the $\mathbb{Z}_2$ symmetry.
The alternative is to assume that $s_{2\beta}\neq 0$ and instead 
impose $m_{11}^2=m_{22}^2$, which
requires an enhanced symmetry of the scalar potential.

The enhanced symmetries of the 2HDM have been classified in Refs.~\cite{Ivanov:2007de,Ferreira:2009wh,Ferreira:2010hy,Ferreira:2010yh,Battye:2011jj}.
Starting from a generic $\Phi_1$--$\Phi_2$ basis, 
these symmetries fall into two separate categories: (i) Higgs family symmetries of the form $\Phi_a\to U_{ab}\Phi_b$, and (ii) Generalized CP (GCP) symmetries of the form  $\Phi_a\to U_{ab}\Phi_b^*$, where $U$ resides in a subgroup (either discrete or continuous) of U($2$).  Although it appears that the number of possible choices for symmetries is quite large, it turns out that in many cases, different choices of $U$ yield the same constraints on the 2HDM scalar potential parameters.  

Note that the gauge covariant kinetic energy terms of the scalar fields are invariant under the 
full global U($2$) Higgs family symmetry transformation.  Moreover, the scalar potential is invariant under a global hypercharge transformation, $\ug_{\rm Y}$, which is a subgroup of U($2$).  Thus, any enhanced Higgs family symmetries that are respected by the scalar potential would be a subset of the U($2$) transformations that are orthogonal to $\ug_{\rm Y}$.    In Tables~\ref{tab:symm1} and \ref{tab:symm2}, we summarize the possible discrete and continuous Higgs family symmetries modulo the
$\ug_Y$ hypercharge symmetry that can impose constraints on the 2HDM scalar potential.
Note that the list of symmetries in Table~\ref{tab:symm1} contains a redundancy. It may appear that the $\mathbb{Z}_2$ and $\Pi_2$ discrete symmetries are distinct (as they yield different constraints on the 2HDM scalar potential parameters in the $\Phi_1$--$\Phi_2$ basis).
Nevertheless,  
starting from the scalar potential of a $\Pi_2$-symmetric 2HDM, one can find a different basis of scalar fields in which the corresponding scalar potential manifestly exhibits the $\mathbb{Z}_2$ symmetry, and vice versa~\cite{Davidson:2005cw}.  In Table~\ref{tab:class},  the constraints of the various possible Higgs family 
symmetries and  GCP symmetries on the 2HDM scalar potential in a  generic $\Phi_1$--$\Phi_2$ 
basis are exhibited.

\begin{table}[b!]
\begin{tabular}{|ll|}
\hline
symmetry &\phantom{xxxxx} transformation law \\
\hline
$\mathbb{Z}_2$ &
$\Phi_1 \rightarrow \Phi_1$,
\hspace{7.5ex}
$\Phi_2 \rightarrow  -\Phi_2$ \\
$\Pi_2$\,\,\, ({\rm mirror symmetry})
&
$\Phi_1 \longleftrightarrow \Phi_2$ \\
$\ug$\,\,\,({\rm Peccei-Quinn symmetry~\cite{Peccei:1977ur}}) & 
$\Phi_1 \rightarrow e^{-i \theta} \Phi_1$,
\hspace{4ex}
$\Phi_2 \rightarrow e^{i \theta} \Phi_2$ \\
$\mathrm{SO}(3)$\,\,\,({\rm maximal Higgs flavor symmetry)} & $\Phi_a\to U_{ab}\Phi_b$\,,\qquad 
\,\, $\mathrm{U}\in{\rm \mathrm{U}(2)}/{\rm \ug}_Y$ \\
\hline
\end{tabular}
\caption{Classification of 2HDM scalar potential Higgs family symmetries in a generic $\Phi_1$--$\Phi_2$ basis~\cite{Ivanov:2007de,Ferreira:2009wh,Ferreira:2010hy,Ferreira:2010yh,Battye:2011jj}.  The corresponding constraints on the 2HDM scalar potential parameters are shown in Table~\ref{tab:class}. \\[-20pt]
\label{tab:symm1}}
\end{table}
\begin{table}[t!]
\begin{tabular}{|ll|}
\hline
symmetry & \phantom{xxxxx}transformation law \\
\hline
GCP$1$ &
$\Phi_1 \rightarrow \Phi_1^*$,
\hspace{7.5ex}
$\Phi_2 \rightarrow \Phi_2^*$ \\
GCP$2$ &
$\Phi_1 \rightarrow \Phi_2^*$,
\hspace{7.5ex}
$\Phi_2 \rightarrow -\Phi_1^*$ \\
GCP$3$ &
$\begin{cases} \Phi_1 \rightarrow \Phi_1^*\cos\theta+\Phi_2^*\sin\theta, & \\
\Phi_2 \rightarrow -\Phi_1^*\sin\theta+\Phi_2^*\cos\theta \end{cases}$,\qquad \text{for $0<\theta<\half\pi$}. \\
\hline
\end{tabular}
\caption{Classification of 2HDM scalar potential generalized CP (GCP) symmetries in a generic $\Phi_1$--$\Phi_2$ basis~\cite{Ivanov:2007de,Ferreira:2009wh,Ferreira:2010hy,Ferreira:2010yh,Battye:2011jj}.  Note that a GCP$3$ symmetry with \text{any} value of $\theta$ that lies between $0$ and $\half\pi$ yields the same constrained 2HDM scalar potential.   The corresponding constraints on the 2HDM scalar potential parameters are shown in Table~\ref{tab:class}. \\[-20pt]
\label{tab:symm2}}
\end{table}
\begin{table}[H]
\begin{tabular}{|ccccccccccc|}
\hline
symmetry &  $m_{11}^2$ & $m_{22}^2$ & $m_{12}^2$ & $\lambda_1$ &
 $\lambda_2$ & $\lambda_3$ & $\lambda_4$ &
$\lambda_5$ & $\lambda_6$ & $\lambda_7$ \\
\hline
$\mathbb{Z}_2$ & &   & $0$
   &  &  &  & &
   & $0$ & $0$ \\
$\Pi_2$  &  &$ m_{11}^2$ & real &&
    $ \lambda_1$ & &  &  
real &  & $\lambda_6^\ast$
\\
$\mathbb{Z}_2\otimes\Pi_2$ & & $m_{11}^2$ & $0$ && $\lambda_1$ &&& real & $0$ & $0$
\\
$\ug$ & &  & $0$ 
 &  & &  & &
$0$ & $0$ & $0$ \\
$\ug$ $\otimes\Pi_2$  & & $m_{11}^2$ & $0$ && $\lambda_1$ &&& $0$ & $0$ & $0$
\\
$\mathrm{SO}(3)$  && $ m_{11}^2$ & $0$
   && $\lambda_1$ &  & $\lambda_1 - \lambda_3$ &
$0$ &$0$ & $0$ \\
GCP$1$   & & & real
 & &  &  &&
real & real & real \\
GCP$2$   && $m_{11}^2$ & $0$
  && $\lambda_1$ &  &  &
   &  & $- \lambda_6$ \\
GCP$3$   && $m_{11}^2$ & $0$
   && $\lambda_1$ &  &  &
$\lambda_1 - \lambda_3 - \lambda_4$ (real) & $0$ & $0$ \\
\hline
\end{tabular}
\caption{Classification of 2HDM scalar potential symmetries and their impact on the coefficients of the scalar potential [cf.~\eq{lambdapotential}] in a generic basis~\cite{Ivanov:2007de,Ferreira:2009wh,Ferreira:2010hy,Ferreira:2010yh,Battye:2011jj}.
Empty entries in Table~\ref{tab:class} correspond to a lack of constraints on the corresponding parameters. Note that $\Pi_2$, $\mathbb{Z}_2\otimes\Pi_2$ and
$\ug\otimes\Pi_2$ are
 not independent symmetries, since a change of scalar field basis can be performed in each case to a new basis in which the $\mathbb{Z}_2$, GCP$2$ and GCP$3$ symmetries, respectively, are manifestly 
realized.  
\label{tab:class}}
\end{table}

One can also consider the possibility of applying two of the symmetries listed above simultaneously in the same basis.  Ref.~\cite{Ferreira:2009wh} showed that no new independent models arise in this way.
For example, applying $\mathbb{Z}_2$ and $\Pi_2$ in the same basis yields a $\mathbb{Z}_2\otimes \Pi_2$ model that is equivalent to CP$2$ when expressed in a different basis.
Similarly, applying $\ug_{\rm PQ}$ 
and $\Pi_2$ in the same basis yields a $\ug\otimes\Pi_2$ model that is equivalent to GCP$3$ when expressed in a different basis.  
The equivalence of GCP$3$ and $\ug\otimes\Pi_2$ is explicitly demonstrated in Appendix~\ref{appequiv}.\footnote{The U(1)$\otimes\Pi_2$-symmetric 2HDM scalar potential was first introduced in Ref.~\cite{Fayet:1974fj}.}

A quick perusal of Table~\ref{tab:class} shows that the Higgs alignment limit, which can be achieved by setting $m_{11}^2=m_{22}^2$ and $m_{12}^2=0$ arises automatically by imposing one of the following Higgs family symmetries: $\mathbb{Z}_2\otimes\Pi_2$, $\ug\otimes\Pi_2$, or $\mathrm{SO}(3)$.   As noted above, one can replace the first two symmetries of this list with
GCP$2$ and GCP$3$, respectively, since a GCP$2$ [GCP$3$] invariant scalar potential exhibits a $\mathbb{Z}_2\otimes\Pi_2$ [$\ug\otimes\Pi_2$] symmetry in a different basis of scalar fields.  
If the Higgs alignment is approximate, then one can tolerate a 
soft breaking of the enhanced symmetries by allowing for $m_{11}^2\neq m_{22}^2$ and $m_{12}^2\neq 0$.   It turns out that it is more convenient to employ the softly-broken Higgs family symmetries.  Thus, we shall focus on the implications of the softly-broken $\mathbb{Z}_2\otimes\Pi_2$, $\ug\otimes\Pi_2$, or $\mathrm{SO}(3)$ symmetries in what follows.

We begin with the case of least enhanced symmetry---the softly-broken $\mathbb{Z}_2\otimes\Pi_2$ model.  As indicated in Table~\ref{tab:class}, this means that $\lambda_1=\lambda_2$ and $\lambda_6=\lambda_7=0$ while taking $\lambda_5$ real.  The softly-broken parameters $m_{11}^2$, $m_{12}^2$ and $m_{12}^2$ are taken to be arbitrary (with $m_{12}^2$ generically complex).    
It is convenient to introduce the parameter,
\beq \label{aredef}
R\equiv \frac{\lambda_3+\lambda_4+\lambda_5}{\lambda}\,.
\eeq
It then follows from \eq{threefourfive} that $\lambda_{345}=\lambda R-2\lambda_5\sin^2\xi$.

Assuming that $v_1$ and $v_2$ are both nonzero, one can use \eqst{min1}{min3} [with $\lambda\equiv\lambda_1=\lambda_2$] to eliminate $m_{11}^2$, $m_{22}^2$ and $\Im(m_{12}^2 e^{i\xi})$.  It then follows that the Higgs basis parameters are given by,
\beqa
&& Y_2=\frac{2\Re(m_{12}^2 e^{i\xi})}{s_{2\beta}}-\half\lambda v^2+\half v^2\bigl[\lambda(1-R)+2\lambda_5 \sin^2\xi\bigr](1-\half s_{2\beta}^2)\,,\label{whytwocp2}\\
&& Z_1=Z_2=\lambda-\half\bigl[\lambda(1-R)+2\lambda_5 \sin^2\xi\bigr]s_{2\beta}^2\,,\label{zeeonecp2} \\
&& Z_3=\lambda_3+\half\bigl[\lambda(1-R)+2\lambda_5 \sin^2\xi\bigr]s_{2\beta}^2\,, \label{zeethreecp2} \\
%\eeqa
%\beqa
&& Z_4=\lambda_4+\half\bigl[\lambda(1-R)+2\lambda_5 \sin^2\xi\bigr]s_{2\beta}^2\,, \label{zeefourcp2} \\
&& Z_5=\bigl\{\half\bigl[\lambda(1-R)+2\lambda_5 \sin^2\xi\bigr]s_{2\beta}^2+\lambda_5(\cos 2\xi+ic_{2\beta}\sin 2\xi)\bigr\} e^{-2i\xi}\,,\label{zeefivecp2} \\
&& Z_6=-Z_7=\bigl\{-\half\bigl[\lambda(1-R)+2\lambda_5 \sin^2\xi\bigr] c_{2\beta}+\half i\lambda_5\sin 2\xi\bigr\}s_{2\beta} e^{-i\xi}\,.\label{zeesevencp2}
\eeqa
One can also check that the minimization conditions of the Higgs basis given by \eq{YZ},
are satisfied as expected.

The scalar sector is CP conserving if and only $\Im(Z_5^* Z_6^2)=0$.    A straightforward computation yields,
\beq \label{CPcond}
\Im(Z_5^* Z_6^2)=-\tfrac14\lambda\lambda_5(\lambda-\lambda_3-\lambda_4-\lambda_5)(\lambda-\lambda_3-\lambda_4+\lambda_5)s_{2\beta}^2 c_{2\beta}\sin 2\xi\,.
\eeq

We shall henceforth impose CP conservation in the scalar sector, which simplifies the model that will be analyzed in this paper.   
In light of \eq{CPcond}, one can achieve a CP conserving scalar sector in a number of different ways.
The case of $s_{2\beta}=0$ corresponds to the IDM which has already been noted above.
The case of $\lambda_1=\lambda_3+\lambda_4+\lambda_5$ corresponds to the case of GCP$3$, whereas the case of $\lambda_5=0$ corresponds to the case of $\ug\otimes\Pi_2$, which is equivalent to GCP$3$ in a different scalar field basis as noted above.   Moreover, one is always free to rephase $\Phi_2\to i\Phi_2$ in the GCP$3$ basis, which changes the sign of the real parameter $\lambda_5$.   Thus, the case of $\lambda_1=\lambda_3+\lambda_4-\lambda_5$ also corresponds to GCP$3$.
These models automatically yield a CP conserving scalar sector.  These considerations motivate us to focus primarily on the softly-broken $\ug\otimes\Pi_2$ model.
Thus, we now examine the scalar sector of this model in more detail.

\section{The softly-broken GCP\texorpdfstring{$3$}{3}-symmetric \texorpdfstring{$2$}{2}HDM scalar potential}
\label{gcpthree}

In this section, we examine in detail the scalar mass spectrum and neutral scalar mixing in the softly-broken GCP$3$-symmetric 2HDM.  As previously indicated, it is more convenient to impose a $\ug\otimes\Pi_2$ Higgs family symmetry in the generic $\Phi_1$--$\Phi_2$ basis, which is equivalent to the realization of a GCP$3$ symmetry in another basis, as shown in Appendix~\ref{appequiv}.
Consider the softly-broken $\ug\otimes\Pi_2$ model, where $\lambda\equiv \lambda_1=\lambda_2$ and $\lambda_5=\lambda_6=\lambda_7=0$, whereas the softly-broken parameters $m_{11}^2$, $m_{22}^2$ and $m_{12}^2$ are arbitrary.    
If we demand that the potential is bounded from below, then the following conditions must be satisfied,
\beq \label{ineq}
\lambda>0\,,\qquad \lambda+\lambda_3>0\,,\qquad \lambda+\lambda_3+\lambda_4>0\,.
\eeq

Assuming that $v_1$ and $v_2$ are both nonzero, \eqst{min1}{min3} yield,
\begin{align}
&m_{11}^2 =\Re(m_{12}^2 e^{i\xi})\tan\beta-\half\lambda v^2 c^2_\beta-\half(\lambda_3+\lambda_4) v^2 s^2_\beta\,,\label{min1a} 
\\&m_{22}^2 = \Re(m_{12}^2 e^{i\xi})\cot\beta-\half\lambda v^2  s_\beta^2-\half(\lambda_{3}+\lambda_4) v^2 c_\beta^2\,, \label{min2a} 
\\&\Im(m_{12}^2 e^{i\xi})=0\,.\label{min3a}
\end{align}

\Eqs{min1a}{min2a} fix the value of $\beta$.  In particular,
\beq \label{betaeq}
\cos 2\beta=\frac{m_{22}^2-m_{11}^2}{m_{11}^2+m_{22}^2+\lambda v^2}\,,
\eeq
where $0<\beta<\half\pi$, under the assumption that $m_{11}^2\neq m_{22}^2$. 

Since $m_{12}^2$ is the only potentially complex parameter, one can rephase one of the two Higgs doublet fields to set $\xi=0$.  
After this rephasing, it follows from \eq{min3a} that $m_{12}^2$ is real. 
 Then, \eqst{whytwocp2}{zeesevencp2} yield,
\beqa
&& Y_2=\frac{2m_{12}^2}{s_{2\beta}}-\half\lambda v^2\bigl[R+\half s_{2\beta}^2(1-R)\bigr]\,,\label{whytwo}\\
&& Z_1=Z_2=\lambda\bigl[1-\half s^2_{2\beta}(1-R)\bigr]\,,\label{zeeone} \\
&& Z_3=\lambda_3+\half\lambda  s^2_{2\beta}(1-R)\,, \\
&& Z_4=\lambda_4+\half\lambda  s^2_{2\beta}(1-R)\,, \\
&& Z_5=\half\lambda s^2_{2\beta}(1-R)\,,\label{zeefive} \\
&& Z_6=-Z_7=-\half s_{2\beta}c_{2\beta}\lambda(1-R)\,,\label{zeeseven}
\eeqa
where 
\beq \label{Rdef}
R\equiv \frac{\lambda_3+\lambda_4}{\lambda}\,.   
\eeq
It is noteworthy that in the limit of $R=1$, the quartic terms of the scalar potential are invariant under the full
global U($2$) Higgs family symmetry, which was denoted by SO($3$) in Table~\ref{tab:symm1} after removing the hypercharge U($1$)$_{\rm Y}$ transformations (which have no effect on the scalar potential parameters). That is, in the limit of $R=1$, we obtain the softly-broken SO($3$)-symmetric 2HDM, where
the conditions $\lambda=\lambda_1=\lambda_2=\lambda_3+\lambda_4$ and $\lambda_5=\lambda_6=\lambda_7=0$ [specified in Table~\ref{tab:class}]
are satisfied for all possible choices of the scalar field basis. 

The squared masses of the neutral Higgs bosons are obtained by computing the eigenvalues of \eq{matrix33}.  In light of \eqs{zeefive}{zeeseven},
it is convenient to take $\eta=0$ in \eq{matrix33}, since this choice yields $\Im(Z_5 e^{-2i\eta})=\Im(Z_6 e^{-i\eta})=0$.   One can then immediately identity the squared mass of the CP-odd neutral scalar,
\beq \label{mha}
m_A^2=\half v^2(Z_3+Z_4-Z_5)+Y_2=\frac{2m_{12}^2}{s_{2\beta}}\,.
\eeq
Note that since $s_{2\beta}>0$, the positivity of $m_A^2$ requires that $m_{12}^2>0$.
One can also combine \eqss{min1a}{min2a}{mha} to obtain an alternative expression,
\beq \label{mha2}
m_A^2=m_{11}^2+m_{22}^2+\half \lambda v^2(1+R)\,.
\eeq

Likewise, the charged Higgs squared mass is given by
\beq \label{mch}
m_{H^\pm}^2= Y_2+\half Z_3 v^2=m_A^2-\half \lambda_4 v^2\,,
\eeq
after making use of \eq{mha}.
Finally, the squared masses of the CP-even neutral scalars, denoted by $h$ and $H$, are the eigenvalues of the $2\times 2$ matrix,
\beq
\mathcal{M}^2_H=\begin{pmatrix} Z_1 v^2 & \quad Z_6 v^2 \\ Z_6v^2 & \quad m_A^2+Z_5 v^2\end{pmatrix} 
=\begin{pmatrix}  \lambda v^2\bigl[1-\half s_{2\beta}^2(1-R)\bigr] & \quad -\half \lambda v^2 s_{2\beta}c_{2\beta}(1-R) \\ 
-\half \lambda v^2 s_{2\beta}c_{2\beta}(1-R)  & \quad m_A^2+\half \lambda v^2 s_{2\beta}^2 (1-R)\end{pmatrix}\,, \label{mhM2}
\eeq
where $\mathcal{M}^2_H$ is expressed with respect to the Higgs basis fields
$\{\sqrt{2}\,\Re H_1^0-v, \sqrt{2}\,\Re H_2^0\}$.
The CP-even neutral scalar mass eigenstates are denoted by $H$ and $h$ (where $m_H>m_h$), which are related to the Higgs basis fields as follows,
\beq \label{Hh}
\begin{pmatrix} H\\ h\end{pmatrix}=\begin{pmatrix} \cbma & \,\,\, -\sbma \\
\sbma & \,\,\,\phantom{-}\cbma\end{pmatrix}\,\begin{pmatrix} \sqrt{2}\,\,{\rm Re}~H_1^0-v \\ 
\sqrt{2}\,{\rm Re}~H_2^0
\end{pmatrix}\,,
\eeq
where $\cbma\equiv\cos(\beta-\alpha)$ and $\sbma\equiv\sin(\beta-\alpha)$ in a convention where $0\leq\beta-\alpha\leq\pi$. In a generic $\Phi_1$--$\Phi_2$ basis,
$\tan\beta=v_2/v_1$ 
 and $\alpha$ is the mixing angle that diagonalizes the CP-even Higgs squared-mass matrix when expressed with respect to $\{\sqrt{2}\,{\rm Re}~\Phi_1^0-v_1\,,\,\sqrt{2}\,{\rm Re}~\Phi_2^0-v_2\}$.  Nevertheless, the quantity $\beta-\alpha$ independent of the choice of the scalar field basis.

The exact Higgs alignment limit corresponds to $Z_6=0$, where the neutral scalar interaction eigenstate corresponding to the SM Higgs boson, $\sqrt{2}\,\Re H_1^0-v$, does not mix with the other neutral scalar interaction eigenstates of the 2HDM.  
We shall henceforth assume that the lighter of the two CP-even Higgs mass eigenstates, $h\simeq\sqrt{2}\,\Re H_1^0-v$, is SM-like and thus should be identified with the observed Higgs boson with $m_h\simeq 125$~GeV.  Under this assumption, it follows that $\cbma\to 0$ in the Higgs alignment limit.

After diagonalizing the matrix $\mathcal{M}_H^2$, the neutral CP-even scalar masses are given by,
\beq \label{cpevenmasses}
m^2_{H,h}=\half\biggl\{\mha^2+\lambda v^2\pm \sqrt{\bigl[\mha^2-\lambda v^2(c_{2\beta}^2+Rs_{2\beta}^2)\bigr]^2+\lambda^2 s_{2\beta}^2 c_{2\beta}^2(1-R)^2 v^4}\,
\biggr\}\,,
\eeq
and
\beq\label{equation:betaalpha}
c_{\beta-\alpha}=\frac{\lambda v^2 s_{2\beta}c_{2\beta}(1-R)}{2\sqrt{(\mhh^2-\mhl^2)\bigl[\mhh^2-\lambda v^2\bigl(1-\half s_{2\beta}^2(1-R)\bigr)\bigr]}}\,.
\eeq

 As noted above, if the Higgs alignment limit is approximately realized, then it follows that $|\cbma|\ll 1$.  
In light of \eq{equation:betaalpha}, which has been derived under the assumption that
$s_{2\beta}\neq 0$, one can achieve $|\cbma|\ll 1$ if either $c_{2\beta}$ is close to $0$ and/or $R$ is close to 1.   In light of \eqs{betaeq}{mha2}, it follows that $|c_{2\beta}|\ll 1$ when 
\beq
|\Delta m^2|\equiv |m_{22}^2-m_{11}^2|\ll m_A^2+\half\lambda v^2(1-R)\,.
\eeq
That is, we shall require that the parameter $\Delta m^2$, which if present (and nonzero) corresponds to a soft-breaking of the U(1)$\otimes\Pi_2$ symmetry, should not be too large.
Alternatively, if $|1-R|\ll 1$, which approaches the SO($3$) symmetry limit noted below \eq{Rdef}, it again follows that the Higgs alignment limit is approximately realized.  

It is noteworthy that there are cases in which the Higgs alignment limit is exactly realized (corresponding to $\cbma=0$) even though soft-symmetry breaking terms are present.
For example, if $c_{2\beta}=0$ then \eq{betaeq} yields
$\Delta m^2=0$ and exact Higgs alignment is achieved even though the U(1)$\otimes\Pi_2$ symmetry remains softly broken if $m_{12}^2\neq 0$. 
Likewise, exact Higgs alignment is achieved when $R=1$ despite the fact that the SO($3$) symmetry remains softly broken if either $\Delta m^2$ and/or $m_{12}^2$ are 
nonzero. One can verify that in these two examples, $Y_3=0$ [cf.~\eq{why3}] when the scalar potential minimum conditions [\eqst{min1a} {min3a}] are imposed. 

A stable minimum requires that the scalar squared-masses should be positive.   Hence, 
\beq 
m_{12}^2> 0\quad \text{and} \quad \lambda_4<2m_A^2/v^2\,.
\eeq
due to the positivity of $m_A^2$ and $m^2_{H^\pm}$.
In addition, we demand that
\beqa
\Tr\mathcal{M}_H^2&=&m_A^2+\lambda v^2> 0\,,\label{ineq1}\\
\frac{1}{v^2}\det \mathcal{M}_H^2&=&\tfrac14\lambda^2v^2 s_{2\beta}^2(1-R^2)+\lambda m_A^2\bigl[1-\half s_{2\beta}^2(1-R)\bigr]> 0\,.\label{ineq2}
\eeqa
Note that \eq{ineq1} is automatically satisfied in light of \eq{ineq}.  On the other hand, \eq{ineq2} is satisfied only if $R$ lies below a critical positive value that depends on $\lambda$, $\beta$ and $m_A^2/v^2$,

\beq \label{bound}
-1< R< \frac{m_A^2}{\lambda v^2}+\sqrt{\left(\frac{m_A^2}{\lambda v^2}-1\right)^2+\frac{4m_A^2}{\lambda v^2s_{2\beta}^2}}\,,
\eeq
after employing \eq{ineq}.\footnote{Apart from the upper bound given in \eq{bound}, one can obtain an independent upper bound by imposing either tree-level unitarity or a perturbativity constraint.   One would then expect $R/(4\pi)\lsim \mathcal{O}(1)$.}
It follows that \eq{ineq2} is satisfied for all values of $\beta$ if
\beq
-1< R< 1+\frac{2m_A^2}{\lambda v^2}\,.
\eeq

The cases of $v_1=0$ or $v_2=0$ should be treated separately and imply that $m_{12}^2=0$ in light of \eqs{inert1}{inert2}. First, suppose that $v_2=0$ and $v_1=v$.   Then, \eqs{mha}{mha2} are replaced by
\beq \label{inertA}
m_A^2=Y_2+\half \lambda v^2R\,,
\eeq
where $Y_2$ is a free parameter of the model that is no longer given by \eq{whytwo}.  In particular, \eq{betaeq} is no longer valid since $Y_2=m_{22}^2$ is independent of the squared mass parameter $m_{11}^2$; only the latter is fixed by the scalar potential minimum condition.

The squared-masses of the CP-even scalars and the charged Higgs scalar are given by,
\beq \label{inertB}
m_h^2=\lambda v^2\,,\qquad\quad m_H^2=m_A^2\,,\qquad\quad m_{H^\pm}^2=m_A^2-\half\lambda_4 v^2\,,
\eeq
where $h$ denotes the neutral CP-even Higgs scalar whose tree-level properties exactly coincide with those of the SM Higgs boson.
\Eqst{zeeone}{Rdef} remain valid after setting $\beta=0$.

Second, suppose that $v_1=0$ and $v_2=v$.   In this case, it follows that $Y_2=m_{11}^2$ is a free parameter and $Y_1=m_{22}^2=-\half Z_1 v^2=-\half \lambda v^2$.   
\Eqst{zeeone}{Rdef} remain valid after setting $\beta=\half\pi$.   Moreover, the neutral Higgs masses given by \eqs{inertA}{inertB} also remain valid.

Let us examine more closely when a vacuum can arise in which one of the two vevs vanishes.   First, we require that $R> -1$ in light of \eq{ineq}.
If $v_1=v$ and $v_2=0$, then \eq{inert2} yields $m_{12}^2 =0$\ and $m_{11}^2=-\half \lambda v^2<0$. The positivity of $m_A^2$ given in \eq{inertA} yields $m_{22}^2+\half\lambda Rv^2>0$.    Hence, it follows that
\beq \label{m22gtRm11}
m_{22}^2>Rm_{11}^2\,.
\eeq
The above inequality is equivalent to
\beq
(1+R)(m_{11}^2-m_{22}^2)<(1-R)(m_{11}^2+m_{22}^2)\,.
\eeq
Since $1+R$ is always positive, it follows that
\beq \label{Rineq2}
m_{22}^2-m_{11}^2> -\left(\frac{1-R}{1+R}\right)(m_{11}^2+m_{22}^2)\,.
\eeq

In the case of $v_1=0$ and $v_2=v$, one simply interchanges the roles of $m_{11}^2$ and $m_{22}^2$.  
In particular,
\beq \label{Rineq3}
m_{22}^2-m_{11}^2< \left(\frac{1-R}{1+R}\right)(m_{11}^2+m_{22}^2)\,.
\eeq

Although the vanishing of one of the two vevs requires that $m_{12}^2=0$, the converse is not necessarily true.   That is, if $m_{12}^2=0$, then two different phases of the 2HDM are 
possible: an inert phase in which either $v_1$ or $v_2$ vanishes and a mixed phase in which both $v_1$ and $v_2$ are nonzero.  To analyze the latter possibility more detail, we note that if $m_{12}^2=0$ and $v_1$, $v_2\neq 0$, then 
\eqs{min1a}{min2a} yield
\beqa
m_{11}^2 &=&-\half\lambda\bigl(v_1^2+R v_2^2\bigr)\,,\label{minonemixed} \\
m_{22}^2 &=& -\half\lambda\bigl(v_2^2+Rv_1^2\bigr)\,.\label{mintwomixed}
\eeqa
It is convenient to eliminate $v_1$ and $v_2$ in favor of the scalar potential parameters.
Using \eqs{minonemixed}{mintwomixed}, one easily obtains,
\beq \label{vineq}
v_1^2=\frac{2}{\lambda}\left(\frac{m_{22}^2 R-m_{11}^2}{1-R^2}\right)\,,\qquad\quad
v_2^2 =\frac{2}{\lambda}\left(\frac{m_{11}^2 R-m_{22}^2}{1-R^2}\right)\,.
\eeq

One feature of the mixed phase with $m_{12}^2=0$ is that $m_A=0$ due to the spontaneous breaking of the global Peccei-Quinn $\ug$ symmetry.   Thus, we will exclude this possibility in our subsequent phenomenological analysis.  Nevertheless, for completeness it is instructive to examine the range of scalar potential parameters that yields this mixed phase scenario.

One can work out a number of inequalities that must be satisfied if the mixed phase is stable.  We again require that $R> -1$ in light of \eq{ineq}. Using \eq{mhM2}, the trace and determinant of the $2\times 2$ neutral CP-even scalar squared-mass matrix yields,

\beq
m^2_h+m^2_H=\lambda v^2\,,\qquad\quad m_h^2 m_H^2=\tfrac14 \lambda^2 v^4 s_{2\beta}^2(1-R^2)\,.
\eeq
Hence, the positivity of the CP-even scalar squared masses implies that $|R|<1$. Next, we employ \eqs{minonemixed}{mintwomixed} along with $|R|<1$ to obtain,
\beqa
m_{11}^2+m_{22}^2&=&-\half\lambda v^2(1+R)<0\,,\\
m_{11}^2+m_{22}^2+\lambda v^2&=&\half\lambda v^2(1-R)>0\,.\label{secondineq}
\eeqa
Finally, the requirement that $v_1^2$ and $v_2^2$ are strictly positive implies that
\beq \label{Rineq}
m_{22}^2 R>m_{11}^2\,,\qquad\quad m_{11}^2 R>m_{22}^2\,,
\eeq
in light of \eq{vineq}.
The above equations are actually equivalent to the requirement that $|c_{2\beta}|<1$ after making use of \eqs{betaeq}{secondineq}.  It then follows that
\beq \label{Rineq1}
\left(\frac{1-R}{1+R}\right)(m_{11}^2+m_{22}^2)< m_{22}^2-m_{11}^2< -\left(\frac{1-R}{1+R}\right)(m_{11}^2+m_{22}^2)\,,
\eeq
which is easily shown to be equivalent to \eq {Rineq}.
Comparing \eq{Rineq1} with \eqs{Rineq2}{Rineq3}, it follows that a stable mixed phase and inert phase never coexist for any choice of the scalar potential parameters of the softly-broken $\ug\otimes\Pi_2$ symmetric 2HDM.\footnote{The same conclusion applies
in the case of a softly-broken $\mathbb{Z}_2\otimes\Pi_2$ symmetric scalar potential, where $\lambda_5$ is a nonzero real number.  In this case  \eqss{Rineq2}{Rineq3}{Rineq1} still apply, where $R$ is now defined as in \eq{aredef}. This corrects an error in Ref.~\cite{Draper:2016cag} which neglected to include the left hand side of the inequality given in \eq{Rineq1} and hence incorrectly concluded that the inert and mixed phases could coexist over part of the parameter space with $m_{12}^2=0$.}

Based on the considerations above, it follows that we can fix the parameter space of the $\ug\otimes\Pi_2$ model by specifying the values of $\lambda$, $\lambda_4$, $R$, $\beta$ and $m_A$ (with $v$ fixed to be $246$~\si{\giga\electronvolt}).   One can always replace $\lambda$ with $m_h$ and $\lambda_4$ with $m_{H^\pm}$, in which case the independent parameters of the  $\ug\otimes\Pi_2$ model can be taken to be $m_h$, $m_A$, $m_{H^\pm}$, $R$ and $\beta$.   If $\beta\neq 0$, $\half\pi$, then one is free to take $m_{12}^2$ (which is assumed to be real and positive) in place of $m_A$ as the independent parameter.

The inert limit of the $\ug\otimes\Pi_2$ model corresponds to setting $Z_6=0$, in which case we have $Y_3=Z_6=Z_7=0$, implying the presence of an exact $\mathbb{Z}_2$ symmetry (despite of the presence of squared-mass parameters that softly break the $\ug\otimes\Pi_2$ symmetry).   The inert limit arises if either $v_1=0$ or $v_2=0$, but is more general.  Indeed, \eq{zeeseven} implies that the inert limit arises if one of the following conditions are satisfied: $\beta=0$, $\tfrac14\pi$, $\half\pi$, or $R=1$.  (We reject the possibility of $\lambda=0$ which results in a massless CP-even scalar.)  In the inert 
limit, $c_{\beta-\alpha}=0$ and the neutral CP-even scalar $h$ with squared-mass $m_h^2=Z_1 v^2$ possesses 
the tree-level properties of the SM Higgs boson.

Finally, we observe that the $\ug\otimes\Pi_2$ symmetry is explicitly preserved by the scalar potential if $m_{11}^2=m_{22}^2$ and $m_{12}^2=0$.  
If both vevs are nonzero then the $\ug\otimes\Pi_2$ symmetry limit arises if $m_{12}^2=0$ and $\beta=\tfrac14\pi$.  In this case, the neutral scalar mass spectrum is $m^2_A=0$, $m^2_{h}=\half\lambda v^2(1+ R)$ and $m^2_{H}=\half\lambda v^2(1-R)$, which corresponds to a stable minimum if $|R|< 1$.  The $\ug$ symmetry is spontaneously broken by the vacuum, resulting in a massless scalar state.
Note that in the special case of $m_{11}^2=m_{22}^2$, $m_{12}^2=0$ and $R=1$, an SO($3$) symmetry is explicitly preserved by the scalar potential [cf.~Table~\ref{tab:class}].   The SO($3$) symmetry is spontaneously broken 
by the vacuum, leaving a residual unbroken U($1$) symmetry, which results in two massless Goldstone bosons, $H$ and $A$.

If only one of the two vevs is nonzero, then
$\sin 2\beta=0$, which implies that $m_{12}^2=0$.  After setting $m_{11}^2=m_{22}^2$, we obtain $m_A^2=m_H^2=\half \lambda v^2(R-1)$ and $m_h^2=\lambda v^2$,
which corresponds to a stable minimum if $R>1$. 
Note that in this case the $\ug$ symmetry is preserved by the vacuum and results in the $H$, $A$ mass degeneracy.  In the limit of $R\to 1$ one again finds an SO$(3$)-symmetric scalar potential where SO$(3$) is spontaneously broken down to U($1$), resulting in two Goldstone boson states $A$ and $H$ as previously noted.

In all of the unbroken $\ug\otimes\Pi_2$ symmetry cases above and in the limiting SO($3$) case in the limit of $R=1$, note that
$c_{\beta-\alpha}=0$, corresponding to a Higgs alignment limit where~$h$ has the tree-level properties of the SM Higgs boson.   Although the $\ug\otimes\Pi_2$ and SO($3$) symmetry limits yield the inert model, the converse does not necessarily hold.  In particular, if $m_{12}^2>0$ and $\beta=\tfrac14\pi$ then $m_A^2=2m_{12}^2$, $m_h^2=\half \lambda v^2(1+R)$  and $m_H^2=m_A^2+\half\lambda v^2(1-R)$ due to an explicit breaking of the $\ug$ symmetry. If $m_{12}^2> 0$, $s_{2\beta}\neq 0$ and $R=1$ then $m_h^2=\lambda v^2$ and $m_A^2=m_H^2= 2m_{12}^2/s_{2\beta}$.   If $\sin 2\beta=0$ and $m_{11}^2\neq m_{22}^2$  then $m_h^2=\lambda v^2$ and $m_A^2=m_H^2$.  In the latter two cases, the $\Pi_2$ symmetry is softly broken, whereas an unbroken $\ug$ symmetry is responsible for the $H$, $A$ mass degeneracy.

%%%%%%%%%%%%%%%%%%%%%%%%%%%%%%%%%%%%%%%%
\section{GCP\texorpdfstring{$3$}{3}-symmetric Yukawa couplings}\label{section:YukawaCouplings}
\label{cp3yukawa}
%%%%%%%%%%%%%%%%%%%%%%%%%%%%%%%%%%%%%%%%

If we wish to employ a GCP$3$-symmetric 2HDM scalar potential (broken at most by dimension-two squared-mass parameters), then we should impose the GCP$3$ symmetry on the Higgs-fermion Yukawa couplings.   Such an attempt was made in Ref.~\cite{Ferreira:2010bm} by extending the GCP$3$ transformation laws to the fermion fields.  Unfortunately, any such extension must relate fermions of different generations, and the resulting phenomenology was incompatible with observed experimental data.   A possible way out of this conundrum was suggested in Ref.~\cite{Draper:2016cag}, which proposed adding new vector-like fermions to the two Higgs doublet extended SM.\footnote{The phenomenology of such models has been examined previously in Ref.~\cite{Arhrib:2016rlj}.}  In this way, one could devise an extension of the GCP$3$ transformation laws to the fermion sector that relates the fermion of the SM to vector-like fermion partners of the same flavor.  It is again convenient to work in the $\ug\times\Pi_2$ basis of scalar fields, and thus all fermion field transformation laws introduced below will be extensions of the $\ug\otimes\Pi_2$ scalar field transformations exhibited in Table~\ref{tab:symm1}.

\subsection{Extending the 2HDM to include vector-like fermions}
\label{sec:mirror}

First, consider the top-quark sector.  
Let $q=(u\,\,\,d)^{\T}$ denote the third generation of color triplet, $\mathrm{SU}(2)$ doublet of two-component quark fields, and $\bar{u}$ denote the color anti-triplet, $\mathrm{SU}(2)$ singlet two-component top quark field.  We now add a mirror two-component top partner field, $\overline{U}$, having the same SM gauge quantum numbers as $\bar{u}$.  
As suggested by the notation, one can easily extend these considerations to three generations of quarks and their mirrored partners by considering the generation indices on the fermion fields defined 
above to be implicit.
Under a $\ug\otimes\Pi_2$ symmetry transformation,
\beqa
\Pi_2:&&  q \longleftrightarrow q, \quad \bar{u} \longleftrightarrow \overline{U}, \qquad \Phi_1 \longleftrightarrow \Phi_2, \label{trans1} \\
{\rm U}(1):&&  q \longrightarrow q,  \quad \bar{u} \longrightarrow e^{-i\theta}\bar{u}, \quad \overline{U} \longrightarrow e^{i\theta}\overline{U}, \quad  \Phi_1\longrightarrow e^{-i\theta}\Phi_1,\quad \Phi_2\longrightarrow e^{i\theta}\Phi_2. \label{trans2} 
\eeqa

It is possible to impose the symmetries on the fermion sector in other ways, for example, by adding a mirror isospin doublet for $q$ either instead of or in addition to the singlet for $\bar{u}$. The choice above is minimal in terms of the additional matter content.  Note that the gauge covariant kinetic energy terms of the fermions and their mirror partners are automatically invariant under $\ug\otimes\Pi_2$, whereas the form of the Yukawa couplings is constrained.  In particular,
the Yukawa couplings invariant under $\ug\otimes\Pi_2$ transformations now take the form,
\beq \label{eq:TopYukawa}
-\mathscr{L}_{\rm Yuk}\, \supset \, y_t \left(q \Phi_2 \bar{u} + q \Phi_1 \overline{U} \right) + \rm{h.c.},
\eeq
where $q\Phi_i\equiv \epsilon^{ab}q_a\Phi_{ib}$ (for $i=1$, $2$) and $a$ and $b$ are SU($2$) gauge group indices.
The antisymmetric epsilon symbol defined such that $\epsilon^{12}=-\epsilon^{21}=1$.
In order to avoid gauge anomalies, we shall add a two-component color triplet, SU($2$) singlet field ${U}$ with a weak hypercharge that is opposite in sign to that of its conjugate field, $\ol{U}$.
This new fermion $U$ transforms under $\ug\otimes\Pi_2$ as,
\beqa
\Pi_2:&&   {U} \longleftrightarrow {U} \label{trans3} \\
{\rm U}(1): &&   {U} \longrightarrow e^{\pm i\theta} {U}\,,  \label{PQlaw}
\eeqa
where one of the two signs in \eq{PQlaw} should be selected (either sign choice is equally valid). 
No additional Yukawa interaction involving ${U}$ is allowed by the symmetry.  The mirror top partner $\ol{U}$ together with $U$ can be combined into a Dirac fermion that possesses vector-like couplings to gauge bosons.  Henceforth we will  refer to $U$ and $\ol{U}$ as the vector-like partners of the top quark.

As indicated in Table~\ref{tab:class}, if $m_{11}^2=m_{22}^2$, $m_{12}^2=0$, $\lambda=\lambda_1=\lambda_2$ and $R\equiv (\lambda_3+\lambda_4)/\lambda=1$ then the scalar potential is invariant under the U($2$) Higgs flavor symmetry, $\Phi_a\to U_{ab}\Phi_b$, where $U\in{\rm U}(2)$ and $a,b=1,2$.
Moreover, the Yukawa Lagrangian specified in \eq{eq:TopYukawa} is also invariant under the U($2$) Higgs family symmetry.  In particular, we can combine $\bar{u}$ and $\overline{U}$ into a U($2$) multiplet,
$
\mathcal{U}^\dagger\equiv \begin{pmatrix} \overline{U} & \bar{u}\end{pmatrix}
$,
with a transformation law under U($2$) given by $\mathcal{U}^\dagger_a\to\mathcal{U}^\dagger_b U^\dagger_{ba}$.   
We can then rewirte \eq{eq:TopYukawa} to exhibit its invariance under U($2$),
\beq \label{eq:TopYukawa2}
-\mathscr{L}_{\rm Yuk}\, \supset \, y_t q \mathcal{U}^\dagger_a\Phi_a + \rm{h.c.}
\eeq
Furthermore, we recognize that the  $\mathrm{U}(1)$ and $\Pi_2$ symmetry transformations specified in \eqs{trans1}{trans2} are special elements of the two-dimensional representation of the $\mathrm{U}(2)$ symmetry with
\begin{align}
\mathrm{U}(1):~U =
\begin{pmatrix} e^{-i \theta} & 0 \\
0 & e^{i \theta}
\end{pmatrix},
\hspace*{1cm}
\Pi_2:~U =
\begin{pmatrix} 0 & 1 \\
1 & 0
\end{pmatrix}.
\end{align}
The fields $u$, $d$ transform as singlets under the U($2$) transformation, whereas $U$ transforms as a nontrivial one-dimensional representation of U($2$) as indicated in \eq{PQlaw}.

In order to evade the experimental limits on the nonobservation of vector-like fermions at the LHC, we shall add explicit $\ug_Y$ gauge invariant mass terms that softly break the
U($2$) symmetry,
\beq \label{Vmass}
-\mathscr{L}_{\rm mass}=  M_U \ol{U}U +(M_u \bar{u} U + \rm{h.c.}).
\eeq 
The U($1$)$\otimes\Pi_2$ subgroup of U($2$) is also softly broken once \eq{Vmass} is introduced.
The vector-like mass terms explicitly break the $\Pi_2$ symmetry if $M_U\neq M_u$, whereas the $\Pi_2$ symmetry is preserved if $M_U=M_u$.   
In contrast, if $M_U M_u\neq 0$ then one of the two mass terms above must explicitly break the $\ug$ symmetry for either sign choice of the $\ug$ transformation law given in \eq{PQlaw}.
Indeed, as noted at the end of \sect{gcpthree}, one must avoid the spontaneously breaking of the $\ug$ symmetry by the vacuum, which yields an undesirable massless scalar.

Having introduced the soft symmetry breaking of \eq{Vmass}, it then follows that the soft symmetry breaking squared-mass parameters of the scalar potential will be automatically generated in the low energy effective 2HDM once the mirror fermions are integrated out.   For example,  
because of the breaking of the $\Pi_2$ symmetry, quantum corrections spoil the symmetry protected degeneracy, $m_{11}^2=m_{22}^2$.  However, due to the soft nature of the symmetry breaking, $m_{22}^2 - m_{11}^2$ is protected from quadratic sensitivity to the cutoff scale $\Lambda_c$.  
Likewise, due to the soft breaking of the $\ug$ symmetry, we expect the following contributions to $m_{12}^2 \propto M_u M_U$ and $m_{11}^2-m_{22}^2\propto (M_U^2-M_u^2)$. We return to the effects of soft symmetry breaking in \sect{sec:BreakDiscrete}.

Other SM fermion partners can be included analogously: $\mathrm{SU}(2)$ doublets of two-component lepton fields are denoted by $\ell=(\nu\,\,\,e)^{\T}$; and the remaining two-component SU($2$) singlet fermion fields of the SM, $\bar{d}$, and $\bar{e}$, acquire mirror partners $\ol{D}$ and $\ol{E}$. These mirror fields pair up with their
conjugate fields $D$ and $E$ (generation indices are implicit) to
yield vector-like mass terms.  The U(1)$\otimes\Pi_2$ symmetries are taken to act as,
\beqa
\Pi_2:&&  \bar{d} \Longleftrightarrow \ol{D}, \quad \ell\Longleftrightarrow\ell, \quad \bar{e} \Longleftrightarrow \ol{E}, \quad {D} \Longleftrightarrow {D}, \quad E \Longleftrightarrow E \label{trans4} \\
{\rm U}(1): &&  \bar{d} \Longrightarrow e^{i\theta\xi_d}\bar{d}, \quad \ell\Longrightarrow \ell,\quad \bar{e} \Longrightarrow e^{i\theta\xi_e} \bar{e},\quad  \ol{D}\Longrightarrow e^{-i\theta\xi_d}\ol{D}, \quad {D}\Longrightarrow e^{\pm i\theta}{D},  \nonumber \\
&&\ol{E}\Longrightarrow e^{-i\theta\xi_e}\ol{E}, \quad {E}\Longrightarrow e^{\pm i\theta}{E}\label{trans5}
\eeqa
As discussed below \eq{PQlaw}, in the transformation laws of $D$ and $E$, one of the two sign choices should be selected, although any one of the four possible sign choices is equally valid.
The factors $\xi_d$ and $\xi_e$ are also sign factors that can be chosen in four different ways.  For example, if $\xi_d=\xi_e=1$, then the Yukawa couplings are 
Type-I Higgs-quark and Higgs-lepton couplings~\cite{Haber:1978jt,Hall:1981bc},
\beq \label{eq:BotYukawa}
-\mathscr{L}_{\rm YUK}  \supset y_b \, \left( \Phi_2^\dagger q \bar{d} +  \Phi_1^\dagger q \ol{D} \right)  + y_\tau \, \left(\Phi_2^\dagger \ell \bar{e} +  \Phi_1^\dagger\ell \ol{E} \right).
\eeq
Likewise, if $\xi_d=\xi_e=-1$, then one must switch $\Phi_1\leftrightarrow\Phi_2$ in \eq{eq:BotYukawa}, which yields Type II Higgs-quark and Higgs-lepton couplings~\cite{Donoghue:1978cj,Hall:1981bc}.   Alternatively, one could choose $\xi_d=-\xi_e$, in which case, $\xi_d=1$ corresponds to \mbox{Type~X} Higgs-quark and Higgs-lepton couplings and $\xi_d=-1$ corresponds to Type Y Higgs-quark and Higgs-lepton couplings~\cite{Barger:1989fj,Aoki:2009ha}.  In a multi-generational model, there are no FCNCs mediated by tree-level neutral Higgs boson exchange in models with 
Type I, II, X or Y Yukawa couplings.

Once again, the Yukawa Lagrangian specified in \eq{eq:BotYukawa} is invariant under the U($2$) Higgs family symmetry.  We can combine $\bar{d}$ and $\overline{D}$ 
and likewise $\bar{e}$ and $\overline{E}$ into U($2$) multiplets, $\mathcal{D}^\dagger\equiv \begin{pmatrix} \overline{D} & \bar{d}\end{pmatrix}$ and
$\mathcal{E}^\dagger\equiv \begin{pmatrix} \overline{E} & \bar{e}\end{pmatrix}$, with transformation laws under U($2$) given by $\mathcal{D}^\dagger_a\to\mathcal{D}^\dagger_b U^\dagger_{ba}$ and  $\mathcal{E}^\dagger_a\to\mathcal{E}^\dagger_b U^\dagger_{ba}$.   That is, we can rewirte \eq{eq:BotYukawa} to exhibit its invariance under U($2$),
\beq \label{eq:BotYukawa2}
-\mathscr{L}_{\rm Yuk}\, \supset \, y_b q \mathcal{D}^\dagger_a\Phi_a +y_\tau \ell \mathcal{E}^\dagger_a\Phi_a + \rm{h.c.}
\eeq
The fields $\nu$, $e$ transform as singlets under the U($2$) transformation, whereas $D$ and $E$  transform as a nontrivial one-dimensional representation of U($2$) as indicated in \eq{trans5}.

As in \eq{Vmass}, we add vector-like fermion mass terms to softly break the U($2$) symmetry,
\beq \label{Vmass2}
-\mathscr{L}_{\rm mass}  \supset  M_D \ol{D}D + M_E  \ol{E}E+(M_d \bar{d} {D}+M_e \bar{e} {E}+{\rm h.c.}).
\eeq
Once again, the U($1$)$\otimes\Pi_2$ subgroup of U($2$) is also softly broken.
The vector-like mass terms explicitly break the $\Pi_2$ symmetry if $M_D\neq M_d$ and/or $M_E\neq M_e$, whereas the $\Pi_2$ symmetry is preserved if $M_D=M_d$ and $M_E=M_e$.   
In contrast, if $M_D M_d\neq 0$ [or $M_E M_e\neq 0$] then one of the two mass terms appearing in $M_D \ol{D}D +M_d \bar{d} {D}$  [or $M_E \ol{E}E +M_e \bar{e} {E}$]
above must explicitly breaks the $\ug$ symmetry for either sign choice in the corresponding $\mathrm{U}(1)$ transformation law 
given in \eq{trans5}.

Given vector-like mass parameters $M_f$ and $M_F$, there is a one-loop correction to $m_{22}^2 - m_{11}^2$.  Requiring this correction to be smaller than the electroweak scale implies a bound of order
\beq
\frac{y_f^2}{16\pi^2}|M_F^2-M_f^2| \lesssim v^2.
\eeq
Assume for simplicity that $M_f\ll M_F$, which  suppresses the mixing of $f$ with its vector-like partners. Then
we require $M_{F=b,\tau}\lesssim 100$ TeV and $M_{F=e}\lesssim 10^8$~\si{\giga\electronvolt}. 
Therefore, if the cutoff scale~$\Lambda_c$ is not too high, then the simplest, most minimal new field content  needed to enforce the symmetries in a natural way is a vector-like right-handed top partner near the electroweak scale. 
Integrating out the top partner at its threshold, the low-energy effective theory is that of a 2HDM with a scalar potential governed by an approximate (softly-broken) 
$\mathrm{U}(1)\otimes\Pi_2$ symmetry.

\subsection{Soft Symmetry Breaking Effects}
\label{sec:BreakDiscrete}
Vector-like masses softly break the discrete mirror $\Pi_2$ symmetry and the top sector dominates as previously noted. We imagine that above a cutoff scale $\Lambda_c$, the symmetry is restored;  below $\Lambda_c$, explicit $\Pi_2$-breaking enters with a characteristic scale,
$M^2\equiv M_U^2+M_u^2$.
Following \eqs{eq:TopYukawa}{Vmass}, we consider the Lagrangian,
\begin{align}
-\mathscr{L}\subset y_t \left(q \Phi_2 \bar{u} + q \Phi_1 \ol{U} \right) + (M_U \ol{U}U+M_u \bar{u} {U}+\rm{h.c.})\,. \label{minuslag}
\end{align}

The one generation model possesses four potentially complex parameters: $m_{12}^2$, $y_t$, $M_u$ and $M_U$ (where we are only including top partners among the vector-like quarks).  However, one can remove all complex phases by absorbing them into the definition of the scalar and fermion fields.  In particular, given $\Phi_1$, $\Phi_2$, $q$, $\bar{u}$, $\ol{U}$ and $U$, the Lagrangian is invariant under a $\mathrm{U}(1)_Y$ transformation.  This leaves five additional global $\mathrm{U}(1)$ transformations that can be used to absorb phases.\footnote{Including one generation of vector-like down-type quarks and charged leptons introduces additional potentially complex parameters but also adds additional $\mathrm{U}(1)$ transformations to remove those phases.}  Hence, without loss of generality, one can assume that $m_{12}^2$, $y_t$, $M_u$ and $M_U$ are real positive parameters.

The mass terms appear in the Lagrangian in the following form,
\beq \label{Dmassmat}
-\mathscr{L}_{\rm mass}=\mathcal{M}^i{}_j\hat{\chi}_i\hat{\eta}^j+{\rm h.c.},
\eeq
where $\mathcal{M}$ is (in general) a complex matrix with matrix elements $\mathcal{M}^i{}_j$.  Following Ref.~\cite{Dreiner:2008tw}, we have denoted the two-component fermion interaction eigenstates by hatted fields, $\hat{\chi}_i$ and
$\hat{\eta}^i$,  which are related to the unhatted
fermion mass eigenstate fields, $\chi_i$ and $\eta^i$, via
\beq
\label{lrdef}
\hat\chi_i=L_i{}^k\chi_k\,,\qquad\qquad
\hat\eta^i=R^i{}_k\eta^k\,,
\eeq
where the unitary matrices $L$ and $R$, with matrix elements given respectively by $L_i{}^k$ and $R^i{}_k$, are chosen such that
$\mathcal{M}^i{}_j L_i{}^k R^j{}_\ell = m_k\delta^k_\ell$ (no sum over $k$),
such that the $m_k$ are real and nonnegative. Equivalently, in matrix notation with suppressed indices,
$\hat\chi=L\chi\,,\,\hat\eta=R\eta$ and
\beq
\label{svd}
L^{\T} \mathcal{M} R= {\boldsymbol{m}}={\rm diag}(m_1,m_2,\ldots),
\eeq
where the real and nonnegative $m_k$ can be identified as the physical masses of the fermions.

The singular value
decomposition of linear algebra
states that for any complex matrix $\mathcal{M}$, unitary matrices $L$ and $R$
exist such that \eq{svd} is satisfied.  It then follows that:
\beqa
L^{\T}(\mathcal{M}\mathcal{M}^\dagger)L^* \,=\,
R^\dagger(\mathcal{M}^\dagger \mathcal{M}) R \,=\, {\boldsymbol{m}}^2 .
\eeqa
That is, since $\mathcal{M}\mathcal{M}^\dagger$
and $\mathcal{M}^\dagger \mathcal{M}$ are both hermitian, they can be diagonalized by unitary matrices.
The diagonal elements of $\boldsymbol{m}$ are therefore
the nonnegative square roots of the
corresponding eigenvalues of $\mathcal{M}\mathcal{M}^\dagger$ (or equivalently, $\mathcal{M}^\dagger \mathcal{M}$).
In terms of the fermion
mass eigenstate fields,
\beq
\label{lagDiracdiag}
-\mathscr{L}= \sum_i m_i\chi_i\eta^i +{\rm h.c.}
\eeq
The mass matrix now consists of $2\times 2$ blocks
$\bigl(\begin{smallmatrix}0 & m_i \\ m_i & 0\end{smallmatrix}\bigr)$
along the diagonal.  For $m_i\neq 0$, each $\chi_i$--$\eta^i$ pair describes a charged
Dirac fermion.

In our present application,
$\mathcal{M}$ is a real $2\times 2$ matrix in the convention where the mass parameters of \eq{minuslag} are real and nonnegative, in which case
the matrices $L$ and $R$ can be taken to be real orthogonal matrices.  Thus, we shall employ the singular value decomposition of an arbitrary real $2\times 2$ matrix,
whose explicit form is given in Appendix~\ref{app:svd}.

We identify the interaction eigenstates as follows: $\hat\chi_i=(u\quad U)$ and $\hat\eta^j=(\bar{u}\quad \ol{U})$,  where $u\equiv q_1$.  Hence,
prior to electroweak symmetry breaking, \eq{minuslag} yields,
\beq \label{CalMat}
\mathcal{M}=\begin{pmatrix} 0 & \,\,\, 0 \\ M_u &  \,\,\, M_U\end{pmatrix}.
\eeq
Using the \eqs{eq:OLR}{rcoslimit}, it follows that $L=\mathds{1}_{2\times 2}$ and 
\beq
R=\begin{pmatrix} \phm \cos\gamma & \,\,\, \sin\gamma \\ -\sin\gamma & \,\,\, \cos\gamma \end{pmatrix},
\eeq
where
\beq \label{angle}
s_\gamma\equiv\sin\gamma=\frac{M_u}{M}\,,\qquad\quad c_\gamma\equiv\cos\gamma=\frac{M_U}{M}\,,\qquad\quad \text{where $M\equiv (M_U^2+M_u^2)^{1/2}$ },
\eeq
and $0\leq\gamma\leq\half\pi$ in the convention adopted below \eq{minuslag} where $M_U$ and $M_u$ are taken to be nonnegative quantities. The two-component fields $u$ and $U$ do not mix, whereas $\bar{u}$ and $\ol{U}$ mix to form two-component fermion mass eigenstates that we shall denote by
$\eta^k=(\bar{x}_0\quad \ol{X}_0)$, where the subscript~$0$ indicates a mass-eigenstate field prior to electroweak symmetry breaking.   In particular, 

\beq \label{xx}
\bar{x}_0=-s_\gamma\ol{U}+c_\gamma \bar{u}\,,\qquad\quad \ol{X}_0=c_\gamma\ol{U}+s_\gamma\bar{u}\,.
\eeq

Rewriting \eq{minuslag} in terms of the two-component fermion mass-eigenstate fields yields,
\beq \label{twover}
-\mathscr{L}\subset y_t \bigl\{q(\Phi_2 s_\gamma+\Phi_1c_\gamma)\ol{X}_0+q(\Phi_2 c_\gamma-\Phi_1 s_\gamma)\bar{x}_0\bigr\}+M\ol{X}_0U+{\rm h.c.}
\eeq
One can introduce four-component fermions fields,
\beq
t_0=\begin{pmatrix} u \\ \bar{x}_0\end{pmatrix}\,,\qquad T_0\equiv \begin{pmatrix}  U \\ \ \ol{X}_0\end{pmatrix}\,.
\eeq
Then the four-component fermion version of \eq{twover} contains the following mass terms and couplings to the neutral Higgs fields,
\beq
-\mathscr{L}\subset y_t \bigl\{(\Phi^0_2 c_\gamma-\Phi^0_1 s_\gamma)\bar{t}_0t_0 +(\Phi^0_2 s_\gamma+\Phi^0_1c_\gamma)(\overline{T}_0P_Lt_0+{\rm h.c.)}\bigr\}+
M\overline{T}_0T_0\,,
\eeq
where $P_L\equiv\half(1-\gamma_5)$.

The $\Pi_2$ symmetry is broken if $M_u\neq M_U$, which corresponds to $s_\gamma\neq c_\gamma$.  When evolved down from the UV theory, this breaking generates a mass splitting $m_{11}^2-m_{22}^2\neq 0$. In the infrared the mass splitting is approximately 
\beq
\Delta m^2 \equiv m_{22}^2-m_{11}^2\sim  \kappa_{\Delta m^2} (M_U^2-M_u^2)-\frac{3y_t^2(M_U^2-M_u^2)}{4\pi^2}\log(\Lambda_c/M)\;.
\label{eq:Deltam2}
\eeq
There is a similar contribution to $m_{12}^2$,
\beq 
m_{12}^2\sim \kappa_{m_{12}^2} M_U M_u+\frac{3 y_t^2 M_U M_u}{4 \pi^2}\log\left(\frac{\Lambda_c}{M}\right). \label{eq:m12}
\eeq
The first terms exhibited on the right hand sides of \eqs{eq:Deltam2}{eq:m12} represent threshold corrections at the UV scale $\Lambda_c$, where $\kappa_{\Delta m^2},~\kappa_{\Delta m_{12}^2}$ are dimensionless couplings, and the second terms represent a radiative correction from the quark loops below $\Lambda_c$.  
Thus, the framework under consideration is an example of a partially natural 2HDM introduced in Ref.~\cite{Draper:2016cag}, where only one fine-tuning of scalar parameters is necessary to determine the electroweak scale.

In estimating the numerical values of $\Delta m^2$ and $m_{12}^2$ above, one must determine the value of the parameter~$y_t$.
In particular, $y_t$ is not the physical top-quark coupling, but it is related to the physical top-quark mass via,
\beq \label{whytop}
 m_t\simeq (y_t v/\sqrt{2})|s_{\beta-\gamma}|\,,
 \eeq
after electroweak symmetry breaking is taken into account.  This relation implies that $|s_{\beta-\gamma}|$ should not be too small; otherwise \eq{whytop} would require $y_t\gg 1$ leading to a non-perturbative top Yukawa coupling (as well as a Landau pole in the running of $y_t$ that is uncomfortably close to the TeV scale).  
Since we expect that realistic values of $\sin\gamma$ should be rather small compared to unity in order to avoid significant shifts in the top quark couplings away from their SM values, it follows that the preferred parameter regime will correspond to values of $\tan\beta$ above~1.

Note that $m_{12}^2$ given by \eq{eq:Deltam2} vanishes if either $M_U$ or $M_u$ vanish due to an unbroken Peccei-Quinn $\mathrm{U}(1)$ symmetry.
However, in contrast to the model of Ref.~\cite{Draper:2016cag} where $M_u$ was assumed to vanish, we expect that both $M_U$ and $M_u$ are generically nonzero, as these parameters are presumably generated by physics that lies above the UV cutoff scale $\Lambda_c$.  
It is still possible that $m^2_{12}=0$ accidentally due to the a cancellation of the two terms on the right hand side of \eq{eq:m12}.  However, we would regard such a cancellation as an unnatural fine-tuning of the model parameters.  Thus, we conclude that $m_{12}^2$ is generically non-zero, which implies that the inert limit of the model (where Higgs alignment is exact) is not realized.  That is, in the scenario presented in this paper, the Higgs alignment is expected to be approximate,   implying that deviations from SM Higgs couplings should eventually be detected in future Higgs precision experiments.

Below the scale $M$, the vector-like fermions can be integrated out, in which case the parameters of the scalar potential (evaluated at the electroweak scale) will shift, 
$\lambda_i\to\lambda_i+\delta\lambda_i$, due to the evolution of the scalar potential parameters from the scale $M$, which characterizes the mass scale of the vector-like top quarks, 
down to the top quark mass $m_t$.  The parameter shifts in the one-loop approximation are roughly given by,\footnote{ There are additional symmetry-preserving contributions to the running of the $\lambda_i$ between $\Lambda_c$ and $M$, which should be understood to be absorbed into the symmetry-preserving values of the $\lambda_i$ at the scale $M$. Also, compared with Eq.~(\ref{eq:Deltam2}), we give only the leading-log correction to the $\lambda_i$ below the scale $M$, neglecting finite symmetry-breaking threshold corrections. The logarithmic terms give a qualitative estimate for the size of the corrections.}
\beq \label{deltalams}
\delta\lambda_i \sim  \frac{3y_t^4 }{4 \pi^2}k_i\log\left(\frac{M}{m_t}\right)\,,
\eeq
for $i=1,2,\ldots,7$ with $k_i\equiv (-s_\gamma)^{p_1}(c_\gamma)^{p_2}$, where $p_i$ is equal to the number of times the scalar field $\Phi_i$ or its complex conjugate appears in the $i$th quartic term of the scalar potential.
In particular,
\beq \label{kays}
k_1=s_\gamma^4\,,\qquad k_2=c_\gamma^4\,,\qquad k_3=k_4=k_5=s_\gamma^2 c_\gamma^2\,,\qquad k_6=-s^3_\gamma c_\gamma\,,\qquad
k_7=-s_\gamma c^3_\gamma\,,
\eeq
as a consequence of the mixing between $\overline{U}$ and $\bar{u}$ [cf.~\eqs{xx}{twover}].

The size of the parameter shifts can be estimated by employing \eq{whytop} for $y_t$.
For example, if we take $M=1.5~\si{\tera\electronvolt}$, $\gam=0.2$, and $\tan \beta >5$,
then we obtain a splitting between $\lambda_1$ and $\lambda_2$, 
\begin{align}
|\lam_2-\lam_1|\sim \frac{3y_t^4 }{4 \pi^2}\log\left(\frac{M}{m_t}\right)|\cos 2\gamma |<0.22\,.
%(c_\gam^4-s_\gam^4). 
\label{lam12diff}
\end{align} 
Likewise nonzero values of $\lam_5,~\lam_6$, and $\lam_7$ are also generated,
\begin{align}
&|\lam_5|\sim \frac{3y_t^4 }{16 \pi^2}\log\left(\frac{M}{m_t}\right) s_{2\gam}^2<0.01\,, \label{lamfive}  \\
&\left|\lam_6\right|\sim \frac{3y_t^4 }{4 \pi^2}\log\left(\frac{M}{m_t}\right) s_\gam^3 c_\gam <0.0019\,,  \\
&\left|\lam_7\right|\sim \frac{3y_t^4 }{4 \pi^2}\log\left(\frac{M}{m_t}\right) s_\gam c_\gam^3 <0.045\,.
\end{align}
Note that $\lam_5,~\lam_6$, and $\lam_7$ vanish when $s_\gamma c_\gamma=0$ since this limit corresponds to an unbroken Peccei-Quinn $\ug$ symmetry. 

In the case of $R=1$, the tree-level theory is a softly-broken SO($3$)-invariant 2HDM. Below the scale $M$, a shift $\delta R\equiv R-1$ will be generated after integrating out the vector-like fermions.   However, some care is needed in defining what we mean by $R$ since in the one-loop approximation we no longer have $\lambda\equiv \lambda_1=\lambda_2$ and $\lambda_5=0$ in light of \eqs{lam12diff}{lamfive}.  Thus, we redefine,
\beq \label{Rdefgen}
R\equiv\frac{\lambda_3+\lambda_4+\lambda_5}{\half(\lambda_1+\lambda_2)}\,,
\eeq
which reduces to our previous definitions of $R$ in the case of $\lambda_1=\lambda_2$ [cf.~\eq{aredef}] and $\lambda_5=0$ [cf.~\eq{Rdef}].
Using \eqst{deltalams}{kays}, we then find that the shift in the $R$ parameter in the one-loop approximation is roughly given by,\footnote{Note that in the one-loop approximation, $\delta R=0$ if $\cos 4\gamma=0$.  Nevertheless, this limiting case does not correspond to the presence of an unbroken SO($3$) symmetry given that $\lambda_1\neq\lambda_2$ and $\lambda_5\neq 0$  if $\cos 4\gamma=0$, in light of \eqs{lam12diff}{lamfive}.  The vanishing of $\delta R$ when $\cos 4\gamma=0$ must be regarded as
an accidental cancellation that is not expected to persist at higher orders in the loop expansion.} 
\begin{align}
\lambda \left| \delta R \right| = \left| \delta\bigl(\lambda_3+\lambda_4+\lambda_5-\half\lambda_1-\half\lambda_2\bigr)\right| \sim\frac{3 y_t^4}{8 \pi^2}\log\left(\frac{M}{m_t}\right) |\cos 4\gamma |< 0.083\,,
%+\mathrm{O}\left(y_t^8\right) \right|< 0.05.
\end{align}
for the same choice of parameters employed below \eq{lam12diff}.

Numerically, in the parameter regions of interest, the corrections to the relations $\lambda_1=\lambda_2$, $\lambda_5=\lambda_6=\lambda_7=0$ (and the deviation of $R$ from 1 in the case of a softly-broken SO($3$)-invariant 2HDM) are  small.  Hence, in our present study, we shall simply neglect these effects as they will have a negligible numerical impact on our final results.

\subsection{Top quark mixing after EWSB}\label{section:TopQuarkMixing}

Additional mixing of fermionic states can occur once the electroweak symmetry breaking effects are taken into account~\cite{Arhrib:2016rlj,Aguilar-Saavedra:2013qpa}.\footnote{Of course, this analysis could have been carried out in one step by first employing \eq{vevshift} in \eq{minuslag} and then computing the mixing of the top quark with its vector-like partners.  Details can be found in Appendix~\ref{twostep}.}
After inserting
\beq \label{vevshift}
\Phi_i^0 = \frac{v_i}{\sqrt{2}}+\overline{\Phi}_i\llsup{0}\,,\qquad \text{for $i=1,2$},
\eeq
in \eq{twover}, we denote $\hat{\chi}=(u \quad U)$ as before and $\hat{\eta}^j=(\bar{x}_0\quad \ol{X}_0)$.  
These states are now considered to be interaction eigenstates. The two-component Dirac fermion mass matrix defined in \eq{Dmassmat} is now given by,
\beq \label{calmtwo}
\mathcal{M}=\begin{pmatrix}  Ys_{\beta-\gamma} & \,\,\, Y c_{\beta-\gamma}\\  0 & \,\,\,M \end{pmatrix},
\eeq
where 
\beq \label{Ydef}
Y\equiv \frac{y_t v}{\sqrt{2}}\,.  
\eeq

Note that if $c_{\beta-\gamma}=0$ then $\mathcal{M}$ is diagonal, and no additional mixing between the top quark and its vector-like partners is generated.  However, since $\beta$ and $\gamma$ are independent parameters, the generic case yields additional mixing effects.
Using the results of Appendix~\ref{app:svd}, the fermion mass spectrum consists of two Dirac fermions 
with squared-masses,
\beq \label{msquared}
m^2_{T,t}=\half\biggl\{M^2+Y^2\pm\sqrt{(M^2+Y^2)^2-4Y^2 M^2 s_{\beta-\gamma}^2}\,\biggr\},
\eeq
where $s_\gamma$ and $c_\gamma$ are defined in \eq{angle} and $m_T>m_t$.    Note that,
\beq \label{prodsum}
m_T^2+m_t^2=M^2+Y^2\,,\qquad\quad m_T^2 m_t^2=Y^2 M^2 s_{\beta-\gamma}^2\,.
\eeq

The singular value decomposition of $\mathcal{M}$ 
[\eq{calmtwo}]
yields 
two mixing angles, $\theta_L$ and~$\theta_R$, 
\beqa
\sin 2\theta_L&=&\frac{2YMc_{\beta-\gamma}}{m_T^2-m_t^2}\,,\qquad\qquad\,\,\, \cos 2\theta_L=\frac{M^2-Y^2}{m_T^2-m_t^2}\,,\label{phione} \\[8pt]
\sin 2\theta_R&=&\frac{Y^2 \sin 2(\beta-\gamma)}{m_T^2-m_t^2}\,,\qquad\, \cos 2\theta_R=\frac{M^2+Y^2\cos 2(\beta-\gamma)}{m_T^2-m_t^2}\,.\label{phitwo}
\eeqa
Note that \eqs{phione}{phitwo} determine both $\theta_L$ and $\theta_R$ modulo~$\pi$.  
In addition to the two mixing angles, the matrices $L$ and $R$ given in \eq{eq:OLR} also depend on $\varepsilon_L$ and $\varepsilon_R$, where $\varepsilon_L\varepsilon_R={\rm \sgn}(s_{\beta-\gamma})$.  

One can make use of \eq{phione} to obtain the following convenient expression,
\beq \label{tanthetaL}
\tan\theta_L=\frac{\sin 2\theta_L}{1+\cos 2\theta_L}=\frac{2YMc_{\beta-\gamma}}{m_T^2-m_t^2+M^2-Y^2}=\frac{YMc_{\beta-\gamma}}{M^2-m_t^2}\,.
\eeq
after using \eq{prodsum} in the last step above.  One can then employ \eq{amusingid} to determine $\theta_R$, which shows that the mixing angles $\theta_L$ and $\theta_R$ are not independent quantities~\cite{Aguilar-Saavedra:2013qpa},
\beq \label{tanid}
\tan\theta_R=\varepsilon_L\varepsilon_R\frac{m_t}{m_T}\tan\theta_L\,.
\eeq 

Since no vector-like top quarks have been observed so far at the LHC, it follows that $m_T\gg m_t$.  Thus, we can obtain useful approximations to the relationship between the physical masses and the parameters $M$ and $Y$ as well as approximations for $\theta_L$ and $\theta_R$.   For example, \eq{msquared} yields,
\beqa
m_t^2&=&Y^2 s^2_{\beta-\gamma}\left[\left(1-\frac{Y}{M} c_{\beta-\gamma}\right)^2+\mathcal{O}\left(\frac{Y^2}{M^2}\right)\right]\,,\\
m_T^2&=&M^2+Y^2\left[c^2_{\beta-\gamma}+\mathcal{O}\left(\frac{Y^2}{M^2}\right)\right]\,.
\eeqa
In a convention where $Y$, $M_u$, $M_U$ and the vevs $v_1$ and $v_2$ are positive, it follows that $0\leq\beta, \gamma\leq\half\pi$, which implies that
$0\leq c_{\beta-\gamma}\leq 1$ and $-1\leq s_{\beta-\gamma}\leq -1$.  Hence,

\beqa
m_t &\simeq &Y|s_{\beta-\gamma}|\left(1-\frac{Y}{M}c_{\beta-\gamma}\right)\,,\label{tmass}\\
M_T &\simeq & M\left[1+\frac{m_t^2}{2M^2}\cot^2(\beta-\gamma)\right]\,.\label{vtmass}
\eeqa
Likewise, we can use \eqs{tanthetaL}{tanid} to obtain,
\beq \label{thetaellare}
\theta_L \simeq \frac{m_t}{M_T}|\cot(\beta-\gamma)|\,, \qquad\quad
\theta_R \simeq  \frac{m_t^2}{M_T^2}\cot(\beta-\gamma)\,.
\eeq

The two-component fields $u$ and $U$ mix to form two-component fermion mass eigenstates that we shall denote by $\chi_k=(x \quad X)$.  Likewise, 
$\bar{u}$ and $\ol{U}$ mix to form two-component fermion mass eigenstates that we shall denote by $\eta^k=(\bar{x}\quad \ol{X})$.  
Note that nothing depends on the separate values of $\varepsilon_L$ and $\varepsilon_R$; only its product is determined.  Henceforth, we shall 
take $\varepsilon_R=1$ with no loss of generality.
Then, the fermion mass eigenstates are explicitly given by
\beqa
x&=&-s_L U+c_L u\,,\qquad\quad X=\varepsilon_L(c_L U+s_L u)\,,\\
\bar{x}&=&-s_R\ol{X}_0+c_R \bar{x}_0\,,\qquad\quad \ol{X}=c_R\ol{X}_0+s_R\bar{x}_0\,,
\eeqa
where $s_{L,R}\equiv \sin\theta_{L,R}$ and $c_{L,R}\equiv \cos\theta_{L,R}$.

Plugging the above results back into \eq{twover} yields the following mass terms and interactions among the fermions and the neutral Higgs fields,
\beq \label{twover2}
-\mathscr{L}\subset y_t \bigl\{(c_L x+\varepsilon_L  s_L X)\bigl[(\overline{\Phi}^0_2 s_{\gamma+\theta_R}+\overline{\Phi}^0_1c_{\gamma+\theta_R})\ol{X}+(\overline{\Phi}^0_2 c_{\gamma+\theta_R}-\overline{\Phi}^0_1 s_{\gamma+\theta_R})\bar{x}\bigr]\bigr\}+m_T\ol{X}X+m_t\bar{x}x+{\rm h.c.}
\eeq
One can introduce four-component fermions fields,
\beq
t=\begin{pmatrix} x \\ \bar{x}\end{pmatrix}\,,\qquad T\equiv \begin{pmatrix}  X \\ \ \ol{X}\end{pmatrix}\,,
\eeq
where $t$ is the physical top quark field.
Then the four component fermion version of \eq{twover2} is,
\beqa
&& \hspace{-0.45in}
 -\mathscr{L}\subset m_t\bar{t}{t}+m_T\overline{T}T+  y_t \bigl\{c_L(\ol{\Phi}^0_2 c_{\gamma+\theta_R}-\ol{\Phi}^0_1 s_{\gamma+\theta_R})\bar{t}t
+\varepsilon_L s_L(\ol{\Phi}^0_2 s_{\gamma+\theta_R}+\ol{\Phi}^0_1c_{\gamma+\theta_R})\ol{T}T\bigr\} \nonumber \\
&& \hspace{-0.15in}
+ y_t \bigl\{c_L(\ol{\Phi}^0_2 s_{\gamma+\theta_R}+\ol{\Phi}^0_1c_{\gamma+\theta_R})(\ol{T}P_L t+{\rm h.c.)}+s_L(\ol{\Phi}^0_2 c_{\gamma+\theta_R}-\ol{\Phi}^0_1 s_{\gamma+\theta_R})(\bar{t}P_L T+{\rm h.c.})\bigr\}.
\eeqa

\vspace*{-0.2cm}
\subsection{Relaxing the GCP\texorpdfstring{$3$}{3} symmetry}
\label{relax}

At the end of \sect{sec:enhanced}, we motivated our study of the softly-broken GCP$3$-symmetric model by noting that it provided a useful simplification by removing the possibility of CP violation in the scalar potential. The absence of CP violation is maintained when including
the coupling of the scalars to one generation of fermions and their vector-like partners.  

Of course, the current Higgs data does not yet rule out the possibility of new sources of CP violation in the scalar sector. Our choice to do so is a matter of convenience, since the neutral mass-eigenstates of a CP-conserving 2HDM are eigenstates of CP consisting of the SM-like Higgs boson, its CP-even scalar partner and a CP-odd scalar. This avoids the necessity of diagonalizing a $3\times 3$ neutral scalar squared-mass matrix and the introduction of additional mixing angles that would be necessary to fully treat the neutral Higgs scalar phenomenology.

We now briefly discuss the possibility of relaxing the softly-broken GCP$3$ symmetry to a softly-broken GCP$2$ symmetry. One way to maintain the CP invariant scalar potential is to assume that $\xi=0$ in \eq{CPcond}.\footnote{A more complete discussion of the softly-broken GCP$2$-symmetric scalar potential where CP invariance is maintained can be found in Ref.~\cite{Haber:2021zva}.}  In this case, it is easy to extend the results of \sect{gcpthree}.  
As discussed in \sect{sec:enhanced}, it is sufficient to employ a basis where the discrete $\mathbb{Z}_2\otimes\Pi_2$ symmetry is manifest in the quartic terms of the scalar potential.   In this basis, $\lambda_5$ is real and nonzero and $m_{12}^2$ is either purely real or purely imaginary (if the latter, then one can rephase, $\Phi_2\to i\Phi_2$ to obtain a real $m_{12}^2$, while flipping the sign of $\lambda_5$).    
Then, all the formulae obtained in \sect{gcpthree} are still valid with the following simple modifications: $R$ is now given by \eq{aredef}, and $m_A^2$ is replaced by $m_A^2+\lambda_5 v^2$.

Under the $\mathbb{Z}_2\otimes\Pi_2$ symmetry transformations, we can impose the transformation laws specified in eqs.~(\ref{trans1}), (\ref{trans2}), (\ref{trans3}) and (\ref{PQlaw}) [and likewise in \eqs{trans4}{trans5}] by setting $\theta=\pi$, in which case a $\ug$ transformation reduces to a $\mathbb{Z}_2$ transformation. Indeed, the imposition of the $\mathbb{Z}_2\otimes\Pi_2$ symmetry on the Yukawa sector automatically yields Yukawa couplings that are invariant under $\ug\otimes\Pi_2$. Consequently, the general structure of the $\ug\otimes\Pi_2$-symmetric Yukawa couplings obtained previously remain unchanged. We can therefore conclude that the numerical analysis presented in \sect{section:Results} for the softly-broken GCP$3$ model also would apply to a softly-broken GCP$2$ model with $\xi=0$ by simply reinterpreting the parameters $R$ and $m_A^2$ as indicated above.

The special case of the softly-broken GCP$2$ model with $\xi=0$ was previously treated in Ref.~\cite{Draper:2016cag}, where it was further assumed that $m_{12}^2=0$ and $M_u=0$.
However, as the next subsection shows, these additional parameter assumptions may be too constraining, and in this paper we have argued that there is no motivation for imposing such additional parameter restrictions.

\vspace*{-0.2cm}
\subsection{Fine-tuning and electroweak precision}\label{section:FineTuning}
\vspace*{-0.2cm}
In \sect{sec:BreakDiscrete} we argued that soft $\mathrm{U}(1)\otimes \Pi_2$ symmetry breaking terms in the scalar sector given by $\Delta m^2\equiv m_{22}^2-m_{11}^2$ and $m_{12}^2$ can be generated from soft-symmetry breaking in an extended Yukawa sector. This extended sector includes a new vector-like top-partner $T$ with a mass of order the $\si{\tera\electronvolt}$ scale. Furthermore, \eqs{eq:Deltam2}{eq:m12} show that by evolving down from a UV theory at the cutoff scale $\Lambda_c$ to the mass scale of the top-quark partner that is characterized by $m_T$, non-zero values for the scalar squared mass parameters $\Delta m^2$ and $m_{12}^2$ are generated,
\beqa
	&&\Delta m^2 \sim  \kappa_{\Delta m^2} M^2 c_{2\gam}-\frac{3y_t^2 M^2 c_{2\gam}}{4\pi^2}\log\left(\frac{\Lambda_c}{m_T}\right),\label{eq:RadGeneratedParam} \\
	&&m_{12}^2\sim \kappa_{m_{12}^2}M^2 s_{2\gamma}+\frac{3 y_t^2 M^2 s_{2\gamma}}{8 \pi^2}\log\left(\frac{\Lambda_c}{m_T}\right),\label{eq:RadGeneratedParam2}
\eeqa
where $\gamma$ and $M$ are defined in \eq{angle}. We would expect in the absence of fine-tuning that $\Delta m^2$ and $m_{12}^2$ are of the same order as the logarithmic terms. However, if the vector-like quark mass is large compared to the weak scale ($m_T\gg v$), a tuning of the $\kappa$ parameters in \eqs{eq:RadGeneratedParam}{eq:RadGeneratedParam2} is needed to keep the scalar squared mass parameters small.

To make the degree of tuning more transparent we approximate $M^2\approx m_T^2$ (in the limit of $M^2\gg Y^2$) by using \eq{msquared}.  
Specific numerical results will depend on the choice of $\Lambda_c$, which should lie sufficiently above $m_T$  so that the (renormalizable) 2HDM, extended to include a vector-like top quark partner, is a good effective field theory, but not so far above $m_T$ that the presence of
the logarithmically enhanced terms in 
\eqs{eq:RadGeneratedParam}{eq:RadGeneratedParam2} requires significant fine-tuning as noted above. In order to provide a concrete example for our subsequent numerical studies, we shall choose $\log(\Lambda_c/M)=3$. 
Using \eqs{eq:RadGeneratedParam}{eq:RadGeneratedParam2}, we then find two possible estimates of~$m_T^2$,
\beq\label{eq:mTEstimate}
	m_T^2\sim \frac{4 \pi^2}{9y_t^2}\left|\frac{\Delta m^2}{c_{2\gamma}}\right| \equiv  |c_\Delta\Delta m^2|  \qquad\text{or}\qquad m_T^2\sim \frac{8 \pi^2}{9 s_{2\gam}y_t^2}m_{12}^2\equiv c_{12}m_{12}^2\,,
\eeq
in a convention where $m_{12}^2$ and $s_{2\gamma}$ are both positive quantities.

If the true value of $m_T$ is significantly larger than these estimates, then there must be a tuning of $\kappa_{\Delta m^2}$ and $\kappa_{m_{12}^2}$ in \eqs{eq:RadGeneratedParam}{eq:RadGeneratedParam2}.
To estimate this tuning one can compare the estimated value of $m_T^2$ in \eq{eq:mTEstimate} to the true value. In particular, a tuning of one part in $N$ corresponds to a squared 
top-quark partner mass $m_T^2=(N-1)|c_\Delta \Delta m^2|$ or $m_T^2=(N-1)c_{12} m_{12}^2$, where
\beqa 
&& \phantom{line}\nonumber  \\[-18pt]
&& \frac{1}{N}=\frac{|c_\Delta\Delta m^2|}{|c_\Delta\Delta m^2|+m_T^2} \qquad \text{or}\qquad
\frac{1}{N}=\frac{c_{12}m_{12}^2}{c_{12}m_{12}^2+m_T^2}\,, \label{eq:TuningMeasure}
\eeqa
corresponding to the two possible estimates of \eq{eq:mTEstimate}, respectively.

These two tuning measures depend on scalar and Yukawa sector parameters and are in general quite different. Yet we can get a feeling for the tuning measures by recalling the formulas for $\Delta m^2$ and $m_{12}^2$ given in section \ref{gcpthree}. Thus we take $m_A$,~$R$, and $\beta$ as free parameters and investigate the impact of each individually. First, \eqs{betaeq}{mha} show that by increasing $m_A$, while holding the other two parameters fixed, the tuning is reduced. This is quite natural since a larger $m_A$ implies a smaller hierarchy between scalar masses and $m_T$. Second, by varying $\beta$ alone we see that $\Delta m^2$ vanishes when $\beta=\tfrac14\pi$, and $m_{12}^2$ vanishes when $\beta=0,~\half\pi$. In each of these limits one of the tunings in equation \eq{eq:TuningMeasure} becomes large. Lastly, the $R$ dependence is quite weak compared to the $m_A$ and $\beta$ dependence.

Next, consider the Yukawa-sector parameters, which now include two additional free parameters, $m_T$ and $\gamma$, in the top quark sector. The $m_T$ dependence of the tuning measures is given by \eq{eq:TuningMeasure}, while the $\gamma$ dependence is more interesting.  As previously noted below \eq{whytop}, we shall avoid the region of $\gamma\sim\beta$ where 
$y_t\gg 1$ in order to maintain the perturbativity of the top quark Yukawa coupling. Furthermore, $\gamma$ is constrained by electroweak precision measurements. Namely, an analysis of electroweak precision constraints\footnote{See, e.g., J.~Erler and A.~ Freitas, \textit{Electroweak Model and Constraints on New Physics}, in Ref.~\cite{Zyla:2020zbs}.} (which determine the allowed range of the electroweak oblique parameters\cite{Peskin:1990zt,Peskin:1991sw}) shows that $\gamma$ must satisfy $\sin\theta_L \lsim 0.1$ for $m_T=1.5~\si{\tera\electronvolt}$. This, together with equation \eq{thetaellare}, implies that $\left|\cot\left(\beta-\gamma\right)\right| \leq 0.85$~\cite{Dawson:2012di,Arhrib:2016rlj}. 
Solving this inequality results in two allowed regions, 
\begin{align}
	 0\leq\beta\leq \gamma-0.87 \quad \text{or} \quad\beta \geq \gamma+0.87\,,
\end{align}
in our convention where $0\leq\beta, \gamma\leq\half\pi$.
These allowed regions correspond to the ranges of $\tan\beta$,
\begin{align}
	\tan\beta\leq 0.84 \quad \text{or}\quad \tan \beta \geq 1.19\,.
\end{align}
It is noteworthy that
$\tan\beta=1$ is not allowed in this model.  Moreover the range of $\tan\beta$ above~1 is the preferred one in light of the remarks below \eq{whytop}.

While this paper focuses on the GCP$3$ model, the above discussion also applies to the CP-conserving GCP$2$ model, as noted in \sect{relax}. Indeed, much of the parameter space for the softly-broken GCP$2$ model treated in Ref.~\cite{Draper:2016cag} is ruled out by experimental limits on the mixing of the top quark with its vector-like partner. In particular, the model in Ref.~\cite{Draper:2016cag} corresponds to $\gamma=0$, and the mixing constraints implies that $\tan \beta \geq 1.19$ (or equivalently, $\epsilon \equiv \cos 2\beta \lsim -0.17$). This means that most parameter space is ruled out if $m_{12}^2=0$, which provides further motivation for analyzing the more generic case where $m_{12}^2\neq 0$.

\vspace*{-0.5cm}
\section{Survey of the parameter space consistent with LHC Higgs data and searches}
\label{analysis}

\vspace*{-0.3cm}
\subsection{LHC constraints}

In this section we assess the experimental constraints on the softly-broken GCP$3$ model. In addition to new scalar particles ($A$,~$H$,~{$H^\pm$}), the model contains a vector-like top-quark partner. This top partner constrains the model directly through collider searches, and indirectly through tuning of $\Delta m^2$ and $m_{12}^2$ as discussed in \sect{section:FineTuning}.

There are a few variations of the model depending on the Yukawa sector. Recall that to naturally avoid tree-level Higgs-mediated FCNCs, the structure of the Higgs-fermion Yukawa couplings of the 2HDM must be of Type-I,~II,~X, or Y; in this work we focus on Type-I and Type-II. Because the SM-like Higgs-boson mass, $m_h=125$~GeV, is known, the scalar sector has 4 free parameters, which we 
 choose to be $m_A,~R,~m_{H^\pm},~\beta$. One of these parameters can be dropped by assuming $m_{H^{\pm}}=m_A$.  This choice minimizes the Higgs-mediated radiative corrections to the tree-level value of the electroweak $\rho$-parameter, leaving $m_A,~R$, and $\beta$ as the remaining free parameters.\footnote{In the Type-II scenario, in light of the most recent theoretical analysis of the SM prediction for the decay rate of $b\to s\gamma$ at NNLO, one can deduce that $m_{H^\pm}\gsim 800$~GeV~\cite{Misiak:2020vlo}.   The corresponding constraint in the Type-I scenario is far less severe~\cite{Arbey:2017gmh}, and allows for a charged Higgs mass in our parameter region of interest for values of $\tan\beta\gsim 2$.  
 Unless the Type-II prediction for the $b\to s\gamma$ decay rate is modified by loop contributions from the vector-like quarks~\cite{Vatsyayan:2020jan} or some other new physics phenomena,  the end result will be to favor the Type-I scenario and strongly disfavor the Type-II scenario in regions where approximate Higgs alignment without decoupling can be achieved.}
We also assume that $m_A>m_h$ since no CP odd Higgs scalar has been found in LHC searches.   Since the goal of this paper is to achieve Higgs alignment without decoupling, we consider 
$m_A\in\left[150,500\right]~\si{GeV}$ (whereas the mass of $H$ can be slightly larger). 

The mass $m_H$ and the scalar potential parameter $\lambda$ can be expressed in terms of $m_A$, $R$ and~$\beta$ by using the relations in \sect{gcpthree}. For example, one can derive a quadratic equation for~$\lambda$
 by multiplying \eq{ineq1} by $m_h^2$ and then making use of \eq{ineq2} to rewrite the product $m_h^2 m_H^2$ in terms of $m_A^2$, $R$ and $\beta$.  The end result is
 \beq
 \lambda^2 v^4s_{2\beta}^2 (1-R^2)-4\lambda v^2\bigl\{m_A^2\bigl[1-\half s^2_{2\beta}(1-R)\bigr]-m_h^2\bigr\}-4m_h^2(m_A^2-m_h^2)=0\,.
 \eeq
 Under the assumption that $m_A>m_h$, the roots of this quadratic equation are real and their product is negative. Since $\lambda>0$, one must choose the positive root.
 
The extended Yukawa sector yields additional free parameters as described in \sect{section:TopQuarkMixing}. Prominent amongst these are the mass of the vector-like top partner, $m_T$, and the top quark mixing angles $\theta_L$ and~$\theta_R$. Other free parameters such as $Y$ and $M$ are related to $m_T$ and the angles $\beta$ and $\gamma$. For example, in the limit of $m_T\gg m_t$, the masses of the top quark and its vector-like partner are given by \eqs{tmass}{vtmass}, and the top quark mixing angles are given by \eq{thetaellare}.

Because the vector-like top quark partner mixes with the SM top quark, it can decay into $tZ$, $th$, and $bW$. Experimental searches at $13~\si{\tera\electronvolt}$~\cite{Sirunyan:2018qau,Aaboud:2018pii} constrain a vector-like quark decaying predominately to these particles to have a mass greater than $1.3$--$1.4~{\rm TeV}$ depending on the relative size of the branching ratios. In our model, these bounds are likely too strong since the vector-like quark can also decay to $tH,~tA$, and $t H^\pm$. These decays are unsuppressed by top quark mixing and are expected to dominate, so the true experimental lower bound on the mass of the vector-like quark may even lie somewhat below $1~\si{\tera\electronvolt}$.

In this paper we focus on the scalar sector.  We take the mass of the vector-like top partner to be $m_T=1.5~\si{\tera\electronvolt}$ in order to safely evade any collider bounds. Moreover, as remarked in \sect{section:FineTuning}, the measured values of the electroweak oblique parameters constrain $\sin \theta_L \leq 0.1$ for $m_T=1.5~\si{\tera\electronvolt}$.\footnote{This bound can be softened by taking $m_{H^{\pm}} \neq m_A$, because the Higgs-mediated contribution to the electroweak oblique $T$ parameter (in the one-loop approximation) are of opposite sign to the vector-like quark contribution~\cite{Lavoura:1992np,Dawson:2012di,Grimus:2007if,Haber:2010bw}.} Finally, the vector-like top partner can contribute to scalar production and decays through loops. This effect is quite small with our chosen vector-like top quark mass and will be neglected in the analysis presented below. 

Scalar parameters are constrained from precision measurements and collider searches. On the precision side, the measured couplings of the observed SM-like  $125~\si{\giga\electronvolt}$ Higgs boson greatly limit the allowed regions of the $\tan \beta$ vs.~$\cos(\beta-\alpha)$ parameter space~\cite{Haller:2018nnx,Chowdhury:2017aav,Aad:2019mbh,Sirunyan:2018koj}. These constraints are particularly severe for the Type-II Yukawa coupling scenario, where only a small deviation from $\cos(\beta-\alpha)=0$ is allowed.

Cross sections and branching ratios of new
heavy scalars are also constrained by direct collider searches. For the low-mass region of interest\te $m_A$, $m_H\in [150,500]~\si{\giga\electronvolt}$, leptonic decay channels ($A/H\rightarrow \tau \tau)$ are strongly constrained.  Experimental limits obtained by CMS and ATLAS~\cite{Sirunyan:2018zut,CMS:2019hvr,Aaboud:2017sjh,Aad:2020zxo} restrict the small $\tan \beta$ region in a Type-I scenario, and likewise place limits on the large $\tan \beta$ region in a Type-II scenario. 
Other channels like $A\rightarrow Z h$~\cite{Aaboud:2017cxo,Sirunyan:2019xls,Sirunyan:2019xjg,ATLAS:2020pgp}
and $A\rightarrow \gamma \gamma$~\cite{Aaboud:2017yyg,Khachatryan:2016hje,Aad:2021yhv} are most relevant for $A$ masses above $\sim 220~\si{\giga\electronvolt}$. These two channels are important for small $\tan\beta$ values in both the Type-I and Type-II scenarios. The diphoton channel is of particular interest since not only small masses ($m_A\lsim 250~\si{\giga\electronvolt}$) are constrained, but also large masses ($m_A\gsim 300~\si{\giga\electronvolt}$).
We have also considered channels such as $H\to ZA$ and $A\to ZH$~\cite{Sirunyan:2019wrn,Aad:2020ncx}, which are not suppressed in the Higgs alignment limit.
 Rates for all of these processes are computed with SusHi~\cite{Harlander:2012pb,Harlander:2016hcx,Harlander:2002wh,Harlander:2003ai,Aglietti:2004nj,Bonciani:2010ms,Eriksson:2009ws,Harlander:2005rq,Chetyrkin:2000yt} and 2HDMC~\cite{Eriksson:2009ws}. For all calculations we neglect the contribution of the vector-like quark in loops since it is expected to be small for $m_T=1.5~{\rm TeV}$.
 
\clearpage

\subsection{Results}\label{section:Results}

\begin{figure}[b!]
     \centering
           \captionsetup[subfigure]{oneside,margin={-0.9cm,0cm}}
     \begin{subfigure}[b]{0.48\textwidth}
         \centering
         \includegraphics[width=\textwidth]{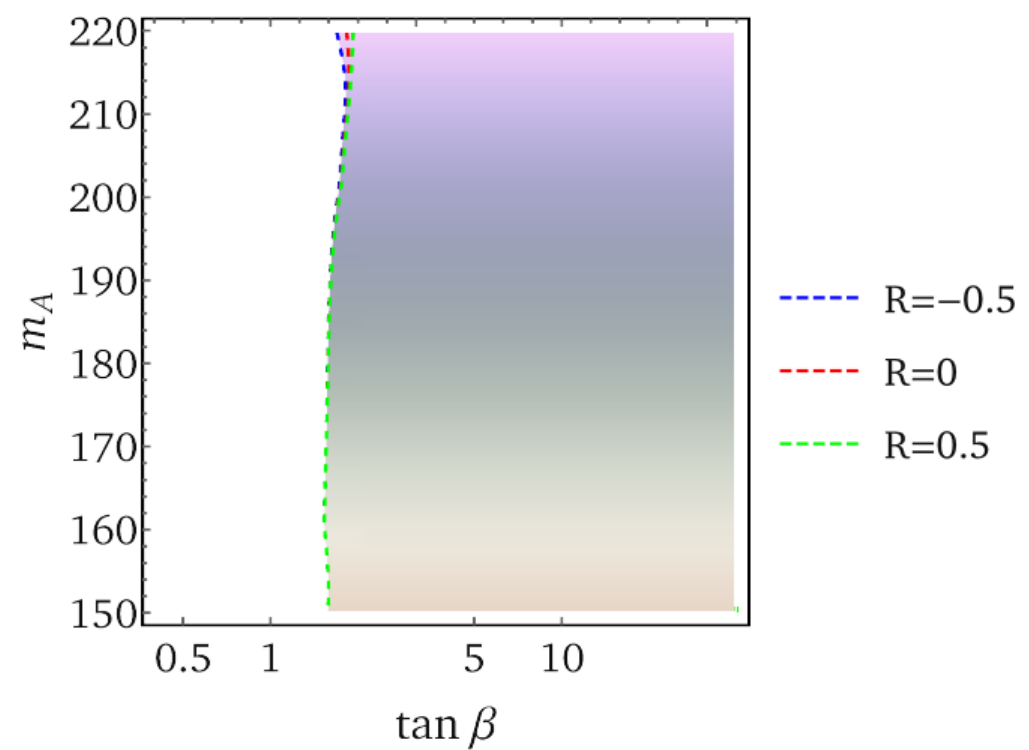}
         \caption{}
         \label{figure:TypeIBoundsA}
     \end{subfigure}
     \hfill
           \captionsetup[subfigure]{oneside,margin={-0.8cm,0cm}}
     \begin{subfigure}[b]{0.48\textwidth}
         \centering
         \includegraphics[width=\textwidth]{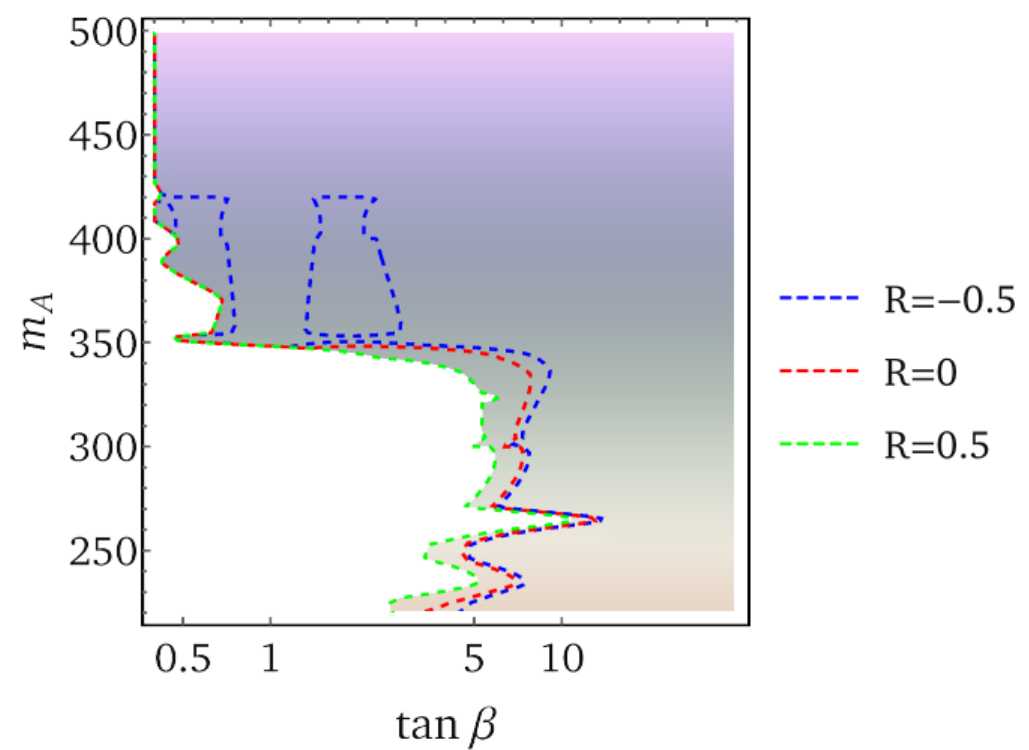}
         \caption{}
         \label{figure:TypeIBoundsB}
     \end{subfigure}
     \newline
           \captionsetup[subfigure]{oneside,margin={-0.9cm,0cm}}
    \begin{subfigure}[b]{0.48\textwidth}
         \centering
         \includegraphics[width=\textwidth]{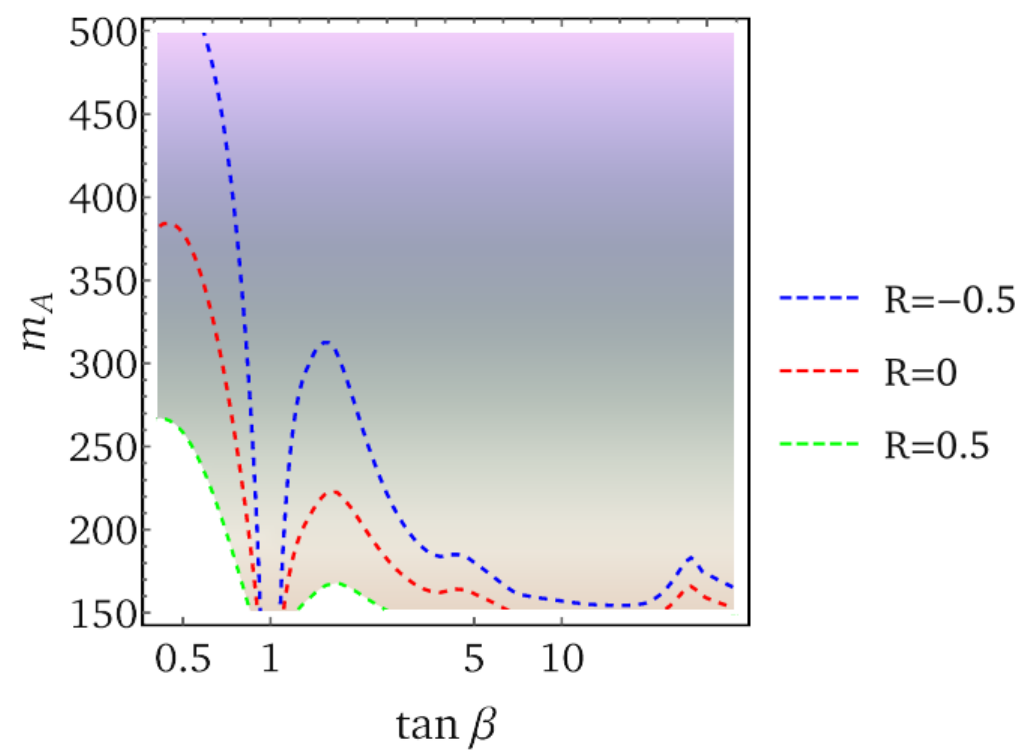}
         \caption{}
         \label{figure:TypeIBoundsC}
     \end{subfigure}
     \hfill
     \captionsetup[subfigure]{oneside,margin={-0.8cm,0cm}}
     \begin{subfigure}[b]{0.48\textwidth}
         \centering
         \includegraphics[width=\textwidth]{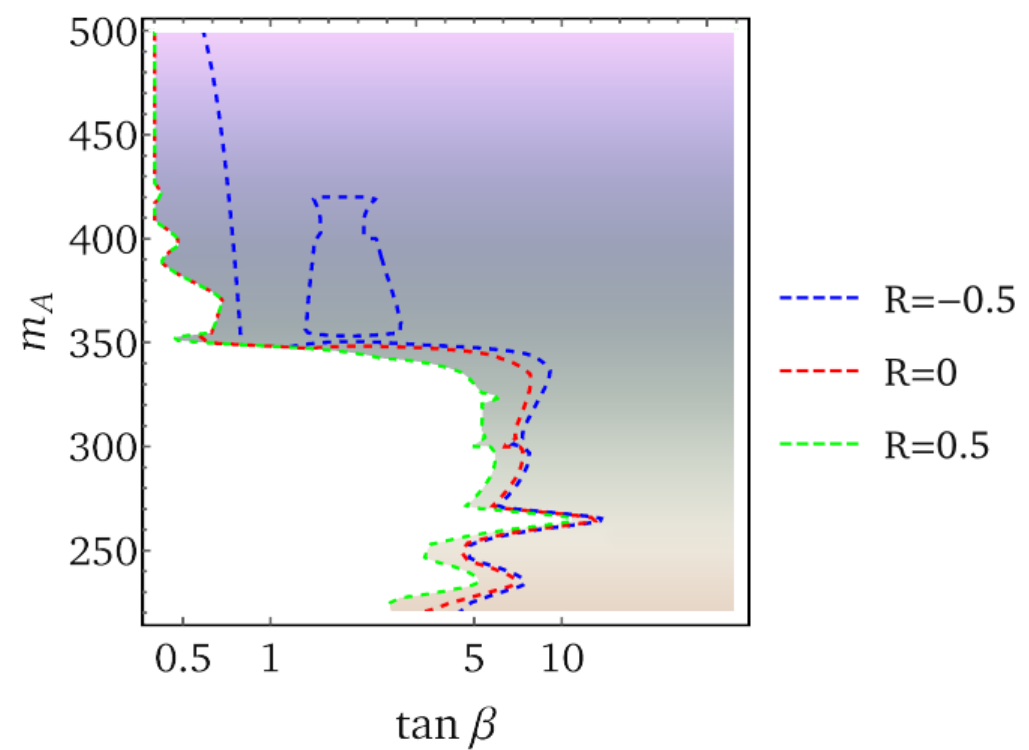}
         \caption{}
         \label{figure:TypeIBoundsD}
     \end{subfigure}
 	\caption{\small Bounds for Type-I Yukawa couplings.  Regions ruled out by (a) $A/H\rightarrow \tau \tau$ data, (b)~combination of collider constraints, including $A/H\rightarrow \tau \tau$, (c) precision Higgs global fits, and (d)~combination of collider bounds and global fits of Higgs precision data. Each panel shows three different $R$ curves; the white regions of the parameter space are ruled out.  In panels (b)--(d) the ruled out areas expand somewhat as $R$ decreases, with the borders of the allowed shaded regions indicated by the corresponding contours.  For $R=-0.5$, the area enclosed by
	the closed dashed blue contour in panels (b) and~(d) is also ruled out.
	There is a different $m_A$ scale in panel (a) as compared to the other three panels because the $A\rightarrow \gamma \gamma$ and $A\rightarrow Zh$ bounds are restricted to $m_A \gsim 220~\si{\giga\electronvolt}$.  The color of the shaded regions bounded by the outermost contour in all figures is chosen solely for its aesthetic allure.}
	\label{figure:TypeIBounds}
\end{figure}

\begin{figure}[t!]
     \centering
     \captionsetup[subfigure]{oneside,margin={-0.9cm,0cm}}
     \begin{subfigure}[b]{0.48\textwidth}
         \centering
         \includegraphics[width=\textwidth]{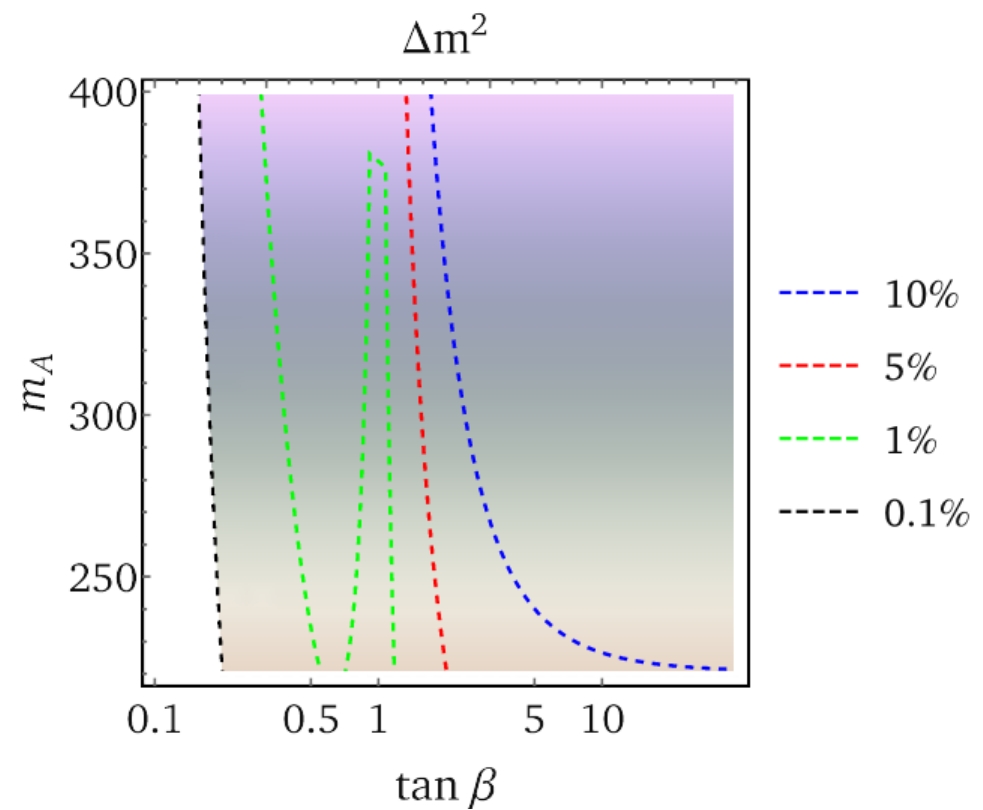}
         \caption{}
         \label{figure:TuningA}
     \end{subfigure}
     \hfill
      \captionsetup[subfigure]{oneside,margin={-0.1cm,0cm}}
     \begin{subfigure}[b]{0.48\textwidth}
         \centering
         \includegraphics[width=\textwidth]{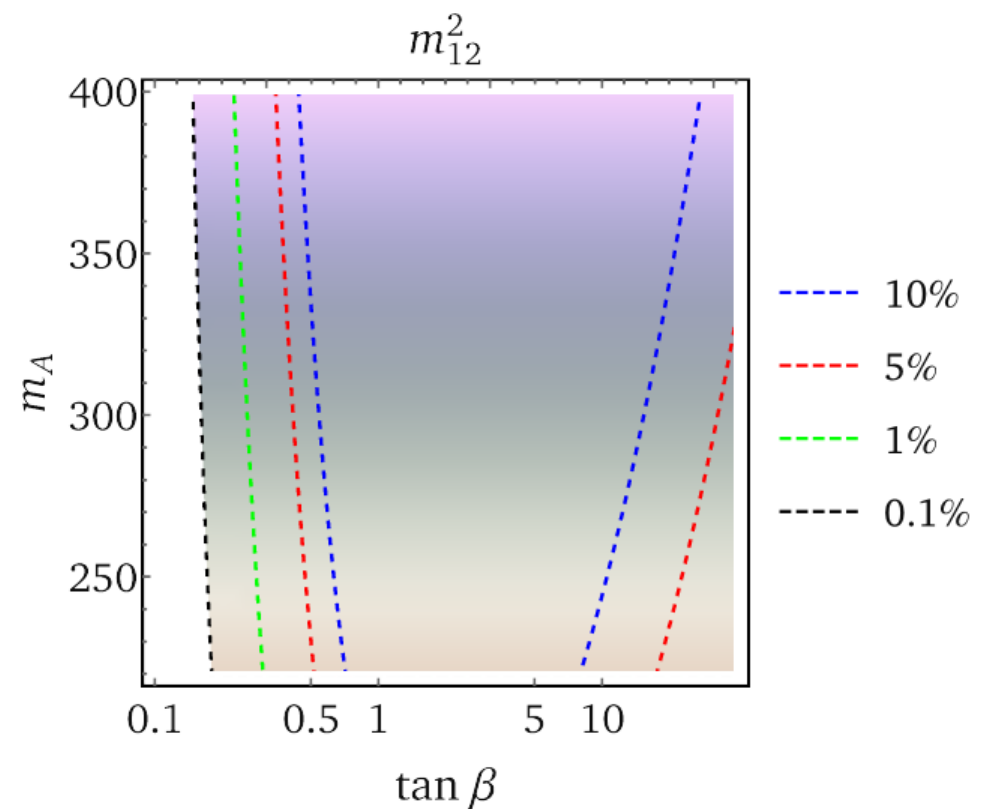}
         \caption{}
         \label{figure:TuningB}
     \end{subfigure}
     \newline
     \captionsetup[subfigure]{oneside,margin={-0.9cm,0cm}}
    \begin{subfigure}[b]{0.48\textwidth}
         \centering
         \includegraphics[width=\textwidth]{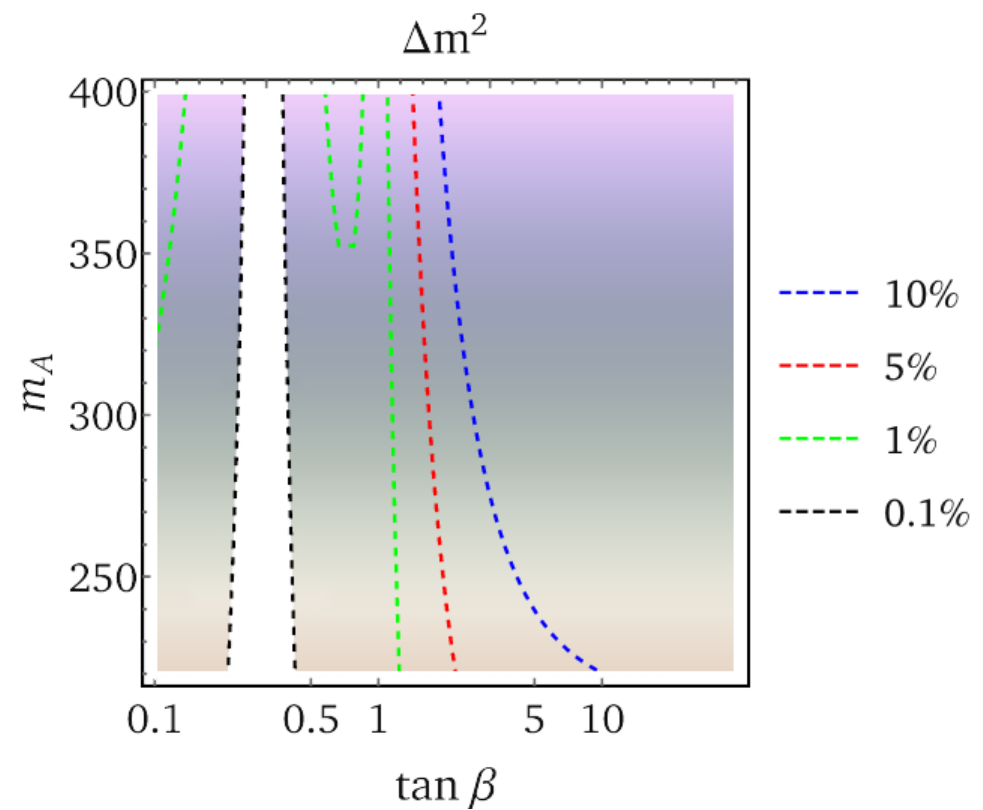}
         \caption{}
         \label{figure:TuningC}
     \end{subfigure}
     \hfill
     \captionsetup[subfigure]{oneside,margin={-0.1cm,0cm}}
     \begin{subfigure}[b]{0.48\textwidth}
         \centering
         \includegraphics[width=\textwidth]{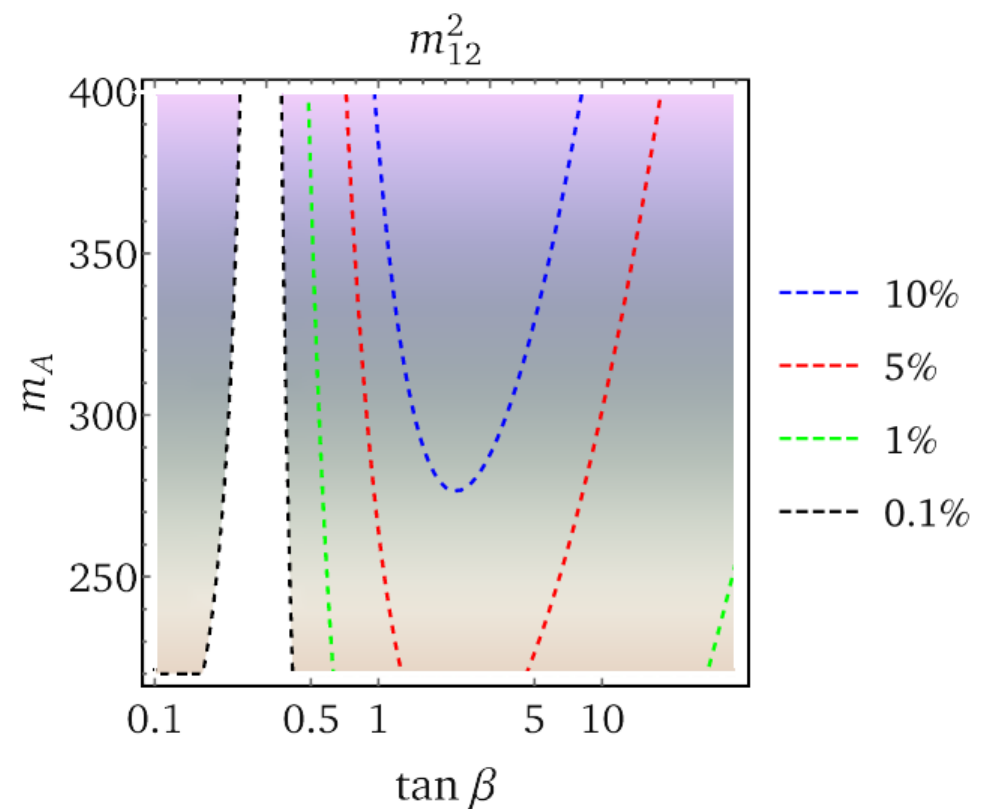}
         \caption{}
         \label{figure:TuningD}
     \end{subfigure}
	\caption{\small Contours of fine-tuning as defined in \eq{eq:TuningMeasure}. Panels (a) and (b) employ the benchmark points $R=0$ and $\gamma=0.1$, whereas panels (c) and (d) employ  the benchmark points $R=0$ and $\gamma=0.3$.  The shaded region 
	inside and/or above each respective contour satisfies the corresponding tuning constraint.
}
	\label{figure:Tuning}
\end{figure}

\begin{figure}[t!]
     \centering
     \captionsetup[subfigure]{oneside,margin={-1.0cm,0cm}}
     \begin{subfigure}[b]{0.48\textwidth}
         \centering
         \includegraphics[width=\textwidth]{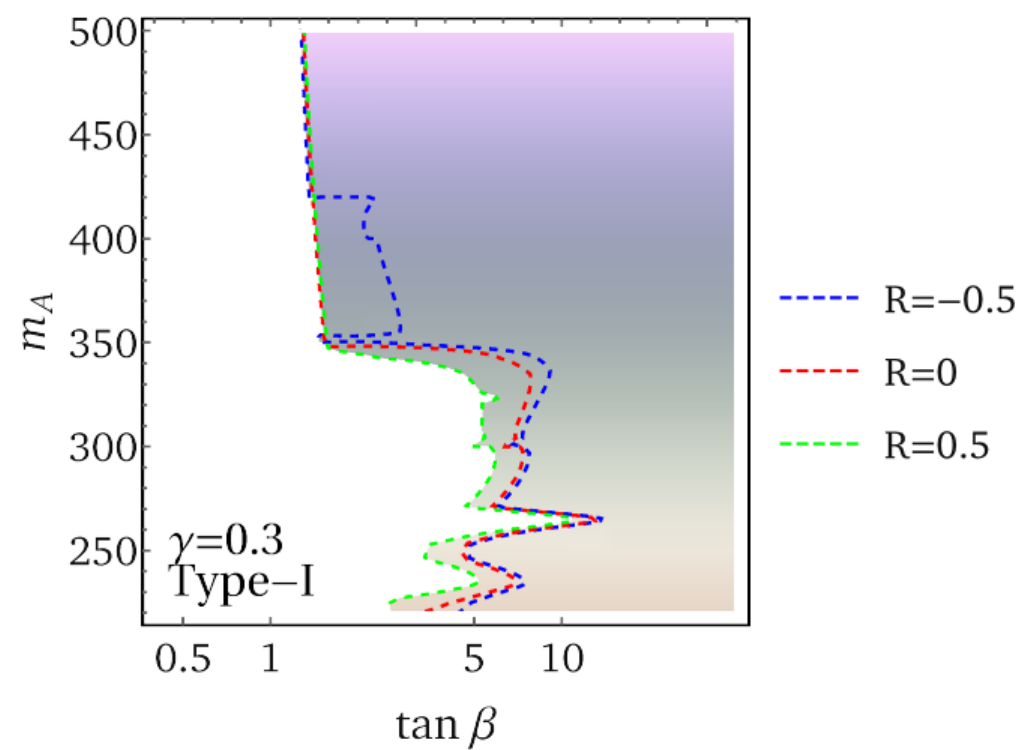}
         \caption{}
         \label{figure:TuningConstraintsA}
     \end{subfigure}
     \hfill
     \captionsetup[subfigure]{oneside,margin={-0.9cm,0cm}}
     \begin{subfigure}[b]{0.48\textwidth}
         \centering
         \includegraphics[width=\textwidth]{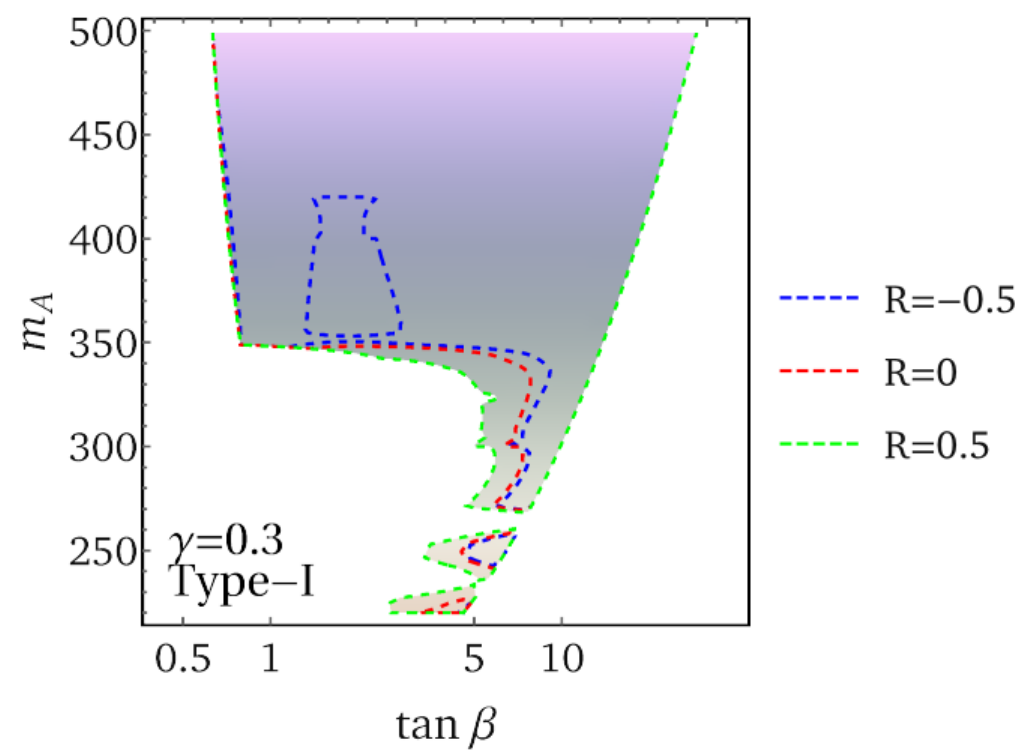}
         \caption{}
         \label{figure:TuningConstraintsB}
     \end{subfigure}
     \newline
      \captionsetup[subfigure]{oneside,margin={-1.0cm,0cm}}
    \begin{subfigure}[b]{0.48\textwidth}
         \centering
         \includegraphics[width=\textwidth]{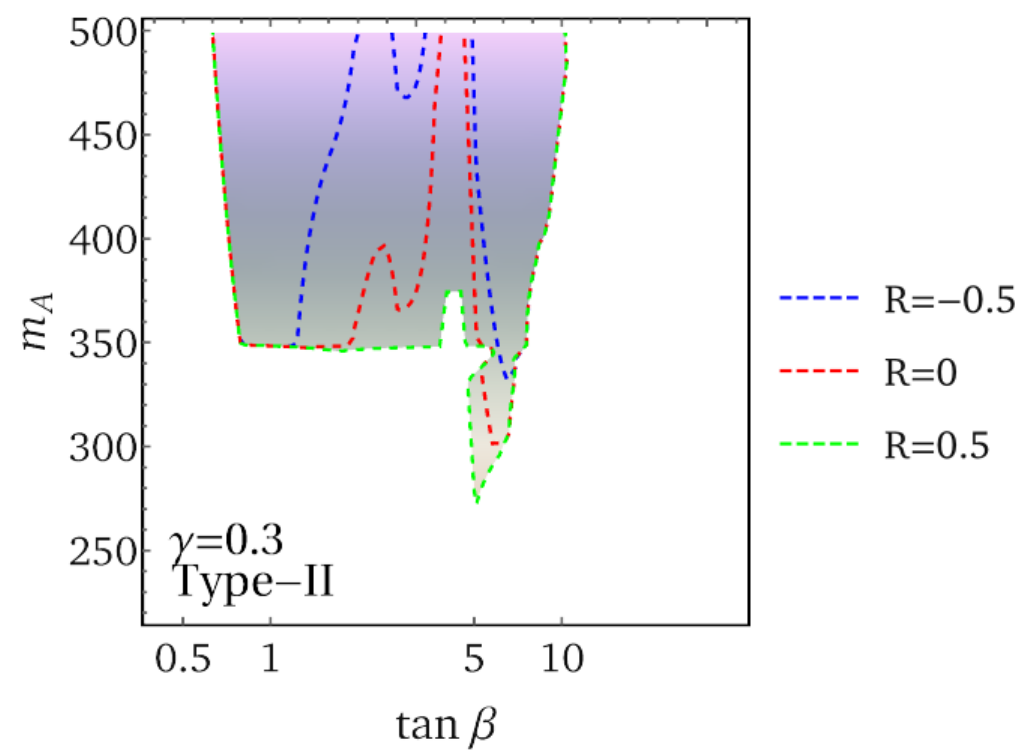}
         \caption{}
         \label{figure:TuningConstraintsC}
     \end{subfigure}
     \hfill
      \captionsetup[subfigure]{oneside,margin={-0.9cm,0cm}}
     \begin{subfigure}[b]{0.48\textwidth}
         \centering
         \includegraphics[width=\textwidth]{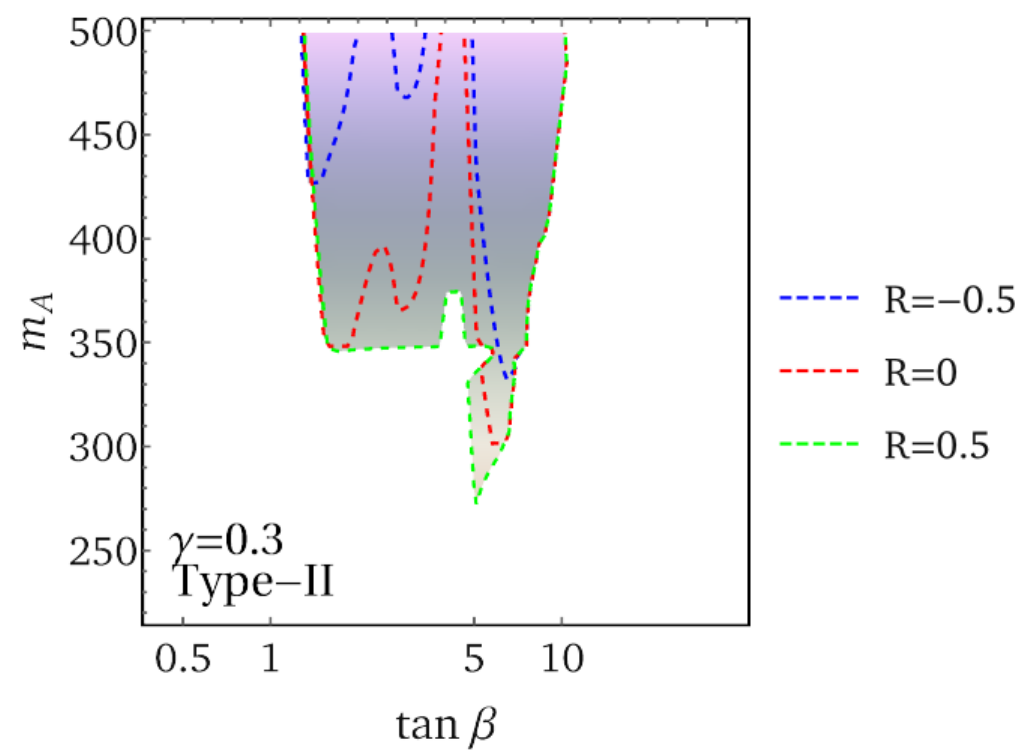}
         \caption{}
         \label{figure:TuningConstraintsD}
     \end{subfigure}
	\caption{\small Experimental and tuning bounds for different $R$ values. The allowed shaded regions inside and/or above each respective contour satisfy all experimental bounds in addition to exhibiting a tuning of at most $5\%$  in  $\Delta m^2$  (left) and $m_{12}^2$ (right). Panels (a) and (b) show the results for Type-I Yukawa couplings, and panels (c) and (d) show the results for Type-II Yukawa couplings. All panels employ the $\gamma=0.3$ benchmark.}
	\label{figure:TuningConstraints}
\end{figure}

\begin{figure}[t!]
     \centering
      \captionsetup[subfigure]{oneside,margin={-1.0cm,0cm}}
       \begin{subfigure}[b]{0.48\textwidth}
         \centering
         \includegraphics[width=\textwidth]{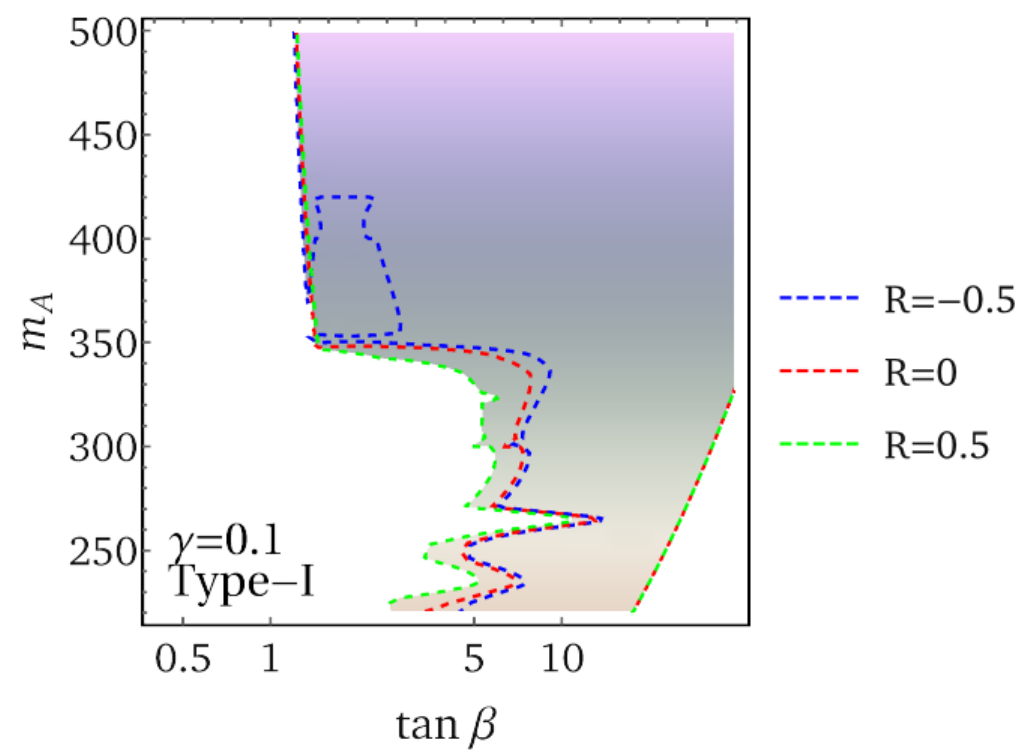}
         \caption{}
         \label{figure:TuningCombinationC}
     \end{subfigure}
     \hfill
     \begin{subfigure}[b]{0.48\textwidth}
         \centering
         \includegraphics[width=\textwidth]{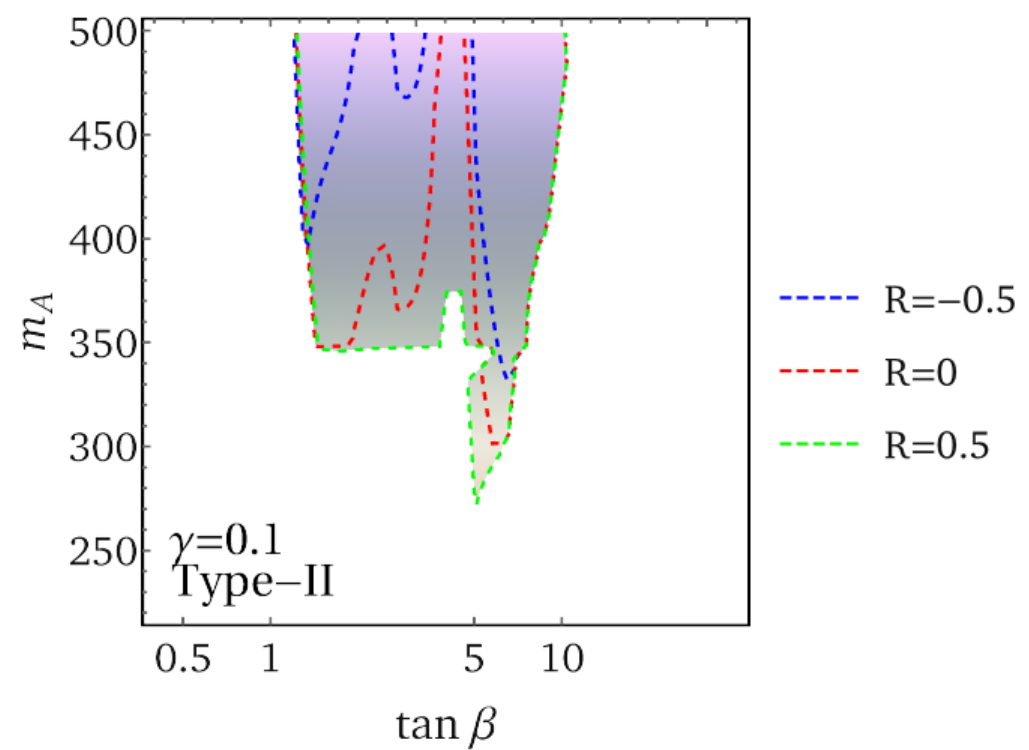}
         \caption{}
         \label{figure:TuningCombinationD}
     \end{subfigure}
      \newline
            \begin{subfigure}[b]{0.48\textwidth}
         \centering
         \includegraphics[width=\textwidth]{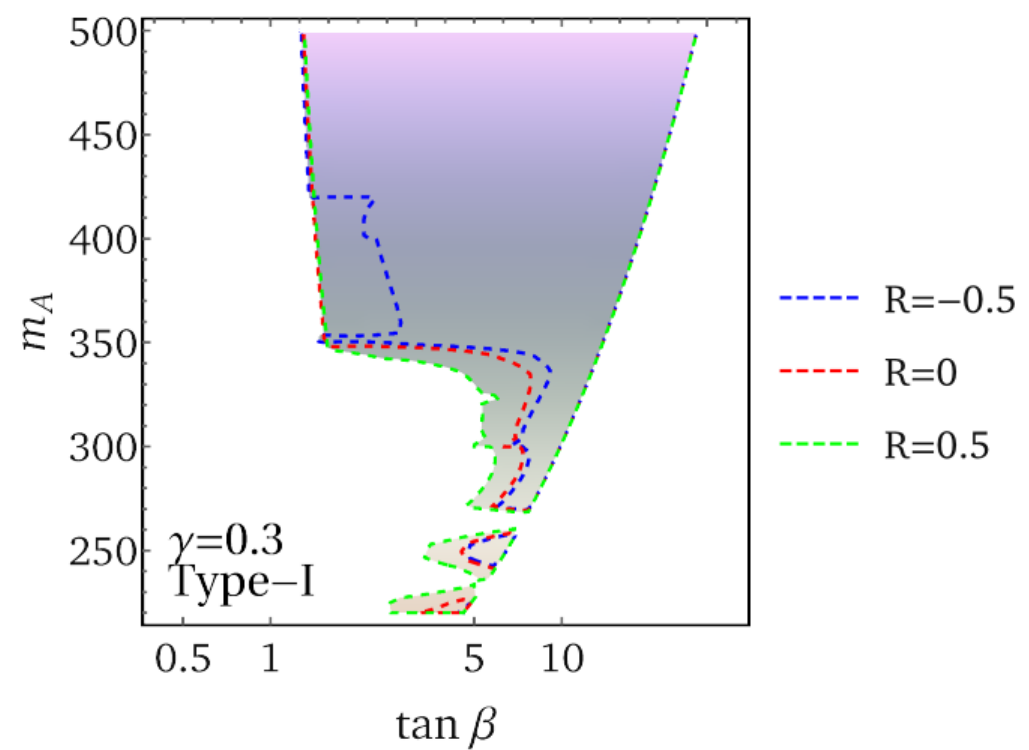}
         \caption{}
         \label{figure:TuningCombinationA}
     \end{subfigure}
     \hfill
     \begin{subfigure}[b]{0.48\textwidth}
         \centering
         \includegraphics[width=\textwidth]{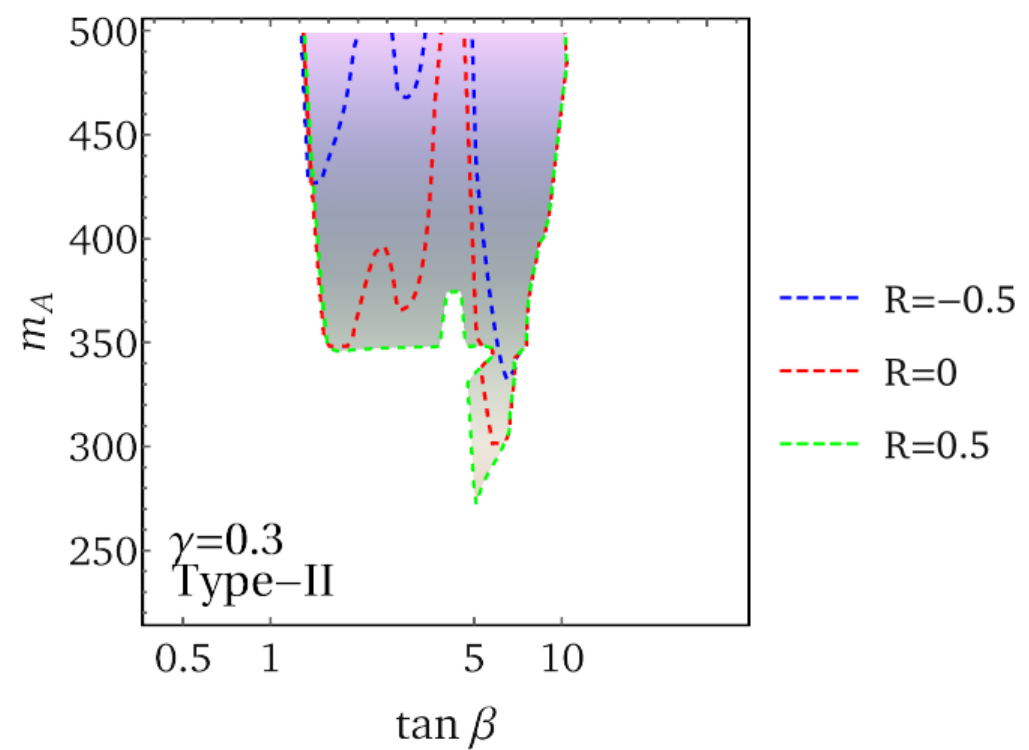}
         \caption{}
         \label{figure:TuningCombinationB}
     \end{subfigure}

		\caption{\small Regions allowed by experimental bounds and tuning constraints for different $R$, with an $m_{12}^2$ and $\Delta m^2$ tuning of at most $5\%$. 
		 Each panel shows three different $R$ curves; the white regions of the parameter space are ruled out.  The ruled out areas expand somewhat as $R$ decreases, with the borders of the allowed shaded regions indicated by the corresponding contours.  For $R=-0.5$, the area enclosed by
	the closed dashed blue contour in panel (a) is also ruled out.
	 Panels (a) and (c) correspond to Type-I Yukawa couplings, and panels (b) and (d) correspond to Type-II Yukawa couplings. Panels (a) and (b) employ the $\gamma=0.1$ benchmark, whereas panels (c) and (d) employ $\gamma=0.3$.} 
\label{figure:TuningCombination}
\end{figure}

\begin{figure}[t!]
     \centering
      \captionsetup[subfigure]{oneside,margin={1.4cm,0cm}}
       \begin{subfigure}[b]{0.48\textwidth}
         \centering
         \includegraphics[width=0.75\textwidth]{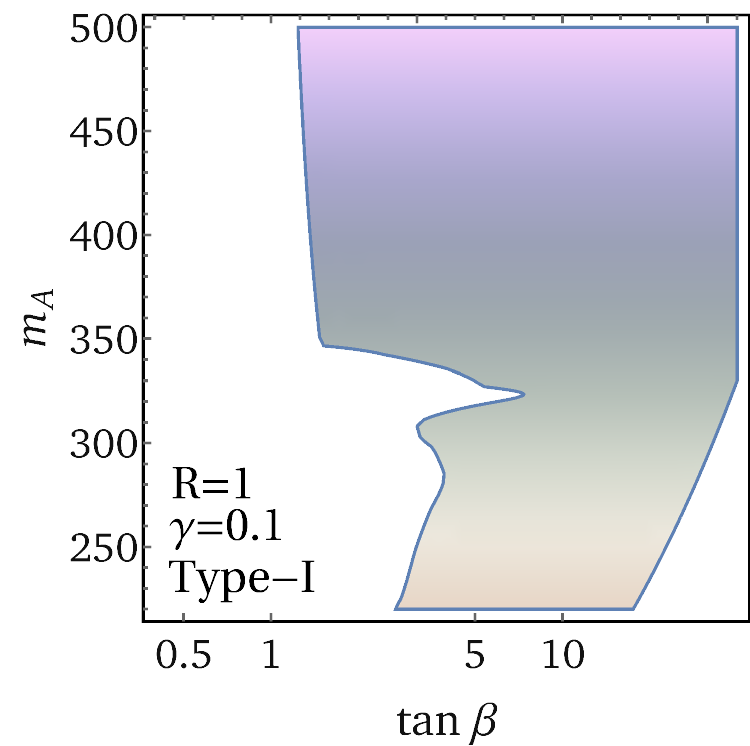}
         \caption{}
         \label{figure:TuningAndColliderSO3C}
     \end{subfigure}
     \hfill
     \begin{subfigure}[b]{0.48\textwidth}
         \centering
         \includegraphics[width=0.75\textwidth]{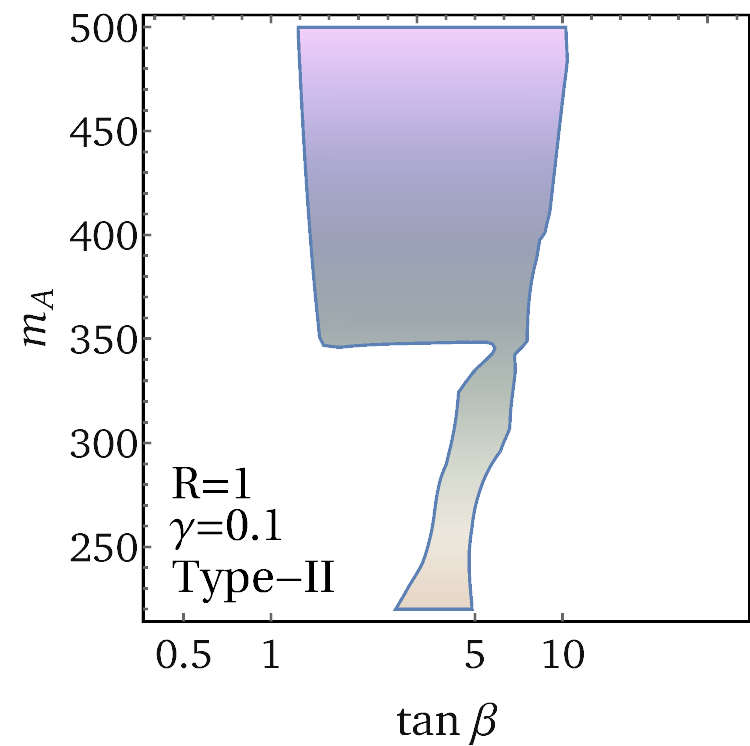}
         \caption{}
         \label{figure:TuningAndColliderSO3D}
     \end{subfigure}
   \newline
         \begin{subfigure}[b]{0.48\textwidth}
         \centering
         \includegraphics[width=0.75\textwidth]{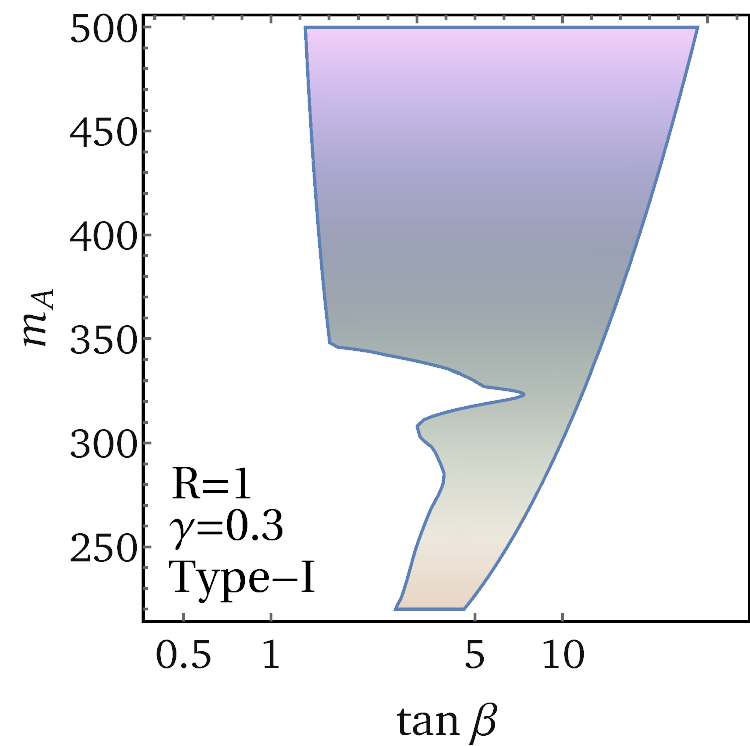}
         \caption{}
         \label{figure:TuningAndColliderSO3A}
     \end{subfigure}
     \hfill
     \begin{subfigure}[b]{0.48\textwidth}
         \centering
         \includegraphics[width=0.75\textwidth]{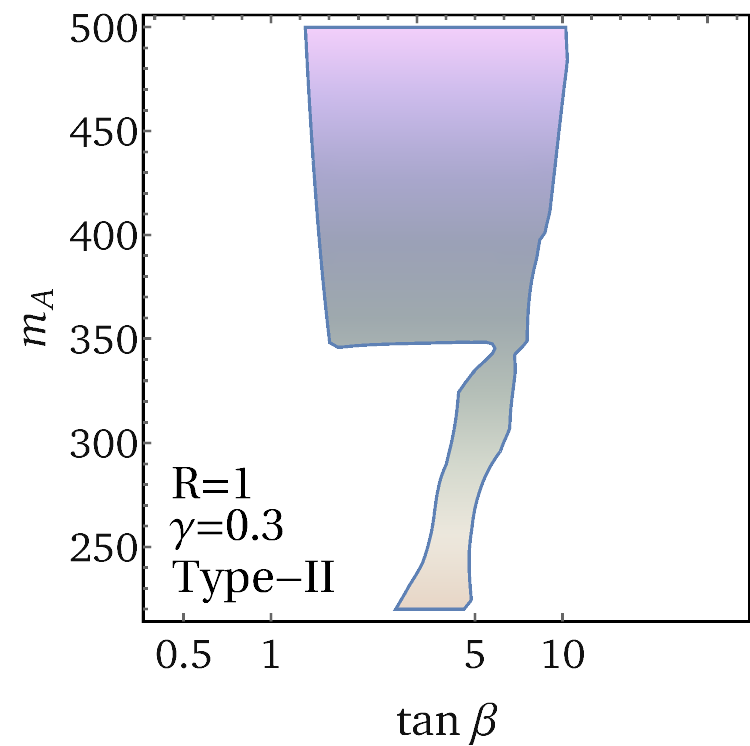}
         \caption{}
         \label{figure:TuningAndColliderSO3B}
     \end{subfigure}
	\caption{\small Experimental and tuning bounds for the softly-broken $\mathrm{SO}(3)$-symmetric 2HDM ($R=1$). The shaded regions in panels (a) and (b) satisfy all experimental bounds in addition to having a tuning of at most $5\%$ for both  $\Delta m^2$  and $m_{12}^2$ in panels (c) and (d). Panels (a) and (c) correspond to Type-I Yukawa couplings, and panels (b) and (d) correspond to Type-II Yukawa couplings. Panels (a) and (b) employ the $\gamma=0.1$ benchmark, whereas panels (c) and (d) employ $\gamma=0.3$. \\[-20pt]}
	\label{figure:TuningAndColliderSO3}
\end{figure}

\begin{figure}[t!]
     \centering
      \captionsetup[subfigure]{oneside,margin={-1.0cm,0cm}}
       \begin{subfigure}[b]{0.48\textwidth}
         \centering
         \includegraphics[width=\textwidth]{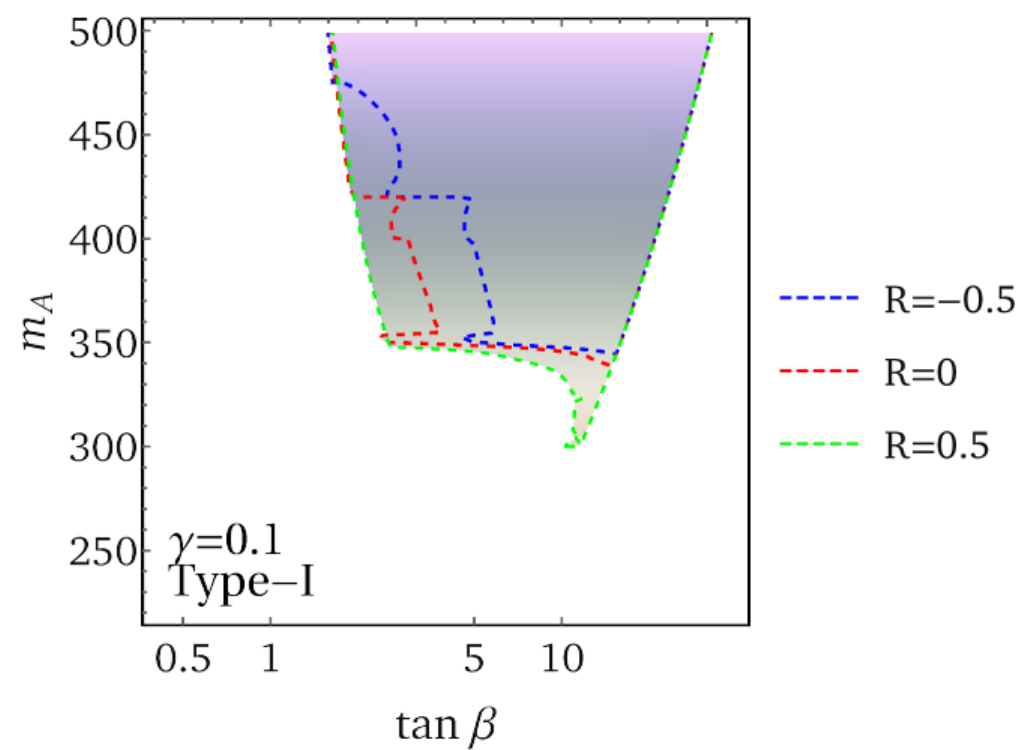}
         \caption{}
         \label{figure:HLLHC_TuningCombinatioC}
     \end{subfigure}
     \hfill
     \begin{subfigure}[b]{0.48\textwidth}
         \centering
         \includegraphics[width=\textwidth]{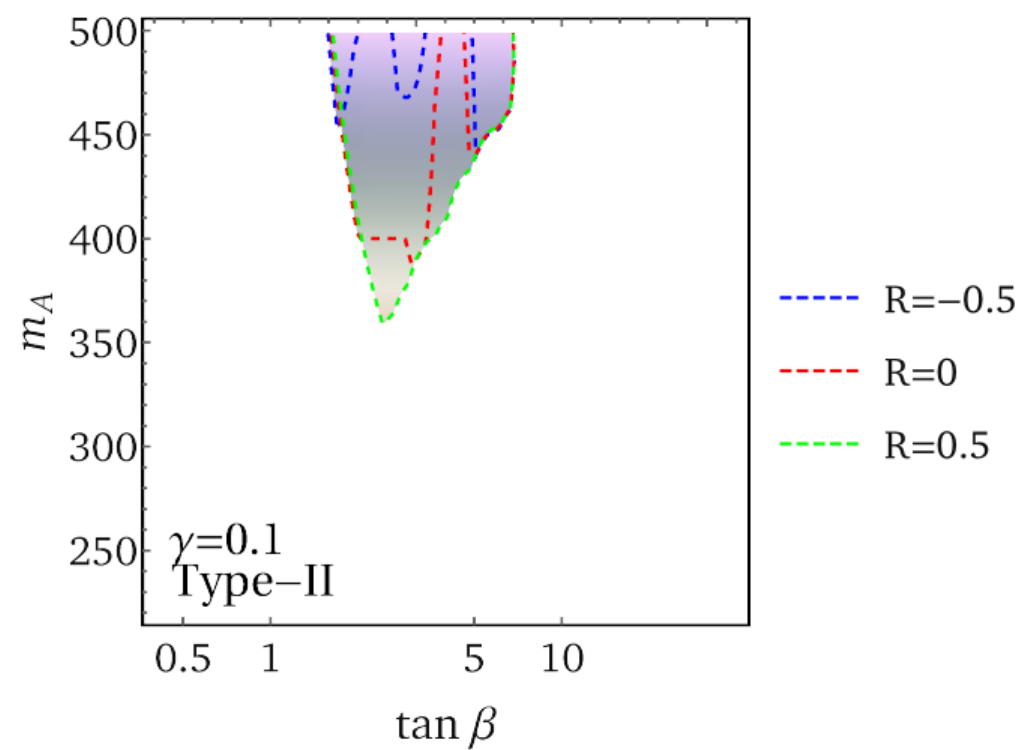}
         \caption{}
         \label{figure:HLLHC_TuningCombinationD}
     \end{subfigure}
      \newline
            \begin{subfigure}[b]{0.48\textwidth}
         \centering
         \includegraphics[width=\textwidth]{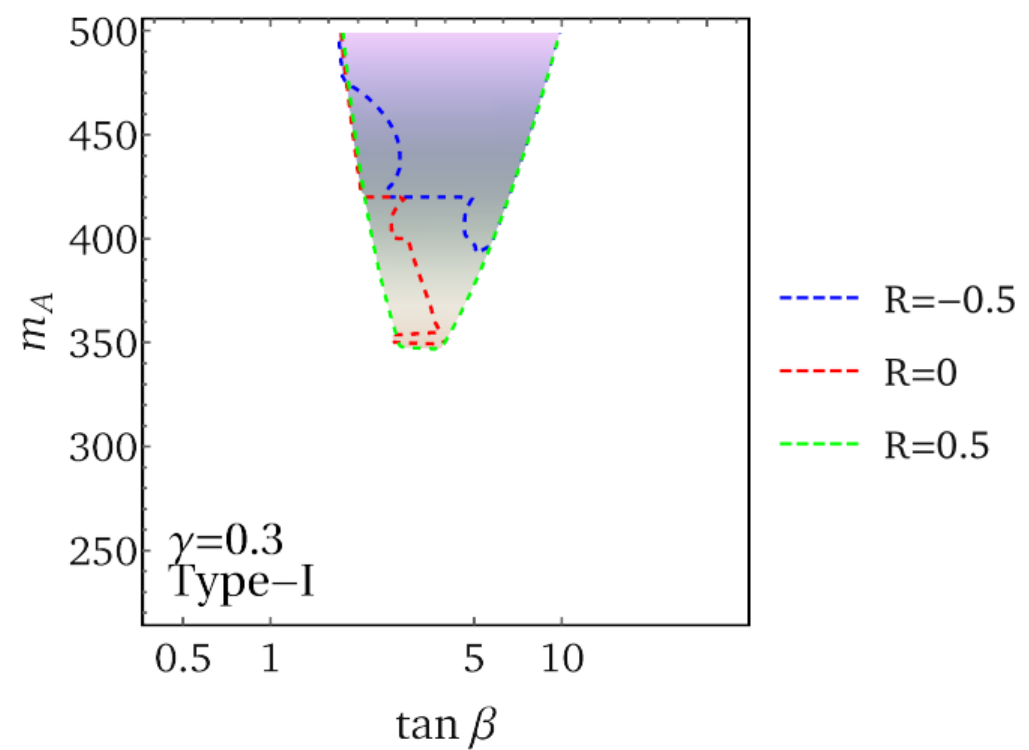}
         \caption{}
         \label{figure:HLLHC_TuningCombinationA}
     \end{subfigure}
     \hfill
     \begin{subfigure}[b]{0.48\textwidth}
         \centering
         \includegraphics[width=\textwidth]{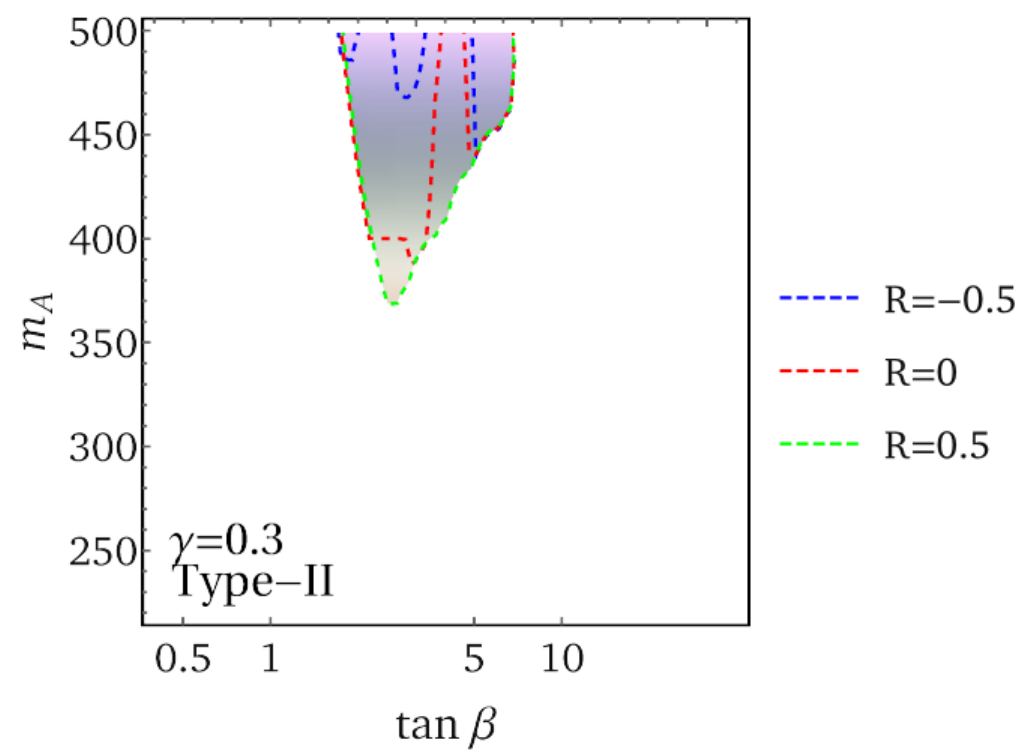}
         \caption{}
         \label{figure:HLLHC_TuningCombinationB}
     \end{subfigure}
		\caption{\small Projected regions allowed by experimental bounds anticipated at the high-luminosity LHC with $3000~\si{\femto\barn^{-1}}$ of data, assuming $m_T=2.5$~TeV with an $m_{12}^2$ and $\Delta m^2$ tuning of at most $5\%$. Each panel shows three different $R$ contours; the white regions of the parameter space are ruled out.  The ruled out areas expand somewhat as $R$ decreases, with the borders of the allowed shaded regions indicated by the corresponding contours. Panels (a) and (c) correspond to Type-I Yukawa couplings, and panels (b) and (d) correspond to Type-II Yukawa couplings. Panels (a) and (b) employ the $\gamma=0.1$ benchmark, whereas panels (c) and (d) employ $\gamma=0.3$.} 
\label{figure:HLLHC_TuningCombination}
\end{figure}

\enlargethispage{\baselineskip}
We now investigate the allowed parameter space taking into account collider, Higgs precision, and fine-tuning constraints\te described in the previous two subsections. Amongst these, collider constraints, based on 95\% CL limits on the cross section times branching ratio of the new heavy scalar states,
depend on whether we employ a Type-I or a Type-II model, whereas tuning constraints are quite insensitive to this choice. Therefore it is natural to assess the impact of the two types of constraints separately.  \figs{figure:TypeIBounds}{figure:TypeIIBounds} show the parameter space allowed by collider constraints, whereas \figs{figure:Tuning} {figure:TuningMT} show the tuning. The combined effects of these constraints are exhibited in Figs.~\ref{figure:TuningConstraints} and \ref{figure:TuningCombination}.  The special case of $R=1$ is presented in
\fig{figure:TuningAndColliderSO3}. Finally, we provide a rough projection of the anticipated sensitivity of the High Luminosity LHC to the parameter space of our model in Fig.~\ref{figure:HLLHC_TuningCombination}.
The color of the shaded regions bounded by the outermost contour in all figures is chosen solely for its aesthetic allure.

\begin{figure}[t!]
     \centering
     \captionsetup[subfigure]{oneside,margin={-1.0cm,0cm}}
     \begin{subfigure}[b]{0.48\textwidth}
         \centering
         \includegraphics[width=\textwidth]{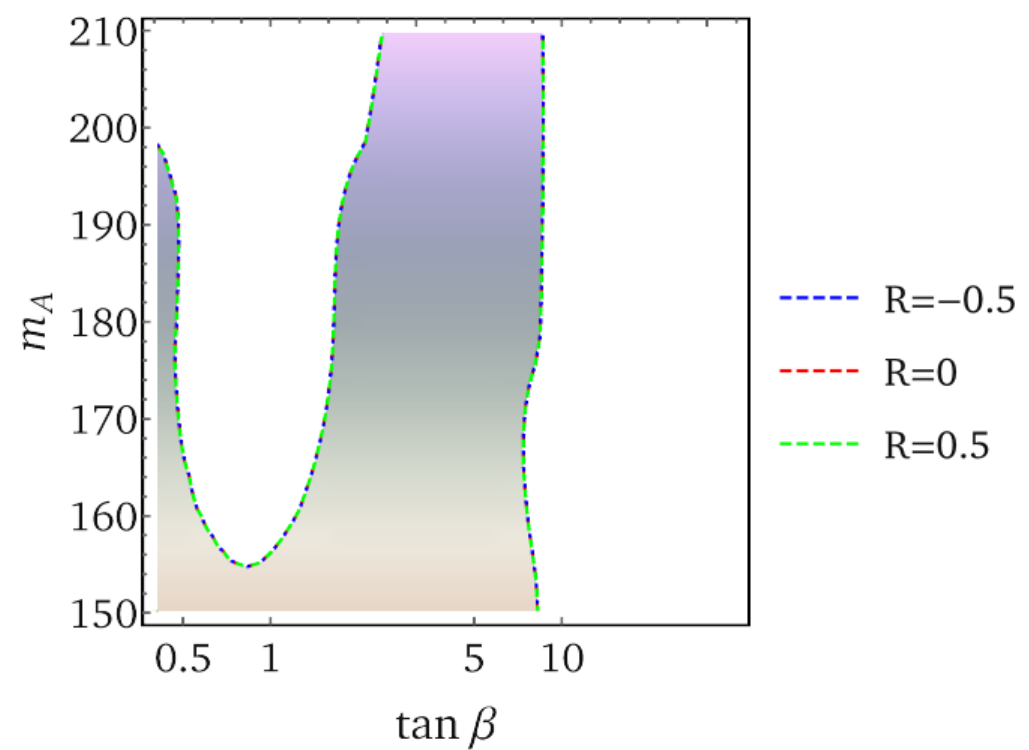}
         \caption{}
         \label{figure:TypeIIBoundsA}
     \end{subfigure}
     \hfill
     \begin{subfigure}[b]{0.48\textwidth}
         \centering
         \includegraphics[width=\textwidth]{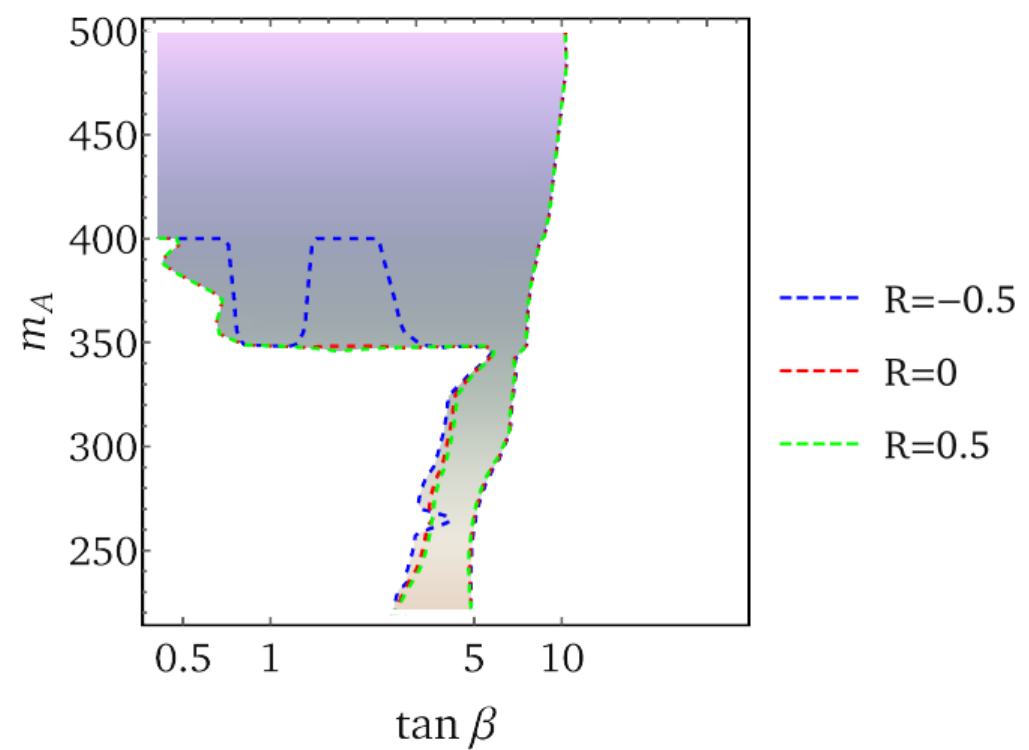}
         \caption{}
         \label{figure:TypeIIBoundsB}
     \end{subfigure}
     \newline
    \begin{subfigure}[b]{0.48\textwidth}
         \centering
         \includegraphics[width=\textwidth]{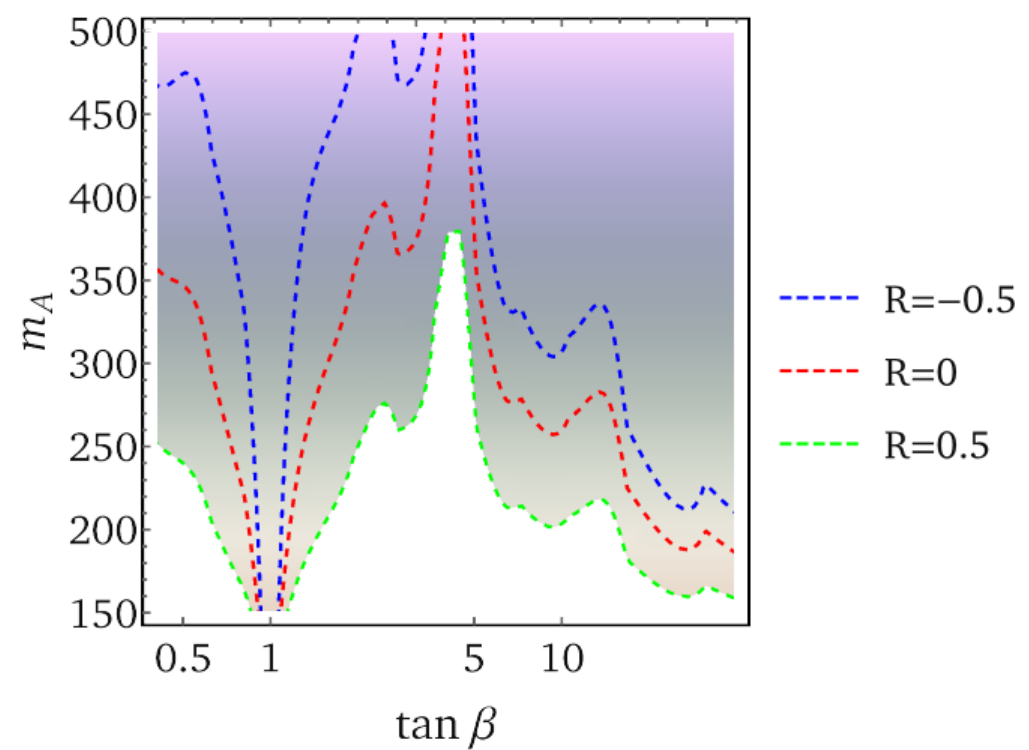}
         \caption{}
         \label{figure:TypeIIBoundsC}
     \end{subfigure}
     \hfill
     \begin{subfigure}[b]{0.48\textwidth}
         \centering
         \includegraphics[width=\textwidth]{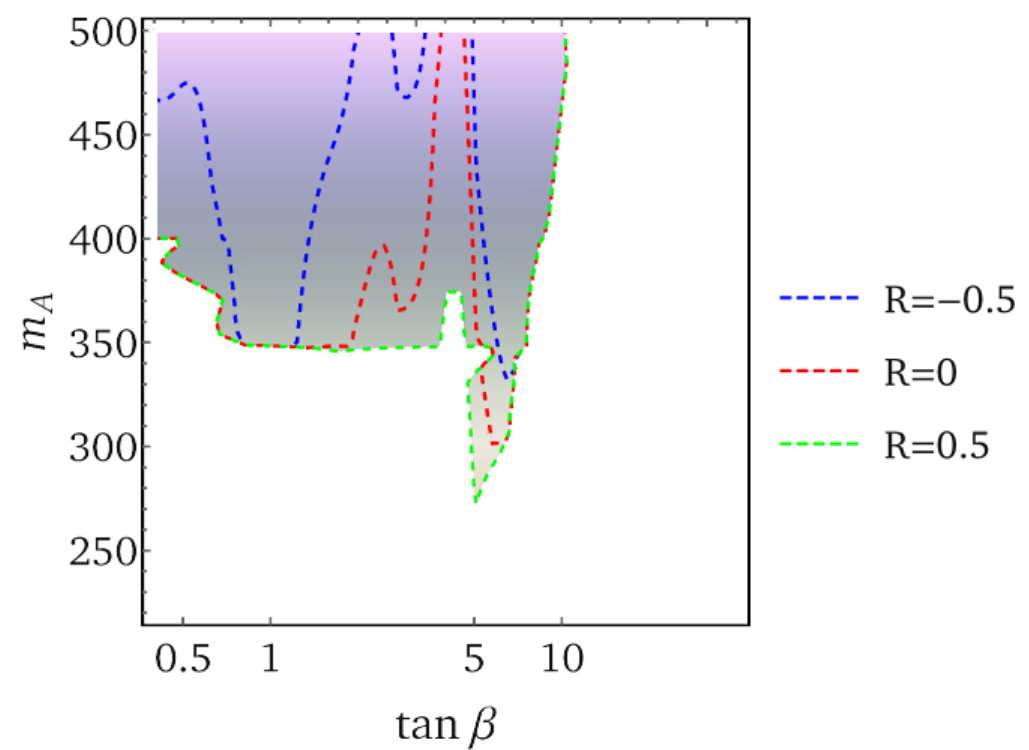}
         \caption{}
         \label{figure:TypeIIBoundsD}
     \end{subfigure}
\caption{\small Bounds for Type-II Yukawa couplings. Regions ruled out by (a) $A/H\rightarrow \tau \tau$ data, (b)~combination of collider constraints, including $A/H\rightarrow \tau \tau$, (c) precision Higgs global fits, and (d)~combination of collider bounds and precision Higgs global fits. 
 Each panel shows three different $R$ curves; the white regions of the parameter space are ruled out.  In panels (b)--(d), the ruled out areas expand somewhat as $R$ decreases, with the borders of the allowed shaded regions indicated by the corresponding contours. 
There is a different $m_A$ scale in panel (a) as compared to the other three panels because the $A\rightarrow \gamma \gamma$ and $A\rightarrow Zh$ bounds are restricted to $m_A \gsim 220~\si{\giga\electronvolt}$.}
\label{figure:TypeIIBounds}
\end{figure}

The collider constraints shown in \figs{figure:TypeIBounds}{figure:TypeIIBounds} are organized as follows: panel~(a) shows the small-mass region\te where $A\rightarrow Zh$ and $A\rightarrow \gamma \gamma$ are not relevant; panel (b) shows the combination of all considered collider  constraints; panel (c) shows the parameter space ruled out from Higgs precision searches; and panel (d) shows the combination of collider and Higgs precision constraints. Note that the $m_A$ scale is different in \figs{figure:TypeIBoundsB} {figure:TypeIBoundsD}.

\begin{figure}[ht!]
     \centering
     \captionsetup[subfigure]{oneside,margin={-1.3cm,0cm}}
     \begin{subfigure}[b]{0.48\textwidth}
         \centering
         \includegraphics[width=\textwidth]{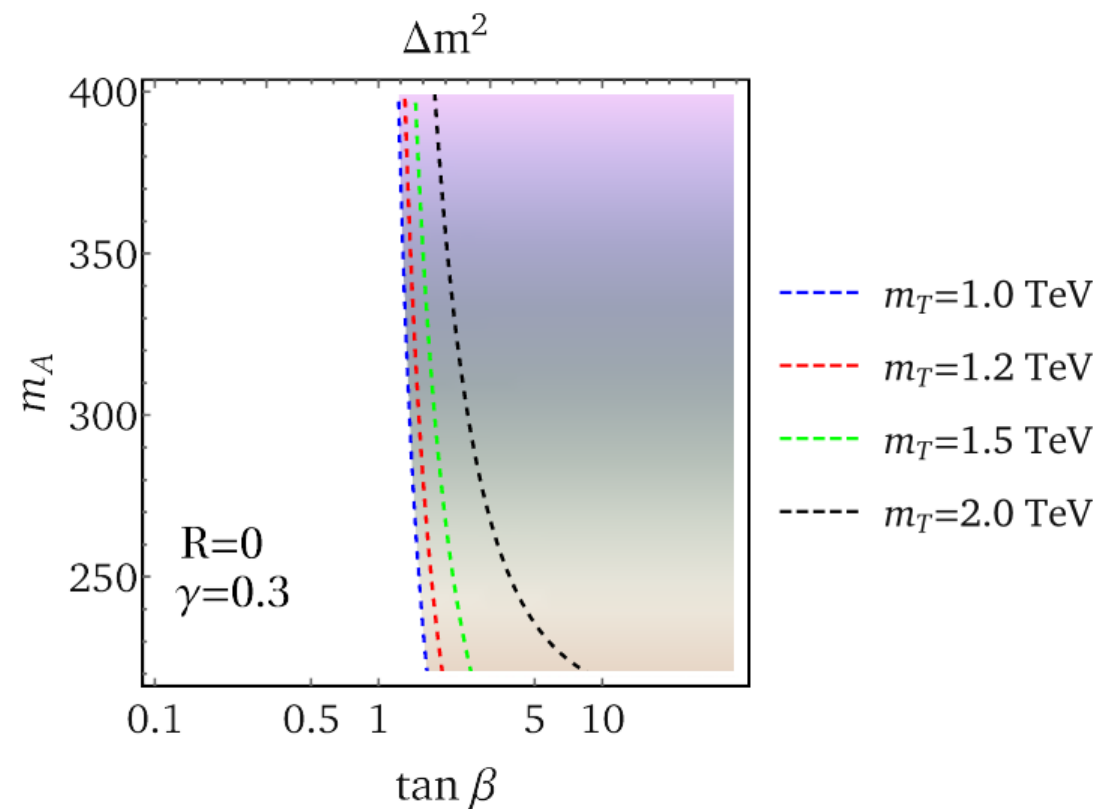}
         \caption{}
         \label{figure:TuningMTA}
     \end{subfigure}
     \hfill
     \captionsetup[subfigure]{oneside,margin={-1.1cm,0cm}}
     \begin{subfigure}[b]{0.48\textwidth}
         \centering
         \includegraphics[width=\textwidth]{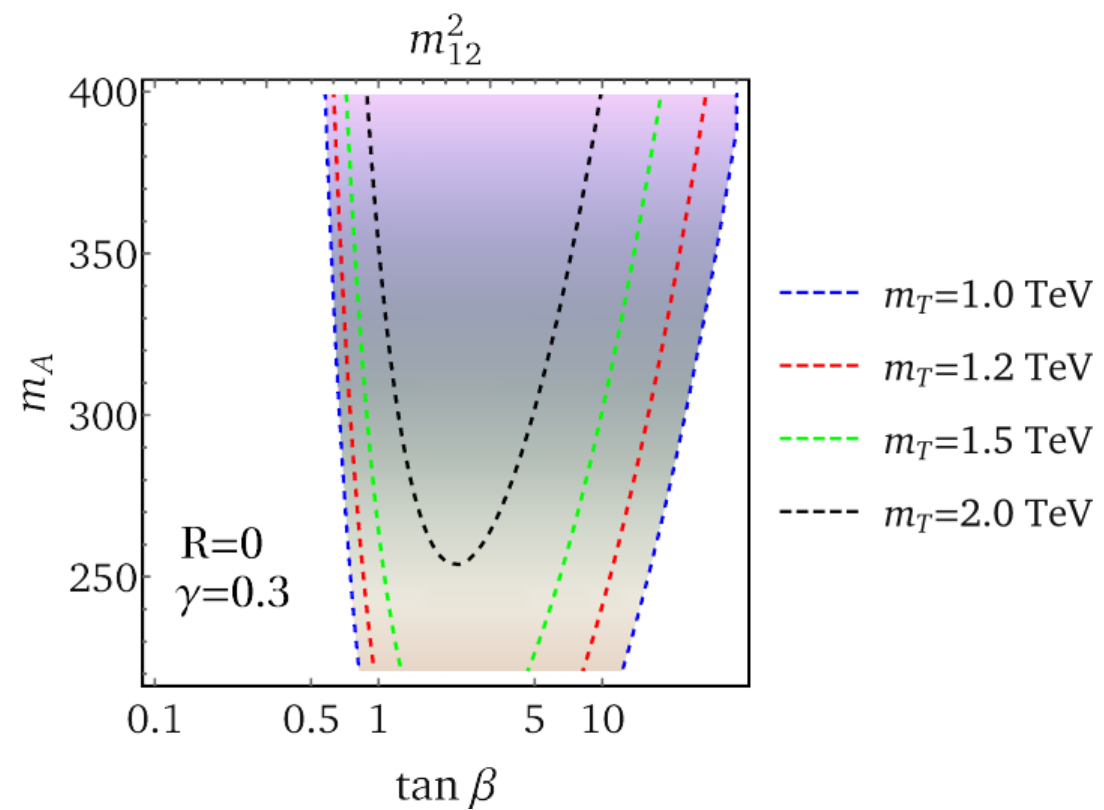}
         \caption{}
         \label{figure:TuningMTB}
     \end{subfigure}
		\caption{\small Dependence of the degree of fine-tuning for various masses of the vector-like top quark partner. 
		The shaded region 
	inside and/or above each respective contour corresponds to a fine-tuning of at most $5\%$. Both panels employ the benchmark points $R=0$ and $\gamma=0.3$.}
		\label{figure:TuningMT}
\end{figure}

\fig{figure:TypeIBoundsA} shows that the $A/H\rightarrow \tau \tau$ channel only restricts small $\tan \beta$ values for the Type-I model. This is expected because the production cross-section rapidly decreases as $\tan\beta \rightarrow \infty$. Other collider searches rule out a sizeable chunk of the low $\tan\beta$ parameter space as seen in  \fig{figure:TypeIBoundsB}. The weak dependence of collider bounds  on $R$ is not shared by the fits to the Higgs precision data. \fig{figure:TypeIBoundsC} shows that larger $R$ values are less constrained than smaller ones. This behavior follows from \eq{equation:betaalpha}, as does the behavior as $\beta \rightarrow \frac14\pi$ and $\beta \rightarrow \half\pi$. The combination of these constraints, in  \fig{figure:TypeIBoundsD}, show that there is plenty of available parameter space at large $\tan \beta$ for the Type-I scenario, even for the smaller values of $m_A$.

The picture is rather different for the Type-II model. The $A/H$ production cross section rises for larger values of $\tan\beta$, and the $A/H\rightarrow \tau \tau$ branching sinks for small $\tan\beta$. This behavior, together with the lepton branching ratios, is reflected in \fig{figure:TypeIIBoundsA}, where the lepton decay channel mainly constrains large $\tan\beta$ values. Nevertheless, \fig{figure:TypeIIBoundsB} shows that the small $\tan\beta$ region is almost entirely ruled by the $A\rightarrow \gamma \gamma$ and $A\rightarrow Zh$ channels. Higgs precision constraints rule out another chunk of parameter space. This is because Higgs precision data force Type-II close to exact Higgs alignment [$\cos(\beta-\alpha)\approx 0$], which is reflected by the strong $R$ dependence.  Indeed, \eq{equation:betaalpha} implies that $\cos(\beta-\alpha)\to 0$ in the limit of $R\rightarrow 1$. All these constraints are combined in \fig{figure:TypeIIBoundsD} which shows, in combination with \fig{figure:TuningAndColliderSO3}, that a light CP-odd scalar ($m_A \leq 350~\si{\giga\electronvolt}$) is only possible for $R$ values close to $1$. 

Moreover, it is noteworthy that all collider bounds are less severe for $m_A\geq 350~\si{\giga\electronvolt}$. This is because the production cross-section drops for energies larger than the two-top-quark threshold. In addition, the area enclosed by the blue dotted line in figures \ref{figure:TypeIBoundsB},~\ref{figure:TypeIBoundsD}, and \ref{figure:TypeIIBoundsB} comes from the $A\rightarrow Zh$ bound; the branching ratio vanishes for $\tan \beta=1$, and the $A\rightarrow Zh$ branching ratio is larger for smaller $R$.

Let us now turn to the degree of fine-tuning. \fig{figure:Tuning} shows the fine-tuning measures as defined in \eq{eq:TuningMeasure} for $\gamma=0.1$ and $\gamma=0.3$ and \fig{figure:TuningMT} shows the tuning for different $m_T$ values. The two tuning measures are complementary: $\Delta m^2$ mainly constrains small and intermediate $\tan \beta$ values, while $m^2_{12}$ constrains small and large $\tan \beta$. 
Note that the white regions in \fig{figure:Tuning} occur when $\beta=\gamma$; that is, at $\beta=0.1$ in \figs{figure:TuningA}{figure:TuningB}, and at $\beta=0.3$ in \figs{figure:TuningC}{figure:TuningD}. This behavior can be traced back to \eq{tmass}, and corresponds to the $y_t \rightarrow \infty$ limit. In addition, there is a region close to $\beta=\frac14\pi$ where $\Delta m^2$ vanishes. This region is quite narrow and is indistinguishable in the figures. Of the two tunings, $m_{12}^2$ depends more strongly on $\gamma$ than $\Delta m^2$, as discussed in \sect{section:FineTuning}.
Moreover, the $m_T$ dependence of the tunings is more pronounced for $m_{12}^2$ than for $\Delta m^2$ as shown in \fig{figure:TuningMT}.

\figs{figure:TuningConstraints}{figure:TuningCombination} show the combination of collider and tuning  constraints.
\fig{figure:TuningConstraintsA}, for Type-I, and \fig{figure:TuningConstraintsC}, for Type-II, allow a $\Delta m^2$ tuning of at most $5\%$ and take into account collider constraints.  Likewise, \figs{figure:TuningConstraintsB}{figure:TuningConstraintsD} are defined analogously and allow a tuning of $m_{12}^2$ of at most $5\%$. Both the $\Delta m^2$ and $m_{12}^2$ tuning constraints are combined in \fig{figure:TuningCombinationA} for \hbox{Type-I,} and \fig{figure:TuningCombinationB} for \mbox{Type-II}. These figures show that tuning constrains a region of parameter space untouched by other constraints. Of the two tuning measures, the $m_{12}^2$ tuning measure is salient\te it restricts the large $\tan \beta$ region that is otherwise unconstrained for Type-I, and likewise but to a smaller extent for Type-II. However, lowering $\gamma$ makes tuning constraints less pronounced, as shown in \figs{figure:TuningConstraintsC}{figure:TuningConstraintsD}.  In summary, tuning constraints are complementary to collider bounds in these models, and moreover are not optional, as the purpose of our models is precisely to achieve approximate Higgs alignment (without decoupling) with minimal tuning.

In Fig.~\ref{figure:TuningAndColliderSO3}, we exhibit the experimental and tuning bounds for $R=1$, corresponding to the softly-broken SO($3$)-symmetric 2HDM.
The allowed parameter regions for Type-I [panels (a) and (c)] and Type-II [panels (b) and (d)] are exhibited for $\gamma=0.1$ and $0.3$, respectively.   This limiting case provides the most robust example of approximate Higgs alignment without decoupling in our framework, with allowed parameter regimes with $m_A$ as low as $200~\si{\giga\electronvolt}$.

With the High-Luminosity LHC upgrade expected to begin taking data later in this decade, it is of interest to estimate the anticipated sensitivity to the parameter space of our model
with $3000~\si{\femto\barn^{-1}}$ of data. 
First, the influx of data will likely tighten the bounds on vector-like quarks. To safely evade any new bounds (in the absence of a discovery)
we increase the vector-like quark mass to $m_T=2.5~\si{\tera\electronvolt}$.  Second, we assume that all limits on experimental signal strengths for $H$ and $A$ production 
decrease by a factor of 4.\footnote{In light of $\sqrt{N}$ statistics, the increased integrated luminosity suggests that limits on experimental signal strengths would improve by roughly a factor of 4.5.  We have chosen a factor of 4 to be conservative and not overestimate the impact of the anticipated data.} The result of these considerations is shown in Fig.~\ref{figure:HLLHC_TuningCombination}. The sub-plots of this figure are defined analogously to those of Fig.~\ref{figure:TuningCombination}.  As expected, the allowed parameter regions are severely reduced in Fig.~\ref{figure:HLLHC_TuningCombination} as compared to Fig.~\ref{figure:TuningCombination}.  Nonetheless, parameter space still exist where
light scalar bosons ($\approx 350$~GeV) cannot be ruled out at the 95\% CL.
 We also observe that the type-I results in Figs.~\ref{figure:HLLHC_TuningCombinatioC} and \ref{figure:HLLHC_TuningCombinationA} are quite sensitive to the value of $\gamma$. In particular, choosing a smaller $\gamma$ increases the allowed parameter space for larger $\tan \beta$ values. Furthermore, the allowed parameter space in the large $\tan\beta$ region increases for Type-I if one is willing to accept more fine-tuning. So it seems unlikely that the entire region of approximate Higgs alignment without decoupling can be ruled out for the Type-I scenario.

\section{Conclusions}
\label{conclusions}

The 2HDM remains one of the simplest and best motivated extensions of the Standard Model. Theoretical and phenomenological studies of the 2HDM have been of great utility in guiding collider searches for new physics phenomena in the Higgs sector and have provided a useful framework for interpreting collider data.

Despite null results thus far from the LHC, the possibility of an extended Higgs sector remains viable. Of particular experimental interest is the case of approximate Higgs alignment without decoupling, where multiple states in the Higgs sector are light, but the neutral scalar interaction eigenstate, whose tree-level properties coincide with the SM Higgs field, does not mix strongly with the other scalar field degrees of freedom. If an approximate Higgs alignment is realized, then the present LHC Higgs data do not rule out this scenario, while the presence of other scalar states with masses not significantly larger than the electroweak scale and a small degree of mixing with the SM Higgs eigenstate provide experimental targets for the long-term LHC program.

In this work we have described an approximate global symmetry structure that results in approximate Higgs alignment. The global symmetry is one of the generalized CP symmetries of the 2HDM, and we extend it to the Yukawa sector by introducing vector-like fermions with soft symmetry-breaking masses. For practical purposes, a minimal addition of vector-like partners for the right-handed top quark is sufficient. The resulting model, although it does not address the ordinary electroweak hierarchy problem, realizes Higgs alignment without decoupling in an otherwise natural way. The structure in the Higgs sector, as well as the extended fermion sector, provide a range of collider signatures that can be accessed by future LHC searches.

We have assessed current experimental and fine-tuning constraints 
on the model parameters relevant for Higgs alignment
 arising from direct searches for vector-like top partners and new Higgs states, the LHC Higgs data,  and precision electroweak observables. The model is viable and the new Higgs states can lie below about 500 GeV if the top partner is between one and a few TeV, $\tan\beta$ is in the range $\sim 1$--$10$,  and the mixing between the top quark and the vector-like top partner is small. These restrictions are the result of an intricate interplay between experimental and tuning constraints. The viable parameter regimes provide attractive targets for the High-Luminosity LHC.

In addition to the generalized CP symmetry GCP$3$ employed in this work, there is another possible generalized CP symmetry of the 2HDM, GCP$2$, that can be used to impose Higgs alignment. Unlike the model studied here, softly breaking the GCP$2$ symmetry allows for the possibility of explicit and/or spontaneous CP violation in the Higgs sector~\cite{Ferreira:2010hy}. Further exploration of this model and its CP-violating phenomenology would be an interesting avenue for future work. 
\clearpage

\section*{Acknowledgments}
P.D.~and H.E.H.~are grateful for the collaboration of Josh Ruderman (see Ref.~\cite{Draper:2016cag}) that provided the framework for the present study. We would also like to thank Francesco D'Eramo for discussions during the early stages of this work.
H.E.H.~also acknowledges collaborations with Pedro Ferreira and Jo\~ao P.~Silva that contributed to some of the material presented in \sects{sec:enhanced}{gcpthree}. 

P.D.~acknowledges support from the National Science Foundation
under Grant No.~\uppercase{PHY}-$1719642$ and from the US Department of Energy under Grant No.~\uppercase{DE-SC}SC$0015655$.
The work of A.E.~is supported by the Grant agency of the Czech Republic, Project No.~$20$-$17490$S and by the Charles University Research Center UNCE/S-CI/$013$, and partly by the Swedish Research Council, Grant No.~$621$-$2011$-$5107$.   
H.E.H.~is supported in part by the U.S. Department of Energy Grant No.~\uppercase{DE-SC}$0010107$. 
Both A.E.~and H.E.H.~acknowledge the support of Grant H$2020$-MSCA-RISE-$2014$
No.~$645722$ (NonMinimalHiggs), which provided funds for a visit by A.E.~to the Santa Cruz Institute for Particle Physics (SCIPP) and for the travel of A.E. and H.E.H. 
to the $2019$ NonMinimalHiggs conference at the University of Helsinki.  Both visits were highly productive in advancing this work, and A.E.~and~H.E.H. are grateful for the hospitality furnished by SCIPP and the University of Helsinki.
In addition,
H.E.H.~also benefited from discussions that took place at the University of Warsaw during visits supported by the HARMONIA  project of  the National Science Centre,  Poland,  under contract UMO-$2015$/$18$/M/ST$2$/$00518$ ($2016$--$2021$).
\bigskip
\begin{appendices}

\section{Equivalence of the softly-broken GCP3 and U(1)\texorpdfstring{$\otimes\Pi_2$}{xPi2} symmetric 2HDMs}

\label{appequiv}
\renewcommand{\theequation}{A.\arabic{equation}}
\setcounter{equation}{0}

To see that the softly-broken $\ug\otimes\Pi_2$-symmetric and GCP$3$-symmetric 2HDMs are in fact the same model expressed with respect to different scalar field bases~\cite{Ferreira:2009wh}, 
we provide the following details taken from Ref.~\cite{Haber:2021zva}. Consider the $\ug\otimes\Pi_2$ basis parameters with $\xi\neq 0$ and $m_{12}^2$ complex [subject to \eq{min3a}].  For convenience, we shall refrain from rephasing the scalar doublet fields to remove the phase~$\xi$.   
Applying the following unitary transformation,
\beq \label{youdef}
\Phi^\prime_a= U_{ab}\Phi_b\,,\quad \text{where $U=\frac{1}{\sqrt{2}}\begin{pmatrix} \phm1 & -i \\ -i & \phm 1\end{pmatrix}$}\,,
\eeq
to the $\ug\otimes\Pi_2$ basis parameters yields the corresponding GCP$3$ basis parameters (denoted with prime superscripts),
\beqa
\lambda^\prime&=& \half\lambda(1+R)\,,\label{lamp}\\
\lambda_3^\prime&=&\lambda_3+\half\lambda (1-R)\,,\\
\lambda_4^\prime&=&\lambda_4+\half\lambda(1-R)\,,\label{lam4p}\\
\lambda_5^\prime&=&-\half\lambda(1-R)\,,\label{lam5p}\\
\lambda_6^\prime&=&-\lambda_7^\prime=0\,.
\eeqa
In particular, 
$\lambda_5^\prime=\lambda_1^\prime-\lambda_3^\prime-\lambda_4^\prime$ is real and $\lambda_6^\prime=\lambda_7^\prime=0$, corresponding to the GCP$3$ basis defined in Table~\ref{tab:class}.

In addition, the corresponding soft-breaking squared-mass parameters are,
\beqa
&& m^{\prime\,2}_{11}=\half (m_{11}^2+m_{22}^2)+\Im m_{12}^2\,, \label{m11p} \\
&& m^{\prime\,2}_{22}=\half (m_{11}^2+m_{22}^2)-\Im m_{12}^2\,,\label{m22p} \\
&& m^{\prime\,2}_{12}=\Re m_{12}^2+\half i (m_{22}^2-m_{11}^2)\,.\label{m12p}
\eeqa
Finally, the complex vevs in the GCP$3$ basis are given by
\beq
v_1^\prime=\frac{1}{\sqrt{2}}\bigl(v_1-iv_2e^{i\xi}\bigr)\,,\qquad\quad v_2^\prime=-\frac{i}{\sqrt{2}}\bigl(v_1+iv_2 e^{i\xi}\bigr)\,.
\eeq
Hence,
\beq
\tan^2\beta^\prime=\left|\frac{v_2^\prime}{v_1^\prime}\right|^2=\frac{v_1^2+v_2^2-2v_1 v_2\sin\xi}{v_1^2+v_2^2+2v_1 v_2\sin\xi}=\frac{1-s_{2\beta}\sin\xi}{1+ s_{2\beta}\sin\xi}\,.
\eeq
which implies that 
\beq \label{sintwob}
s_{2\beta^\prime}^2=1-s_{2\beta}^2\sin^2\xi\,,
\eeq
where $s_{2\beta^\prime}\equiv\sin 2\beta^\prime$ following our usual notation for sines and cosines.
By convention, \mbox{$0\leq\beta^\prime\leq \half\pi$} (or equivalently, $\sin 2\beta^\prime\geq 0$).

The relative phase of the vevs in the GCP$3$ basis, denoted by $\xi^\prime$, is given by,
\beq
e^{i\xi^\prime}\tan\beta^\prime \equiv \frac{v_2^\prime}{v_1^\prime}=\frac{-iv_1+v_2 e^{i\xi}}{v_1-iv_2 e^{i\xi}}
\eeq
Hence, we obtain,
\beq \label{expp}
e^{i\xi^\prime}=\frac{s_{2\beta}\cos\xi-ic_{2\beta}}{(1-s_{2\beta}^2\sin^2\xi)^{1/2}}\,.
\eeq
That is,
\beq \label{sin2xi}
\sin\xi^\prime=\frac{-c_{2\beta}}{(1-s_{2\beta}^2\sin^2\xi)^{1/2}}\,,\qquad\quad \cos\xi^\prime=\frac{s_{2\beta}\cos\xi}{(1-s_{2\beta}^2\sin^2\xi)^{1/2}}\,.
\eeq
Hence \eqs{sintwob}{sin2xi} yield,
\beq \label{cs}
s_{2\beta^\prime}\sin\xi^\prime=-c_{2\beta}\,.
\eeq

It is straightforward to verify that one obtains the same mass spectrum when computed in either scalar field basis.  For example, in the ${\rm U}(1)\otimes\Pi_2$ basis, we have
\beq \label{mhps}
m_A^2=\frac{2\Re(m_{12}^2e^{i\xi})}{s_{2\beta}}\,,
\eeq
prior to a rephasing of the scalar fields to set $\xi=0$ [cf.~\eq{mha}].  To obtain $m_A^2$ expressed in terms of GCP$3$ basis parameters, we first employ 
\eqss{m12p}{sintwob}{expp} to obtain
\beq \label{rem12p}
\frac{2\Re(m_{12}^{\prime\,2}e^{i\xi^\prime})}{s_{2\beta^\prime}}=\frac{2\Re(m_{12}^2)s_{2\beta}\cos\xi+c_{2\beta}(m_{22}^2-m_{11}^2)}{1-s_{2\beta}^2\sin^2\xi}\,.
\eeq
In light of \eq{min3a}, it follows that 
\beq
\Re (m_{12}^2)= \Re(m_{12}^2e^{i\xi})\cos\xi+\Im(m_{12}^2e^{i\xi})\sin\xi=\Re(m_{12}^2e^{i\xi})\cos\xi\,.
\eeq
Hence, after using \eqs{min1a}{min2a} to evaluate $m_{22}^2-m_{11}^2$, it follows that \eq{rem12p} yields,
\beq
\frac{2\Re(m_{12}^{\prime\,2}e^{i\xi^\prime})}{s_{2\beta^\prime}}=\frac{2\Re(m_{12}^2e^{i\xi})}{s_{2\beta}}+\frac{\lambda v^2(1-R)c_{2\beta}^2}{2(1-s_{2\beta}^2\sin^2\xi)}\,.
\eeq
Finally, using \eq{mhps}, we end up with 
\beq \label{mhagcp3}
m_A^2=\frac{2\Re(m_{12}^{\prime\,2}e^{i\xi^\prime})}{s_{2\beta^\prime}}+\lambda_5^\prime v^2\sin^2\xi^\prime\,,
\eeq
after employing \eqs{lam5p}{sin2xi}.    Indeed, one can derive \eq{mhagcp3} directly from the scalar potential expressed in terms of the GCP$3$ basis parameters, as expected.
We have similarly verified that the masses of the other Higgs scalars computed in the GCP$3$ basis match those obtained in the $\ug\otimes\Pi_2$ basis.

As a final check of our computations, one can verify that $Y_2$, $Z_1,\ldots,Z_4$, $|Z_5|$, $|Z_6|$, and $Z_5^* Z_6^2$ are invariant quantities that are independent of the choice of the scalar field basis.   
For example, starting from the GCP$3$ basis and transforming to the Higgs basis,
\beqa
&& Z_5=\lambda_5^\prime e^{-2i\xi^\prime}\bigl(\cos\xi^\prime +ic_{2\beta^\prime}\sin\xi^\prime \bigr)^2\,,\label{zee5}\\
&& Z_6=-Z_7=i\lambda_5^\prime s_{2\beta^\prime}\sin\xi^\prime e^{-i\xi^\prime}\bigl(\cos\xi^\prime+ic_{2\beta^\prime}\sin\xi^\prime\bigr)\,.\label{zee6}
\eeqa
Note that CP is conserved in light of the relation,
\beq \label{fivesix}
Z^2_6=-\lambda_5^\prime s^2_{2\beta^\prime}\sin^2\xi^\prime Z_5\,,
\eeq
which implies that $\Im(Z_5^* Z_6^2)=0$. Hence, \eqs{zee5}{fivesix} yield,
\beq
Z_5^* Z_6^2=-\lambda_5^{\prime\,3} s_{2\beta^\prime}^2\sin^2\xi^\prime(1-s_{2\beta^\prime}^2\sin^2\xi^\prime)^2
=\tfrac18\lambda^3(1-R)^3 c_{2\beta}^2 s^4_{2\beta}\,,
\eeq
in agreement with \eqs{zeefive}{zeeseven}.  One can check that all the other invariant quantities also yield the same values in the GCP$3$ and $ \ug\otimes\Pi_2$ basis.

Likewise, one can invert the transformations above and obtain the $\ug\otimes\Pi_2$ basis parameters starting from the GCP$3$ basis parameters.
For completeness, these results are summarized below.  First, the coefficients of the quartic terms of the scalar potential are given by,
\beqa
\lambda&=& \lambda^\prime -\lambda_5^\prime\,,\\
\lambda_3&=& \lambda_3^\prime +\lambda_5^\prime\,,\\
\lambda_4&=& \lambda_4^\prime +\lambda_5^\prime\,,\\
\lambda R&=&\lambda^\prime+\lambda_5^\prime\,,\\
\lambda_5&=&\lambda_6=\lambda_7=0\,.
\eeqa
Next, the corresponding soft-breaking squared-mass parameters are:
\beqa
&& m^2_{11}=\half (m_{11}^{\prime\,2}+m_{22}^{\prime\,2})-\Im m_{12}^{\prime\,2}\,,  \\
&& m^2_{22}=\half (m_{11}^{\prime\,2}+m_{22}^{\prime\,2})+\Im m_{12}^{\prime\,2}\,, \\
&& m^2_{12}=\Re m_{12}^{\prime\,2}-\half i (m_{22}^{\prime\,2}-m_{11}^{\prime\,2})\,.
\eeqa
Finally, the complex vevs in the $\ug\otimes\Pi_2$ basis are given by
\beq
v_1=\frac{1}{\sqrt{2}}\bigl(v'_1+iv'_2e^{i\xi^\prime}\bigr)\,,\qquad\quad v_2^\prime=\frac{i}{\sqrt{2}}\bigl(v'_1-iv'_2 e^{i\xi^\prime}\bigr)\,.
\eeq
Hence,
\beq
\tan^2\beta=\left|\frac{v_2}{v_1}\right|^2=\frac{v_1^{\prime\,2}+v_2^{\prime\,2}+2v_1 v_2\sin\xi}{v_1^{\prime\,2}+v_2^{\prime\,2}-2v_1 v_2\sin\xi}=\frac{1+s_{2\beta^\prime}\sin\xi^\prime}{1- s_{2\beta^\prime}\sin\xi^\prime}\,.
\eeq
which implies that 
\beq \label{sintwob2}
s_{2\beta}^2=1-s_{2\beta^\prime}^2\sin^2\xi^\prime\,.
\eeq

The relative phase of the vevs in the U(1)$\otimes\Pi_2$ basis, denoted by $\xi$, is given by,
\beq \label{sin2xi2}
\sin\xi=\frac{c_{2\beta^\prime}}{(1-s_{2\beta^\prime}^2\sin^2\xi^\prime)^{1/2}}\,,\qquad\quad \cos\xi=\frac{s_{2\beta^\prime}\cos\xi^\prime}{(1-s_{2\beta^\prime}^2\sin^2\xi^\prime)^{1/2}}\,,\qquad\quad 
\eeq
Hence \eqs{sintwob2}{sin2xi2} yield,
\beq
s_{2\beta}\sin\xi=c_{2\beta^\prime}\,.
\eeq
Of course, once the $\ug\otimes\Pi_2$ basis parameters have been derived, one can perform one further rephasing to remove the complex phase $\xi$ (thereby setting $\xi=0$).

\section{Singular value decomposition of a real \texorpdfstring{$2\times 2$}{2x2} matrix}
\label{app:svd}
\renewcommand{\theequation}{B.\arabic{equation}}
\setcounter{equation}{0}

The material presented in this Appendix is taken from Ref.~\cite{Haber:2020wco}.

For any real $n\times n$ matrix $M$,
real orthogonal $n\times n$ matrices $L$ and $R$ exist such that
\beq
\label{real:svd}
L^{\T} M R= M_D={\rm diag}(m_1,m_2,\ldots,m_n),
\eeq
where the $m_k$ are real and nonnegative.  This corresponds to the singular value decomposition of~$M$ restricted to the space of real matrices. 

The singular value decomposition of a general $2\times 2$ real matrix
can be performed fully analytically.
Let us consider the non-diagonal real matrix,
\beqa
M = \left(\begin{array}{cc}
            a      &  \quad  c  \\
      \tilde{c}    &\quad    b
          \end{array}\right)\,,
\eeqa
where at least one of the two quantities $c$ or $\tilde{c}$ is non-vanishing.  The real singular value decomposition of $M$ is
\beq \label{LMR2r}
L^{\T} MR=\begin{pmatrix} m_1 & \,\,\, 0 \\ 0 & \,\,\, m_2\end{pmatrix}\,,
\eeq
where $L$ and $R$ are real $2\times 2$ orthogonal matrices and $m_1$ and $m_2$ are nonnegative.
In general, one can parameterize $L$ and $R$ in \eq{LMR2r} by %
\beq
 L 
     = \left(\begin{array}{cc}
        \phm \cos\theta_L   &     \sin\theta_L  \\
     - \sin\theta_L &  \cos\theta_L
             \end{array}\right)\begin{pmatrix} 1 & \,\,\,0 \\ 0 & \,\,\, \varepsilon_L\end{pmatrix}\,,\qquad\quad
 R 
     = \left(\begin{array}{cc}
        \phm \cos\theta_R   &   \sin\theta_R  \\
     - \sin\theta_R & \cos\theta_R
       \end{array}\right)\begin{pmatrix} 1 & \,\,\, 0 \\ 0 & \,\,\, \varepsilon_R\end{pmatrix}\,,\label{eq:OLR}
\eeq
where $-\half\pi<\theta_{L,R} \leq \half\pi$, and $\varepsilon_{L,R}=\pm 1$. Note that $\det L=\varepsilon_L$ and
$\det R=\varepsilon_R$, which implies that $\varepsilon_L\varepsilon_R\det M=m_1 m_2$.  Since $m_1$, $m_2\geq 0$, it follows that $\sgn(\det M)=\varepsilon_L\varepsilon_R$.   Thus, only the product of $\varepsilon_L$ and $\varepsilon_R$ is fixed by \eq{LMR2r}.

The diagonal elements of $L^{\T}MR$ 
can be determined by taking the positive square root of the nonnegative
eigenvalues, $m^2_{1,2}$, of the real orthogonal matrix $M^{\T}M$,
\beq \label{msquaredreal}
m^2_{1,2}= \tfrac{1}{2}\bigl[a^2+b^2+c^2+\tilde{c}^2 \mp \Delta\bigr]\,,
\eeq
in a convention where $0\leq m_1\leq m_2$ (i.e., $\Delta\geq 0$), with
\beqa
\Delta &\equiv &\bigl[(a^2-b^2-c^2+\tilde{c}^2)^2
                    +4(ac+b\tilde{c})^2\bigr]^{1/2}\nonumber \\[6pt]
                    &=& \bigl[(a^2+b^2+c^2+\tilde{c}^2)^2
                    -4(a b - c \tilde{c})^2\bigr]^{1/2}\,. \label{Deltasqlong2}
\eeqa
Note that
\beq \label{app:sumprod}
m_1^2+m_2^2=a^2+b^2+c^2+\tilde{c}^2\,,\qquad\quad m_1 m_2=\varepsilon_L\varepsilon_R(ab-c\tilde{c})\,.
\eeq
Moreover, $m_1=m_2$ if and only if $a=\pm b$ and $c=\mp\tilde{c}$, which imply that $ac + b \tilde{c}=0$ and $\Delta=0$.  

We first assume that $m_1\neq m_2$.  Then, if we rewrite \eq{LMR2r} in the form $MR=LM_D$, where $M_D\equiv{\rm diag}(m_1\,,\,m_2)$, then we immediately obtain,
\beqa
m_1 \cos\theta_L&=&a\cos\theta_R-c\sin\theta_R\,,\qquad\quad \varepsilon_L\varepsilon_R m_2\sin\theta_L=a\sin\theta_R+c\cos\theta_R\,,\label{monemtwo}\\
m_1\sin\theta_L&=& b\sin\theta_R-\tilde{c}\cos\theta_R\,,\qquad\quad\,
 \varepsilon_L\varepsilon_R  m_2\cos\theta_L=\tilde{c}\sin\theta_R+b\cos\theta_R\,.\label{monemtwo2}
 \eeqa
 It follows that 
\beq \label{mcs}
m_1^2\cos^2\theta_L+m_2^2\sin^2\theta_L=a^2+c^2\,,\qquad\quad 
m_1^2\sin^2\theta_L+m_2^2\cos^2\theta_L=b^2+\tilde{c}^2\,.
\eeq
Subtracting these two equations, and employing \eq{Deltasqlong2} yields,
\beq \label{c2LR}
\cos 2\theta_L=\frac{b^2-a^2-c^2+\tilde{c}^2}{\Delta}\,,\qquad\quad 
\cos 2\theta_R=\frac{b^2-a^2+c^2-\tilde{c}^2}{\Delta}\,.
\eeq
In obtaining $\cos 2\theta_R$, it is sufficient to note that \eqst{monemtwo}{mcs} are valid under the interchange of $c\leftrightarrow\tilde{c}$ and the interchange of the subscripts $L\leftrightarrow R$.\footnote{One can verify this by rewriting \eq{LMR2r} in the form $L^{\T}M=M_DR^{\T}$, which yields equations of the form given by \eqs{monemtwo}{monemtwo2} with $c\leftrightarrow\tilde{c}$ and the interchange of the subscripts $L\leftrightarrow R$.  Note that $\Delta$ and hence $m_{1,2}^2$ are unaffected by these interchanges.\label{fneleven}}

We can also use \eqs{monemtwo}{monemtwo2} to obtain,
\beqa
m_1^2\cos\theta_L\sin\theta_L=(a\cos\theta_R-c\sin\theta_R)(b\sin\theta_R-\tilde{c}\cos\theta_R)\,,\\
m_2^2\cos\theta_L\sin\theta_L=(a\sin\theta_R+c\cos\theta_R)(\tilde{c}\sin\theta_R+b\cos\theta_R)\,.
\eeqa
Subtracting these two equations yields
\beq \label{s2LR}
\sin 2\theta_L=\frac{2(a\tilde{c}+bc)}{\Delta}\,,\qquad\quad \sin 2\theta_R=\frac{2(ac+b\tilde{c})}{\Delta}\,,
\eeq
after again noting the symmetry under $c\to\tilde{c}$ and the interchange of the subscripts $L\leftrightarrow R$.

Thus, employing \eqs{c2LR}{s2LR}, we have succeeded in uniquely determining the angles $\theta_L$ and $\theta_R$ (where $-\half\pi<\theta_{L,R}\leq\half\pi$).   As noted below \eq{eq:OLR}, the individual signs $\varepsilon_L$ and $\varepsilon_R$ are not separately fixed (implying that one is free to set one of these two signs to $+1$);  only the product $\varepsilon_L\varepsilon_R =\sgn(\det M)$ is determined by the singular value decomposition of $M$.

A useful identity can now be derived that exhibits a direct relation between the angles $\theta_L$ and~$\theta_R$.  First, we note two different trigonometric identities for the tangent function,
\beqa
\tan\theta_L&=&\frac{1-\cos 2\theta_L}{\sin 2\theta_L} =\frac{m_2^2-m_1^2-b^2+a^2+c^2-\tilde{c}^2}{2(a\tilde{c}+bc)}=\frac{a^2+c^2-m_1^2}{a\tilde{c}+bc}\,,\label{tanid1}\\[6pt]
\tan\theta_R&=&\frac{\sin 2\theta_R}{1+\cos 2\theta_R}=\frac{2(ac+b\tilde{c})}{m_2^2-m_1^2+b^2-a^2+c^2-\tilde{c}^2}=\frac{ac+b\tilde{c}}{m_2^2-a^2-\tilde{c}^2}\,,\label{tanid2}
\eeqa
where we have made use of \eqss{app:sumprod}{c2LR}{s2LR}.  It then follows that
\beq \label{tLtR}
\frac{\tan\theta_L}{\tan\theta_R}=\frac{(a^2+c^2-m_1^2)(m_2^2-a^2-\tilde{c}^2)}{(a\tilde{c}+bc)(ac+b\tilde{c})}\,.
\eeq
The numerator of \eq{tLtR} can be simplified with a little help from \eq{app:sumprod} as follows,
\beqa
 (a^2+c^2-m_1^2)(m_2^2-a^2-\tilde{c}^2)&=&a^2(m_1^2+m_2^2)+c^2 m_2^2-\tilde{c}^2m_1^2-(a^2+c^2)(a^2+\tilde{c}^2)-m_1^2m_2^2 \nonumber \\[6pt]
&=& a^2(a^2+b^2+c^2+\tilde{c}^2)-(a^2+c^2)(a^2+\tilde{c}^2) \nonumber \\
&& \qquad\qquad +c^2 m_2^2+\tilde{c}^2 m_1^2-(ab-c\tilde{c})^2 \nonumber \\[6pt]
&=&  c^2 m_2^2+\tilde{c}^2 m_1^2+2(ab-c\tilde{c})c\tilde{c} =(cm_2+\varepsilon_L\varepsilon_R\tilde{c}m_1)^2\,.
\eeqa
Likewise, the denominator of \eq{tLtR} can be simplified as follows,
\beqa
(a\tilde{c}+bc)(ac+b\tilde{c})
&=& (ab-c\tilde{c})(c^2+\tilde{c}^2)+c\tilde{c}(a^2+b^2+c^2+\tilde{c}^2)\nonumber \\[6pt]
&=&\varepsilon_L\varepsilon_R m_1 m_2(c^2+\tilde{c}^2)+c\tilde{c}(m_1^2+m_2^2) \nonumber \\[6pt]
&=&(cm_2+\varepsilon_L\varepsilon_R\tilde{c}m_1)(\tilde{c}m_2+\varepsilon_L\varepsilon_R cm_1)\,.
\eeqa
Hence, we end up with a remarkably simple result,
\beq \label{amusingid}
\frac{\tan\theta_L}{\tan\theta_R}=\frac{cm_2+\varepsilon_L\varepsilon_R\tilde{c}m_1}{\tilde{c}m_2+\varepsilon_L\varepsilon_R cm_1}\,.
\eeq

The case of $m_1=0$, which arises when $\det M =ab - c\tilde{c}=0$, is noteworthy.
It then follows that $\Delta=(a^2+\tilde{c}^2)(b^2+\tilde{c}^2)/\tilde{c}^2$ [cf.~\eq{Deltasqlong2} with $c=ab/\tilde{c}$] and,\footnote{In deriving \eq{rcoslimit}, we have assumed that 
$\tilde{c}\neq 0$.   If $\tilde{c}=0$ then one can repeat the calculation by dividing the equation $ab-c\tilde{c}=0$ by a different nonzero parameter.  For example, if $c\neq 0$ then
$\Delta=(a^2+c^2)(b^2+c^2)/c^2$, in which case it follows that $\tan\theta_L=c/b$ and $\tan\theta_R=a/c$.   The other cases can be similarly treated.}
\beq \label{rcoslimit}
\tan\theta_L=\frac{a}{\tilde{c}}\,,\qquad\quad \tan\theta_R=\frac{\tilde{c}}{b}\,.
\eeq
In particular, after using $ab=c\tilde c$, it follows that
\beq
\frac{\tan\theta_L}{\tan\theta_R}=\frac{c}{\tilde{c}}\,,\qquad\text{for $m_1=0$}.
\eeq
This is indeed the correct limit of 
\eq{amusingid} when $m_1=0$, as expected.
In this case, the signs $\varepsilon_L$ and $\varepsilon_R$ are arbitrary, and  one can choose $\varepsilon_L=\varepsilon_R=1$ without loss of generality.

For completeness, we note that the
case of $m\equiv m_1=m_2\neq 0$ must be treated separately.  In this case, $a=\pm b$ and $c=\mp\tilde{c}$, which yields $m=(a^2+c^2)^{1/2}$.  Since
\eq{LMR2r} implies that $MR=mL$,
one can take $R$ to be an arbitrary $2\times 2$ real orthogonal matrix.   Using \eq{eq:OLR}, the matrix $L$ is now determined,
\beq \label{realsvd}
\cos\theta_L=\frac{a\cos\theta_R-c\sin\theta_R}{\sqrt{a^2+c^2}}\,,\qquad\quad
\sin\theta_L=\pm\left(\frac{c\cos\theta_R+a\sin\theta_R}{\sqrt{a^2+c^2}}\right)\,,
\eeq
subject to the constraint $\varepsilon_L\varepsilon_R=\pm 1$ that determines the sign factor appearing in the expression for $\sin\theta_L$ given in \eq{realsvd}.

\section{Top quark mixing revisited}
\label{twostep}
\renewcommand{\theequation}{C.\arabic{equation}}
\setcounter{equation}{0}

In \sects{sec:BreakDiscrete}{section:TopQuarkMixing}, we determined the mixing of the top quark and its vector-like partners by a two step procedure.  In the first step, the effects of electroweak symmetry breaking were ignored.  The Yukawa interactions and mass terms were then obtained in terms of the mass eigenstate fields, $\bar{x}_0$ and $\ol{X}_0$, resulting in \eq{twover}.
In the second step,  the fields $\bar{x}_0$ and $\ol{X}_0$ were treated as interaction eigenstates, and the new mass eigenstates were determined when the Higgs field vevs 
are taken into account.  

One could have performed the same analysis in one step by treating the effects of electroweak symmetry breaking from the beginning by employing 
$\Phi_i^0 = v_i/\sqrt{2}+\overline{\Phi}_i\llsup{0}$ (for $i=1,2$)
in \eq{minuslag}.  In this case, the interaction eigenstates are given by $\hat\chi_i=(u\quad U)$ and $\hat\eta^j=(\bar{u}\quad \ol{U})$ as in \sect{sec:BreakDiscrete}, whereas the mass matrix given in \eq{CalMat} is modified as follows,
\beq \label{calm}
\mathcal{M}=\begin{pmatrix} Ys_\beta & \,\,\, Yc_\beta \\  M_u &\,\,\, M_U\end{pmatrix}\,,
\eeq
where $Y\equiv y_t v/\sqrt{2}$.
Using the results of Appendix~\ref{app:svd}, one can then directly determine the fermion masses and mixing.

As expected, \eq{msquaredreal} yields 
the squared-masses of the two Dirac fermions previously given in \eq{msquared}.
The singular value decomposition of $\mathcal{M}$ [cf.~\eq{calm}] involves 
two mixing angles, denoted below by $\theta^\prime_L$ and~$\theta^\prime_R$, that are given by
\beqa
\sin 2\theta^\prime_L&=&\frac{2YMc_{\beta-\gamma}}{m_T^2-m_t^2}\,,\qquad\qquad \cos 2\theta^\prime_L=\frac{M^2-Y^2}{m_T^2-m_t^2}\,, \label{tL1}\\[8pt]
\sin 2\theta^{\prime}_R&=&\frac{Y^2 s_{2\beta}+M^2 s_{2\gamma}}{m_T^2-m_t^2}\,,\qquad\, \cos 2\theta^\prime_R=\frac{Y^2 c_{2\beta}+M^2 c_{2\gamma}}{m_T^2-m_t^2}\,,\label{tL2}
\eeqa
which uniquely determine the mixing angles modulo~$\pi$.  
Note that in addition to the mixing angles $\theta^\prime_L$ and $\theta^\prime_R$, the matrices $L$ and $R$ given in \eq{eq:OLR} depend on $\varepsilon_L$ and $\varepsilon_R$, where $\varepsilon_L\varepsilon_R={\rm \sgn}(YMs_{\beta-\gamma})$, which is the same result obtained below \eq{phitwo}.

All the results of this Appendix could have been anticipated given that $\theta^\prime_L$ and $\theta^\prime_R$ are related to the mixing angles $\theta_L$ and $\theta_R$ of \sect{section:TopQuarkMixing} as follows,
\beq \label{anticipate}
\theta^\prime_L=\theta_L\,,\qquad\quad \theta^\prime_R=\gamma+\theta_R\,.
\eeq
Indeed, using \eq{anticipate} and the results of  \eqs{phione}{phitwo}, one can recover the expressions presented in \eqs{tL1}{tL2}.
Finally, \eqs{tanid}{anticipate} yield, 
\beq \label{remarkable2}
\tan\theta^\prime_L=\varepsilon_L\varepsilon_R\frac{m_T}{m_t}\tan(\theta^\prime_R-\gamma)\,.
\eeq
By employing the identities for the tangent function given in \eqs{tanid1}{tanid2}, one can derive \eq{remarkable2} directly starting from \eqs{tL1}{tL2}.

\end{appendices}

	\let\enquote\letmacundefined %Fixes bib error
\bibliographystyle{utphys}
\bibliography{GCP3note}
\end{document}